\documentclass[12pt,3p]{elsarticle}
\usepackage{setspace}
\usepackage{amssymb}
\usepackage{gensymb}
\usepackage{rotating}
\usepackage{tabularx} 
\usepackage{float}
\usepackage{mathrsfs}
\usepackage{booktabs}
\usepackage{url}
\usepackage[displaymath]{lineno}
\usepackage{hyperref}
\hypersetup{colorlinks,linkcolor={[RGB]{0 0 139}},citecolor={[RGB]{0 0 139}},urlcolor={[RGB]{0 0 139}}}
\usepackage{amsmath}
\usepackage{subfigure}
\usepackage{multicol}
\usepackage{wrapfig}
\usepackage{breqn}
\usepackage{array}
\usepackage[dvipsnames]{xcolor}
\biboptions{sort&compress}
\usepackage{nicematrix} 
\usepackage[normalem]{ulem}
\usepackage{soul,xcolor}
\setstcolor{blue}

\makeatletter
\def\ps@pprintTitle{%
	\let\@oddhead\@empty
	\let\@evenhead\@empty
	\let\@oddfoot\@empty
	\let\@evenfoot\@oddfoot
}
\makeatother
\begin{document}
	
	\begin{frontmatter}
		
		\title{\vspace*{0.8cm}A general pressure equation based method for incompressible two-phase flows }

		\author[1]{Hormuzd Bodhanwalla}
		\address[1]{School of Mechanical Sciences, Indian Institute of Technology Goa,  Farmagudi, Goa-403401, India}
		
		\cortext[cor1]{Corresponding author, Email: sudhakar@iitgoa.ac.in}
		
		\author[1]{Dheeraj Raghunathan}

		\author[1]{Y. Sudhakar\corref{cor1}}
		
		\begin{abstract}
			We present a fully-explicit, iteration-free, weakly-compressible method to simulate immiscible incompressible two-phase flows. To update pressure, we circumvent the computationally expensive Poisson equation and use the general pressure equation which is solved explicitly. In addition, a less diffusive algebraic volume-of-fluid approach is used as the interface capturing technique and in order to facilitate improved parallel computing scalability, the technique is discretised temporally using the operator-split methodology. Our method is fully-explicit and stable with simple local spatial discretization, and hence, it is easy to implement. Several two- and three-dimensional canonical two-phase flows are simulated. The qualitative and quantitative results prove that our method is capable of accurately handling problems involving a range of density and viscosity ratios and surface tension effects.
   	\end{abstract}
		
		
	\end{frontmatter}
	
	
	\section{Introduction}
	\label{sec1}
	Many practical applications of fluid mechanics involve two-phase flows in which two immiscible fluids of different densities and viscosities interact and generate complex flow patterns. To accurately simulate such flows, capturing the complex evolution of the interface topology is imperative. The common approaches followed are the volume-of-fluid (VOF)~\cite{hirt1981,saincher2015,bodhanwalla2022}, level-set~\cite{sussman1994,gada2011,gada2012}, and phase-field methods~\cite{allen1979,mirjalili2020,dadvand2021}. The VOF and level-set methods are sharp interface approaches where there exists a jump in the material properties across the interface. On the other hand, in the case of the phase-field method, the interface is assumed to be of a finite thickness across which the properties vary rapidly but smoothly. All these approaches fall under the category of one-fluid formulation~\cite{tryggvason2011}, which is widely used in the simulation of incompressible two-phase flows.
	The numerical treatment of the incompressible Navier-Stokes~(INS) equations requires solving a Poisson equation for pressure at each time step to enforce mass conservation~\cite{ferziger2002}. Although the Poisson equation provides accurate flow fields, it requires a linear algebra solver due to its elliptic nature. This step makes the solver computationally expensive and offers limited parallel computing scalability.  
	
	\subsection{Overview of weakly compressible approaches}
	The scalability issue described above can be overcome by relaxing the incompressibility constraint. There are three such  {\it weakly-compressible} alternatives to overcome this limitation:  Lattice Boltzmann methods, artificial compressibility methods, and techniques based on low Mach number~(Ma) pressure equation. While the first two categories of methods are well-established, the third is a recent development in the field.

	\paragraph{Lattice Boltzmann method~(LBM)} ~ \\
	Due to the attractive features of being explicit in time and instinct kinetic nature,  LBM has been used to develop models for solving the two-phase flow problems~\cite{gunstensen1991,he1999}. However, due to the numerical instability, their application is restricted to low- and moderate-density ratios. Several attempts have been made to develop LBM capable of handling large density ratios~\cite{inamuro2004,wang2015}. Still, due to the well-known fact that such methods utilize many distribution functions and consume large memory, the technique becomes computationally expensive for complex multiphase problems. 

	\paragraph{Artificial compressibility method~(ACM)} ~ \\
	The main idea of the ACM, pioneered by Chorin~\cite{chorin1997}, is to replace the velocity divergence condition with an artificial hyperbolic equation for pressure, which is computationally more convenient to solve than the elliptic equation.  The original ACM is proposed to solve steady flows only and is subsequently extended to simulate unsteady flows by introducing a dual time-stepping procedure~\cite{soh1988,malan2002}.  Kelecy and Pletcher~\cite{kelecy1997} and Shah and Yuan~\cite{shah2011} employed such dual time-stepping ACM techniques to solve two-phase flows.  Although such methods produce time-accurate results,  the necessity of convergence of pseudo-time iterations within each physical time step makes them computationally inefficient. 

    \paragraph{Methods based on low Mach number~\normalfont (Ma) \textit{pressure equation}} ~\\
	In recent years, there has been a surge of interest in obtaining an evolution equation for pressure based on thermodynamic considerations of the compressible Navier-Stokes equations. In contrast to an artificial pressure evolution equation used in ACM, such methods rely on the equation for pressure derived based on the physical arguments at low Ma. Methods belonging to this category retain the primary advantage of ACMs that an explicit time integration can be employed; moreover, the use of a physics-based pressure equation eliminates the need for dual time-stepping. Thus, such methods are computationally more efficient than dual-time schemes for simulating unsteady fluid flows. Three such approaches are proposed in the literature:
	\begin{itemize}
	\item kinetically reduced local Navier Stokes equations~(KRLNS)
	\item entropically damped artificial compressibility method~(EDAC)
	\item a method based on general pressure equation~(GPE)
	\end{itemize}
In the following text, we briefly discuss these three methods, and emphasize their status in simulating two-phase flows.

	KRLNS uses a grand potential as a thermodynamic variable to derive a low Ma description from the compressible Navier-Stokes equations~\cite{ansumali2005}. Although this method solves an evolution equation for the grand potential instead of pressure, due to the algorithmic similarity with the relevant methods,  we grouped it here. Simulations of standard test cases for single-phase flows prove the applicability of KRLNS to unsteady incompressible fluid flows~\cite{karlin2006,hashimoto2013,hashimoto2015,hashimoto2018}. However, the simulation of two-phase flows using this equation is yet to be explored.

	Using entropy to damp the acoustic pressure waves, Clausen~\cite{clausen2013} proposed a parabolic equation for pressure, and the resulting framework is called EDAC. The method is shown to provide accurate results for laminar and turbulent single-phase flows~\cite{delorme2017, kajzer2018}. Kajzer and Pozorski~\cite{kajzer2020} developed an EDAC-based diffuse interface approach to simulate two-phase flows. They turned off the pressure diffusion term in the vicinity of the interface to eliminate pressure oscillations. Qualitative comparisons of droplet problems showed that the method captures the complex topological features well. The same authors proposed an improved method~\cite{kajzer2022} to eliminate the drawbacks of~\cite{kajzer2020}. The interface identification is made Ma independent, and the time-step restriction is relaxed. Both qualitative and quantitative investigations of standard test cases are used to demonstrate that the improved method can accurately capture complex two-phase flows.

	Toutant~\cite{toutant2017} derived the GPE using an asymptotic analysis of compressible Navier-Stokes equations. It has been shown that the GPE-based single-phase flow solver can accurately capture the transient incompressible flows~\cite{toutant2018}. Studies aimed at investigating the applicability of GPE~\cite{dupuy2020,shi2020} concluded that GPE can provide solutions very close to that of INS, even for wall-bounded turbulent flows.   Huang~\cite{huang2020} proposed a method, based on GPE, to simulate two-phase flows. The interface evolution and the surface tension effects are modelled using a phase field approach. The method produces results of two-phase flows that are in good agreement with INS solvers and LBM.
	
	Another recent weakly compressible approach that solves an evolving pressure projection equation to simulate two-phase flows is proposed by Yang and Aoki~\cite{yang2021}. Although they solve a low Ma pressure equation, this approach involves, within each time step, a few iterations of the pressure evolution equation and the associated velocity correction. The inner iterations help in alleviating the acoustic effects. The method is successfully validated against complex two-phase flow problems. 
  
  \subsection{Novelty and a brief overview of the present methodology}
	Methods based on the low Ma pressure equation represent the state-of-the-art weakly compressible techniques to simulate incompressible flows. However, the development of numerical methods to solve two-phase flows under this framework is a fresh topic. To the best of our knowledge, only three research works, two using EDAC~\cite{kajzer2020,kajzer2022} and one (unpublished) using GPE~\cite{huang2020}, reported two-phase flow simulations using low Ma pressure equations. All three of them used a phase-field approach to model the interface dynamics. While the phase-field approach has certain advantages, additional numerical parameters related to the interface width and compression need to be set appropriately, and these parameters noticeably influence the solution accuracy and solver stability. Moreover, existing works introduce additional complicated procedures to achieve accurate results: Kajzer and Pozorski~\cite{kajzer2020,kajzer2022} used a switch parameter and used a distinct discretization scheme in the interface region when compared to the bulk of fluids; Huang~\cite{huang2020} had to use second viscosity term to eliminate the checkerboard oscillations and employed a special averaging procedure to compute pressure gradient without which solution stopped abruptly. Despite these additional procedures,  the method of~\cite{huang2020} can only deal with low density and viscosity ratios. This paper proposes a simple and robust {\it fully-explicit} GPE-based method to simulate complex two-phase flows with large viscosity and density ratios, using VOF to capture the interface.
	
	Due to the ease of implementation and extension to three-dimension, the algebraic VOF scheme is used in the present work. Various algebraic VOF schemes can be found in the literature~\cite{ubbink1999,arote2020,anghan2021}, each having its own advantages. In this work, we employ the modified switching technique for advection and capturing of surfaces (MSTACS) introduced by Anghan et al.~\cite{anghan2021} since it offers low interface diffusivity. Traditionally, the algebraic VOF schemes employ implicit (unsplit) time discretization (except THINC scheme~\cite{xiao2005}), which makes their application computationally expensive, especially in 3D. To overcome this drawback, Saincher and Sriram~\cite{saincher2022} recently proposed the operator-split (OS) framework for algebraic VOF, which provides a fully-explicit treatment of the volume fraction advection equation resulting in faster computations than the traditional approach.
	
	The contributions of this paper are as follows: (i)~we incorporate the operator-split framework with MSTACS formulation as an interface capturing technique which gives us the benefit of explicit time integration and low diffusivity of the interface, and (ii)~we propose a GPE-based iteration-free method to solve two-phase flows with large density and viscosity ratios without any special treatment in the interface region, or ad-hoc procedure for stabilisation. The combined benefit of GPE that eliminates the pressure Poisson equation, and the OS framework for algebraic VOF enables us to employ a \textit{fully-explicit} solution algorithm.
 	
	\subsection{Structure of this paper}
	The rest of the article is structured as follows: The governing equations and details of the GPE-based two-phase solver with the VOF technique are described in \S~\ref{sec2}.  Details of the discretisation, time integration, and algorithmic aspects are elaborated in \S~\ref{sec3}. The validation of the proposed method against existing literature results is reported in \S~\ref{sec4}. Finally, the conclusions drawn from the analysis are reported in \S~\ref{sec5}.

	\section{Mathematical Model}
	\label{sec2}
	The present method is based on `one fluid formulation', in which a single set of equations governs the behaviour of both fluid phases~\cite{tryggvason2011}. In this section, we discuss the governing equations and the details of algebraic VOF.
	
	\subsection{Governing Equations}
	\label{subsec2.1}
	
	The weakly compressible isothermal Navier-Stokes equations considered in the present work are written as
	\begin{linenomath}
		\begin{align}
			\rho \left(\frac{\partial \mathbf{u}}{\partial t} + \mathbf{u}\cdot \nabla \mathbf{u}\right)  &= -\nabla p + \nabla \cdot \left[\mu \left( \nabla \mathbf{u} + \nabla \mathbf{u}^\intercal \right) \right] + \mathbf{F} \label{eqn:NS_dim},\\
			\frac{\partial p}{\partial t} + \rho {c_s}^2 \left( \nabla \cdot \mathbf{u} \right) &= \frac{1}{\rho} \nabla \cdot \left( \mu \nabla p \right) \label{eqn:gpe_dim},
		\end{align}	
	\end{linenomath}
	where equation \eqref{eqn:NS_dim} represents the well-known momentum equation, and equation \eqref{eqn:gpe_dim} is the GPE derived by Toutant~\cite{toutant2017}, with the assumption of $\gamma=\text{Pr}$~(see ~\ref{appendix_gpeDreivation}). The GPE is derived for single phase flows, and we use the standard one-fluid formulation to simulate two-phase flows. $\mathbf{u}$ and $p$ refer to the velocity vector and pressure, respectively.  $c_s$ is the artificial speed of sound, whose value should be sufficiently large to replicate the behaviour of incompressible flows. $\rho$ and $\mu$ refer to the fluid's mixture density and dynamic viscosity. They are evaluated as
	\begin{linenomath}
		\begin{align}
			\begin{split}
				\rho &= C\rho_1 + \left(1-C \right)\rho_2,  \\\
				\mu &= C\mu_1 + \left(1-C \right)\mu_2, 
				\label{eqn:prop_dim}
			\end{split}
		\end{align}
	\end{linenomath}
	where the subscripts 1 and 2 represent the primary and secondary phases, respectively. The volume fraction $C$ is a Heaviside function that jumps across the interface between the two phases.
	The body force $\mathbf{F}$ appearing in equation~\eqref{eqn:NS_dim} is given as
	\begin{linenomath}
		\begin{align}
			\mathbf{F} = \mathbf{F_g} + \mathbf{F_s},
			\label{eqn:forces}
		\end{align}
	\end{linenomath}
	where $\mathbf{F_g}=\rho \mathbf{g}$ is the gravitational force and $\mathbf{F_s}$ is the surface tension force per unit volume given as $\sigma\kappa \hat{\textbf{n}} \delta$.
	$\mathbf{g}$ denotes gravitational acceleration, $\sigma$ is the surface tension coefficient, $\hat{\textbf{n}}$ is the outward unit normal, $\kappa$ is the local curvature of the interface, and $\delta$ is the Dirac delta function.\\
	For completeness, we present the non-dimensional form of the 	GPE (in the case of two-phase flow) below
	\begin{linenomath}
		\begin{align}
			\frac{\partial p}{\partial t} + \frac{\rho^*}{{\text{Ma}}^2} \left( \nabla \cdot \mathbf{u} \right) = \frac{1}{\rho^* \text{Re}} \nabla \cdot \left( \mu^* \nabla p \right),
			\label{eqn:gpe_nondim}
		\end{align}
	\end{linenomath}
	where Re and Ma denote the Reynolds number and the artificial Mach number, respectively. $\rho^*$ and $\mu^*$ refer to the non-dimensional density and viscosity, respectively. Additional parameters ($\gamma$ and $\text{Pr}$) originally present in the GPE do not appear in the above expression, because we set $\gamma=\text{Pr}$ based on~\cite{toutant2018}.
	
	\subsection{Two-phase model}
	\label{subsec2.2}
	Under the VOF framework, the governing equation for the transport of volume fraction $C$ is provided as:
	\begin{linenomath}
		\begin{align}
			\frac{\partial C}{\partial t} + \mathbf{\nabla} \cdot \left( \mathbf{u}C \right)  = C \left( \mathbf{\nabla} \cdot \mathbf{u} \right)
			\label{eqn:vof_transport}
		\end{align}
	\end{linenomath}

	Due to the ease of implementation for 3D applications, the algebraic VOF technique is incorporated. In this method, the interface capturing is accomplished by blending a Compressive Differencing Scheme (CDS) and a High Resolution (HR) scheme. The switching between the two depends on the interface orientation with respect to the flow direction. Traditionally, in the algebraic VOF schemes, the Crank-Nicolson discretization is
	used for the volume fraction transport equation~\cite{ubbink1999,anghan2021}. However, we use the operator-split technique, recently introduced by Saincher and Sriram~\cite{saincher2022}, to make our algorithm free of any linear algebra solver. The OS is a multi-stage approach with the number of stages equal to the problem's spatial dimensions. For a 3D scenario, the volume fraction is updated as follows:	
	\begin{linenomath}
		\begin{align}
			\begin{split}
				x-sweep:&  \\\
				&C_P^* = C_P^{(n)} + \frac{\Delta t}{\Delta x} \left(C_w^{(n)}U_w^{(n)} - C_e^{(n)}U_e^{(n)} \right) + {cf}_P^{(n)}\frac{\Delta t}{\Delta x} \left(U_e-U_w\right)^{(n)} \label{eqn:x_sweep} 
			\end{split} \\\
			\begin{split}
				y-sweep:&  \\\
				&C_P^{**} = C_P^* + \frac{\Delta t}{\Delta y} \left(C_s^*V_s^{(n)} - C_n^*V_n^{(n)} \right) + {cf}_P^{(n)}\frac{\Delta t}{\Delta y} \left(V_n-V_s\right)^{(n)} \label{eqn:y_sweep} 
			\end{split} \\\
			\begin{split}
				z-sweep:&  \\\
				&C_P^{{(n+1)}} = C_P^{**} + \frac{\Delta t}{\Delta z} \left(C_b^{**}W_b^{(n)} - C_f^{**}W_f^{(n)} \right) + {cf}_P^{(n)}\frac{\Delta t}{\Delta z} \left(W_f-W_b\right)^{(n)} \label{eqn:z_sweep}
			\end{split}
		\end{align}
	\end{linenomath}
	where $C^*$ and $C^{**}$ are the updated volume fractions at the intermediate stages. The subscripts $e,w,n,s,f,b$ denote the usual finite volume notation to represent the faces of the cell $P$. $(U, V, W)$ represent components of $\mathbf{u}$ in the Cartesian coordinate system. The notation ${cf}_P$ refers to the colour function term that is active only on cells on which $C>0.5$~\cite{weymouth2010}. The sequence of sweeping is changed after every time step~\cite{saincher2022}.
	
	For a successful time advancement of the volume fraction field, it is essential to compute the flux of volume fraction over the cell face. In the algebraic VOF method, this is calculated based on the value of $C$ in the cells straddling the face. Under the formulation of the donor-acceptor scheme given by Ubbink and Issa~\cite{ubbink1999}, the volume fraction at the face is written as:
	\begin{linenomath}
		\begin{align}
			C_{face} = \left(1-\beta_{face}\right)C_D + \beta_{face} C_A
			\label{eqn:C_face}
		\end{align}
	\end{linenomath}
	where $C_D$ and $C_A$ are the volume fractions of the donor and acceptor cells, respectively, determined based on the velocity direction at that face. The notation $\beta$ refers to the weighting factor which determines the contribution of the donor and acceptor cells based on the gradient of volume fraction and is evaluated as:
	\begin{linenomath}
		\begin{align}
			\beta_{face} = \frac{\widetilde{C}_{face}-\widetilde{C}_D}{1-\widetilde{C}_D}
			\label{eqn:B_face}
		\end{align}
	\end{linenomath}
	where $\widetilde{C}_D$ $=({C_D-C_U})/({C_A-C_U})$ is the normalized volume fraction value of the donor-cell with $C_U$ being the volume fraction of the upwind cell. $\widetilde{C}_{face}$ is normalized volume fraction at the cell face which is given as:
	\begin{linenomath}
		\begin{align}
			\widetilde{C}_{face} = \gamma_{face} \left(\widetilde{C}_{face}\right)_{CDS} + \left(1-\gamma_{face}\right) \left(\widetilde{C}_{face}\right)_{HR}
			\label{eqn:C_tilde_face}
		\end{align}
	\end{linenomath}
	which depicts that the volume fraction fluxed over the cell face is computed using a blend of CDS $\left(\left(\widetilde{C}_{face}\right)_{CDS}\right)$ and HR $\left(\left(\widetilde{C}_{face}\right)_{HR}\right)$ scheme (as mentioned earlier) through a blending function ($\gamma_{face}$) that allows smooth switching between the two based on the orientation of the interface. The interface capturing scheme used in the present study that provides the formulation of CDS, HR and blending function is the modified switching technique for advection and capturing of surfaces (MSTACS) introduced by Anghan et al.~\cite{anghan2021}. The significant advantage that MSTACS offers is the low diffusivity over a wide range of Courant numbers, $Cou_{adv}=U\Delta t/\Delta x$. The scheme is formulated as follows:
	
	\begin{align}
		\left(\widetilde{C}_{face}\right)_{CDS} &= 
		\begin{cases}
			min\left(\frac{\widetilde{C}_D}{Cou_{adv}}, 1 \right), & \text{$0 \le \widetilde{C}_D \le 1$ ; $0 < Cou_{adv} \le 1/3$} \\
			min\left(3\widetilde{C}_D, 1 \right),            & \text{$0 \le \widetilde{C}_D \le 1$ ; $1/3 < Cou_{adv} \le 1$} \\
			\widetilde{C}_D,                                 & \text{$\widetilde{C}_D<0$ ; $\widetilde{C}_D>1$ }
		\end{cases} 
	\label{eqn:CDS}
	\\
		\left(\widetilde{C}_{face}\right)_{HR} &= 
		\begin{cases}
			3\widetilde{C}_D,                                & \text{$0 \le \widetilde{C}_D < 1/5$} \\
			0.5+0.5\widetilde{C}_D,                          & \text{$1/5 \le \widetilde{C}_D < 1/2$} \\
			3/8+3/4\widetilde{C}_D,                          & \text{$1/2 \le \widetilde{C}_D < 5/6$} \\
			1,                                               & \text{$5/6 \le \widetilde{C}_D \le 1$} \\
			\widetilde{C}_D,                                 & \text{$\widetilde{C}_D<0$ ; $\widetilde{C}_D>1$ }
		\end{cases}
		\label{eqn:HR}
		\\
		\gamma_{face} &= min\left[\left(cos\theta\right)^4,1\right]
		\label{eqn:gamma}
	\end{align}
	The $\theta$ is the angle between the outward pointing unit vector normal to the interface ($\hat{\textbf{n}}$) and the unit vector connecting the donor and acceptor cell centers ($\hat{\textbf{d}}$). Hence, $\theta$ is defined as:
		\begin{linenomath}
			\begin{align}
			\begin{split}
				\theta &= cos^{-1}|\hat{\textbf{n}}\cdot \hat{\textbf{d}}|,
			\end{split}
		\label{eqn:theta}
		\end{align}
		\end{linenomath}
	where the interface normal is computed using the Parker-Youngs method~\cite{parker1992}. 
	
	Compared with the unsplit methods, in the presence of shearing velocity fields, the operator-split approach leads to imbalanced velocities on each face of the computational cell. This imbalance leads to a violation of the boundedness criterion, implying undershoots ($C<0$) and overshoots ($C>0$) in the volume fraction. Hirt and Nichols~\cite{hirt1981} employed a truncation step in which a cell with $C<0$ is reset to zero, and the cell with $C>1$ is reset to one, which leads to a conservation error. In the present study, the conservative redistribution algorithm by Saincher and Banerjee~\cite{saincher2015} is utilized to restore the boundedness of the volume fraction field, which conserves mass to zero machine accuracy. The algorithm is employed after each directional sweep of the operator-split method. 
	
	The interface capturing method, OS-MSTACS, used in the present work is fully explicit regarding time advancement. The operator-split framework offers better computational efficiency, Courant independence, and high scalability in terms of parallelization than the iterative-based solvers~\cite{saincher2022}. At the same time, the MSTACS scheme ensures the low diffusivity of the interface~\cite{anghan2021}.

	\section{Numerical Details}
	\label{sec3}
	The governing equations described in section \ref{sec2} are discretized using the finite volume framework. For completeness, in this section, we present the full implementation details. Integrating the equations \eqref{eqn:NS_dim} and \eqref{eqn:gpe_dim} over the control volume $d\forall$ bounded by control surface $dS$ and invoking the Gauss-divergence theorem would result in the integral form of equations:
	\begin{linenomath}
		\begin{align}
			\begin{split}
				\iiint\limits_{C\forall} \frac{\partial \mathbf{u}}{\partial t}\,d\forall + \iint\limits_{CS} \mathbf{u}\mathbf{u} \cdot \mathbf{dS}  = -\frac{1}{\rho} \iiint\limits_{C\forall} \nabla p\,d\forall &+ \frac{1}{\rho} \iint\limits_{CS} \left[\mu \left( \nabla \mathbf{u} + \nabla \mathbf{u}^\intercal \right) \right] \cdot \mathbf{dS} \\ 
				&+ \frac{1}{\rho} \iiint\limits_{C\forall} \mathbf{F}\,d\forall
			\end{split}
			\label{eqn:NS_integral}
		\end{align}
	\end{linenomath}
	
	\begin{linenomath}
		\begin{align}
			\begin{split}
				\iiint\limits_{C\forall} \frac{\partial p}{\partial t}\,d\forall + \rho {c_s}^2 \iint\limits_{CS} \mathbf{u} \cdot \mathbf{dS} = \frac{1}{\rho} \iint\limits_{CS} \mu \nabla p \cdot \mathbf{dS}
			\end{split}
			\label{eqn:GPE_integral}
		\end{align}
	\end{linenomath}

	\subsection{Spatial Discretization}
	\label{subsec3.1}
\begin{figure}[H]
		\centering
		\subfigure[]{\includegraphics[width=7cm]{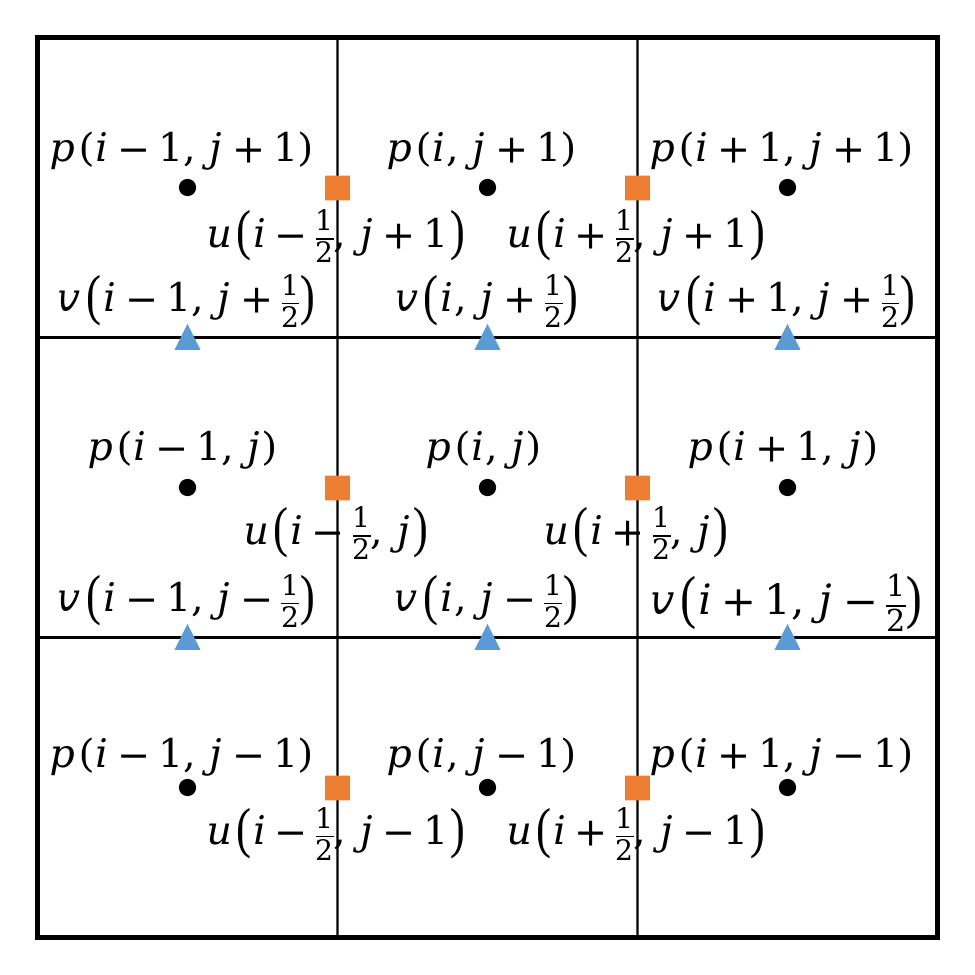}\label{fig_staggered_a}}
		\subfigure[]{\includegraphics[width=4cm]{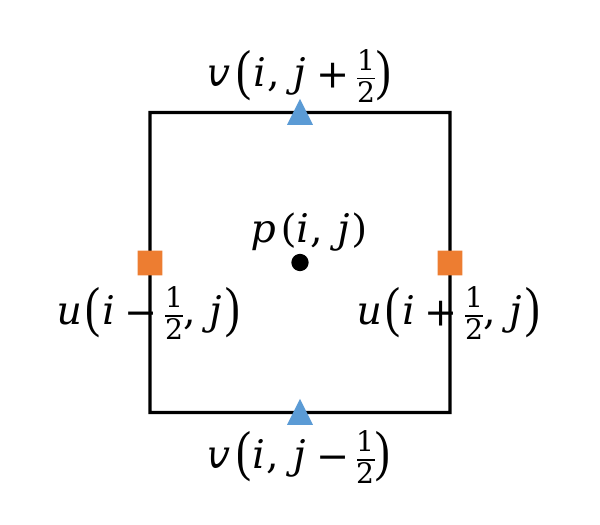}\label{fig_staggered_b}}
		\subfigure[]{\includegraphics[width=4cm]{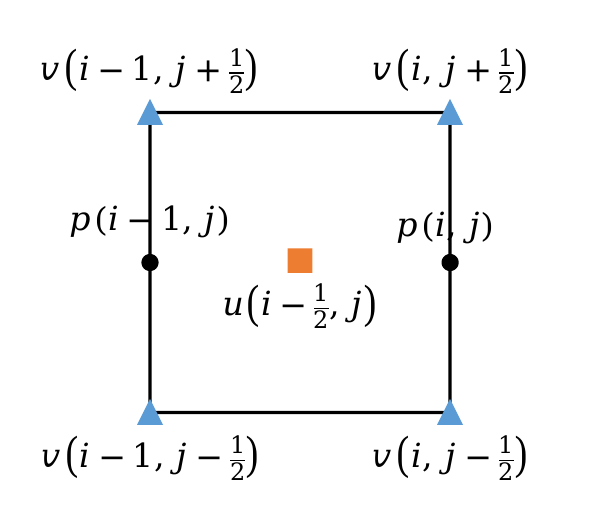}\label{fig_staggered_c}}
		\caption{Staggered grid arrangement. (a) Location of stored quantities, (b) and (c) denote zoomed view of control volume for scalar and $u$, respectively.}
		\label{fig_staggered}
	\end{figure}

	The integral equations are discretized over the staggered grid arrangement to avoid checkerboard oscillations. Here, the velocities are defined at the cell faces and the scalar variables ($p$ and $C$) are determined at the cell centres, as shown in figure \ref{fig_staggered}. For brevity, this section will describe the discretization process for two-dimensional equations; the extension to three-dimensions is straightforward.
	The $x$-component of the advection term in the momentum equations is approximated as:
	\begin{linenomath}
		\begin{align}
			\iint\limits_{CS} u\mathbf{u} \cdot \mathbf{dS} \approx \left[\left(Uu\right)_{i,j} - \left(Uu\right)_{i-1,j}\right] \Delta y + \left[\left(Vu\right)_{i-\frac{1}{2},j+\frac{1}{2}} - \left(Vu\right)_{i-\frac{1}{2},j-\frac{1}{2}}\right] \Delta x
			\label{eqn:adv_comp}
		\end{align}
	\end{linenomath}
	where the notations in the uppercase form represent advecting component of velocity, which is obtained by linear interpolation of the neighboring cell-centered values, as shown below:
	\begin{linenomath}
		\begin{align}
			\begin{split}
				U_{i-1,j} &= \left(u_{i-\frac{1}{2},j}+u_{i-\frac{3}{2},j}\right)/2 \\
				U_{i,j} &= \left(u_{i+\frac{1}{2},j}+u_{i-\frac{1}{2},j}\right)/2\\
				V_{i-\frac{1}{2},j+\frac{1}{2}} &= \left(v_{i,j+\frac{1}{2}}+v_{i-1,j+\frac{1}{2}}\right)/2 \\
				V_{i-\frac{1}{2},j-\frac{1}{2}} &= \left(v_{i,j-\frac{1}{2}}+v_{i-1,j-\frac{1}{2}}\right)/2 
			\end{split}
			\label{eqn:adv_comp_interpol}
		\end{align}
	\end{linenomath}
	The notations in the lowercase form denote the advected velocity component. They can be approximated using various schemes (upwind, central difference, QUICK, TVD, etc.) depending on the advection dominance for the problem at hand. Hence, in the present study, the choice of advection scheme is mentioned for each test case in section \ref{sec4}. 
	
	The viscous term in the momentum equation for 2D space can be expanded as:
	\begin{linenomath}
	\begin{align}
			\mu \left( \nabla \mathbf{u} + \nabla \mathbf{u}^\intercal \right) = \begin{bNiceMatrix}
			2\mu \frac{\partial u}{\partial x} & \mu\left(\frac{\partial v}{\partial x} + \frac{\partial u}{\partial y}\right) \\
			\mu\left(\frac{\partial v}{\partial x} + \frac{\partial u}{\partial y}\right) & 2\mu \frac{\partial v}{\partial y}
		\end{bNiceMatrix}
	\label{eqn:visc_comp}
	\end{align}
	\end{linenomath}
	and the $x$-component of its integral form can be approximated as:
	\begin{linenomath}
		\begin{align}
			\begin{split}
				\frac{1}{\rho} \iint\limits_{CS} \left[\mu \left( \nabla \mathbf{u} + \nabla \mathbf{u}^\intercal \right) \right] \cdot \mathbf{dS}    \approx \frac{1}{\rho} \left\{ \left[\left(2\mu \frac{\partial u}{\partial x}\right)_{i,j} - \left(2\mu \frac{\partial u}{\partial x} \right)_{i-1,j}\right]\Delta y \,\,+ \right. \\
				\left.\left\{\left[\mu\left(\frac{\partial v}{\partial x} +\frac{\partial u}{\partial y} \right)\right]_{i-\frac{1}{2},j+\frac{1}{2}} - \left[\mu\left(\frac{\partial v}{\partial x}+\frac{\partial u}{\partial y} \right) \right]_{i-\frac{1}{2},j-\frac{1}{2}}\right\} \Delta x \right\}
			\end{split}
			\label{eqn:visc_comp_integral}
		\end{align}
	\end{linenomath}
	where the mixture density $\rho$ and viscosity $\mu$ are linearly interpolated during the discretization process. 
	
	The derivatives in the viscous term and the pressure gradient term are discretized using the second-order central difference scheme.
	
	The discretization of the surface tension term should be consistent with the pressure gradient term, so that these two effects are in balance. The surface tension force $\mathbf{F_s}=\sigma\kappa \hat{\textbf{n}} \delta$ is approximated using the Continuum Surface Force (CSF) model~\cite{francois2006}, in which $\hat{\textbf{n}} \delta=\nabla C$. The $x$-component of the surface tension force is discretized as:
	\begin{linenomath}
		\begin{align}
			{\mathbf{F_s}}_{(i-\frac{1}{2},j)} = \sigma \kappa_{i-\frac{1}{2},j} \left( \nabla C \right)_{i-\frac{1}{2},j}
		\label{eqn:surf_tens}
		\end{align}
	\end{linenomath}
	\begin{equation*}
			\text{where $\left( \nabla C \right)_{i-\frac{1}{2},j} = 	\frac{C_{i,j}-C_{i-1,j}}{\Delta x}$ and $\kappa_{i-\frac{1}{2},j}=\frac{\kappa_{i,j}+\kappa_{i-1,j}}{2}$}
	\end{equation*}
	The interface curvature $\kappa$ is calculated using the height function methodology~\cite{cummins2005}.
	
	The velocity divergence term and the diffusion term in the GPE follow the same discretization scheme used for the advection and viscous terms in the momentum equations, respectively. The only difference is that the GPE is discretized over the $p$-control volume. 
	
	\subsection{Time integration}
	\label{subsec3.2}
	As mentioned earlier, the volume fraction transport equation is solved using the operator-split technique, as depicted by Saincher and Sriram~\cite{saincher2022}. For updating the solution of the momentum equation and GPE, we use a three-stage Strong Stability Preserving Runge-Kutta (SSP-RK) scheme~\cite{gottlieb1998}, which is an optimal third order scheme with SSP property. In this method, for a general partial differential equation of the following form,
	\begin{linenomath}
		\begin{align*}
			\frac{\partial \Psi}{\partial t} = L\left(\Psi\right)
		\end{align*}
	\end{linenomath}
	the variable $\Psi^{(n)}$ is updated to $\Psi^{(n+1)}$ in three stages as follows:
	\begin{linenomath}
		\begin{align}
			\begin{split}
			\Psi^{(1)} &= \Psi^{(n)} + \Delta t \, L\left(\Psi^{(n)}\right) \\
			\Psi^{(2)} &= \frac{3}{4}\Psi^{(n)} + \frac{1}{4}\Psi^{(1)} + \frac{1}{4} \Delta t \, L\left(\Psi^{(1)}\right) \\
			\Psi^{(n+1)} &= \frac{1}{3}\Psi^{(n)} + \frac{2}{3}\Psi^{(2)} + \frac{2}{3} \Delta t \, L\left(\Psi^{(2)}\right) \\
		\end{split}
	\label{eqn:RK3}
		\end{align}
	\end{linenomath}
	where $\Psi^{(1)}$ and $\Psi^{(2)}$ are the variables computed at the intermediate stages. SSP schemes for time integration are proposed by Shu and Osher~\cite{shu1988}. They possess large absolute stability region and small error constants~\cite{gottlieb2009}. Moreover, they preserve nonlinear stability properties even with a discontinuous solution. Although they are expensive when compared to the forward Euler method, such methods are widely used in the simulation of incompressible two-phase flows using the weakly compressible framework~\cite{caiden2001,bassano2003,parameswaran2023,kajzer2020,kajzer2022,yang2021}.
		
	\subsection{Solution algorithm}
	\label{subsec3.3}
	For clarity, we summarize the complete solution algorithm to march from time instant $(n)$ to $(n+1)$ below,
	\begin{enumerate}
		\item Compute $C^{(n+1)}$ using $C^{(n)}$ and $\mathbf{u}^{(n)}$ by operator-split techique (eqns.\eqref{eqn:x_sweep}-\eqref{eqn:z_sweep}) using MSTACS (eqns. \eqref{eqn:C_face}-\eqref{eqn:theta}).
		\item Calculate $\rho^{(n+1)}$ and $\mu^{(n+1)}$ using $C^{(n+1)}$ (eqn. \eqref{eqn:prop_dim}).
		\item Compute curvature ($\kappa$) using $C^{(n+1)}$ by height function technique. 
		\item Calculate velocity and pressure at the current time level ($\mathbf{u}^{(n+1)}$ and $p^{(n+1)}$) using third order SSP-RK method (eqn. \eqref{eqn:RK3}), the steps of which are as follows:
			\begin{enumerate} [a.]
				\item Compute $\mathbf{u}^{(1)}$ using $\mathbf{u}^{(n)}$, ${p}^{(n)}$, $\rho^{(n+1)}$ and $\mu^{(n+1)}$.
				\item Evaluate ${p}^{(1)}$ using $\mathbf{u}^{(1)}$ in the divergence term (acting as a source in GPE), ${p}^{(n)}$, $\rho^{(n+1)}$ and $\mu^{(n+1)}$.
				\item Calculate $\mathbf{u}^{(2)}$ using $\mathbf{u}^{(n)}$, $\mathbf{u}^{(1)}$, ${p}^{(1)}$, $\rho^{(n+1)}$ and $\mu^{(n+1)}$.
				\item Compute ${p}^{(2)}$ using $\mathbf{u}^{(2)}$ in the divergence term, ${p}^{(n)}$, ${p}^{(1)}$ $\rho^{(n+1)}$ and $\mu^{(n+1)}$.
				\item Calculate $\mathbf{u}^{(n+1)}$ using $\mathbf{u}^{(n)}$, $\mathbf{u}^{(2)}$, ${p}^{(2)}$, $\rho^{(n+1)}$ and $\mu^{(n+1)}$.
				\item Finally compute ${p}^{(n+1)}$ using $\mathbf{u}^{(n+1)}$ in the divergence term, ${p}^{(n)}$, ${p}^{(2)}$ $\rho^{(n+1)}$ and $\mu^{(n+1)}$.
			\end{enumerate}
	\end{enumerate}
	
	\section{Benchmarking with Literature}
	\label{sec4}
	The present section elucidates the performance of the GPE-based two-phase flow solver proposed in this paper. We simulate two- and three-dimensional canonical two-phase problems involving large density and viscosity ratios, with and without surface tension. Results from our method are rigorously validated with the existing literature qualitatively and quantitatively.

A key computational challenge associated with weakly compressible approaches is the additional time-step restriction introduced by the artificial acoustic speed~($c_s$). This is defined in terms of the acoustic Courant number~($Cou_{acs}=c_s\Delta t/\Delta x$). In order to accurately represent incompressible flows, we need $c_s\gg 1$, but this requires a much smaller time step. Typically, this stability criterion decides the time step of the simulation, and this is more restrictive than the allowable time-step of conventional pressure Poisson equation based approaches~($\Delta t_{INS}$); in general, $\Delta t_{GPE}\approx \text{Ma}\Delta t_{INS}$~\cite{toutant2018}. Hence, on a serial execution, weakly compressible approaches require a longer simulation time. This is a well-known limitation of weakly compressible techniques, including the Lattice Boltzmann method and approaches based on smoothed particle hydrodynamics. However, it is envisaged that these methods can achieve better computational efficiency due to the potential they offer for more effective HPC implementation. The focus of the present work is to propose an accurate GPE-based approach to model two-phase flows. Discussion on the computational benefits of the method with respect to conventional approaches requires careful investigation, which is beyond the scope of the current work.

Analogous to Lattice Boltzmann methods~\cite{huang2020,wang2015} and GPE~\cite{huang2020}, we set the artificial speed of sound, $c_s = \Delta x/(\sqrt{3} \Delta t)$, which corresponds to $Cou_{acs}=1/\sqrt{3}$. However, for problems without surface tension effects, we found that the solver is stable up to $Cou_{acs}\leq 1.25$, and the time-step is chosen appropriately.

 Since the combination of OS framework and MSTACS formulation is used for the first time, it is necessary to validate this algorithm. Hence, to begin with, the ability of the OS-MSTACS scheme to accurately capture the interface is demonstrated.  Then, to verify the correct implementation of the surface tension model, the static droplet test case is simulated. Later, the complete algorithm is tested by simulating several two-phase flow configurations and comparing the results with the published ones. We start with the problem of 2D Rayleigh-Taylor instability, which is a canonical test case with no surface tension effects and produces moderately complex interfacial changes. We simulate the dam break test to validate the solver for a high-density ratio problem. Next, we examine the bubble rise test case that involves a strong influence of surface tension effects. Finally, to extend our study to three dimensions, a 3D Rayleigh-Taylor instability is simulated and benchmarked against the existing results reported in the literature. 
 
	\subsection{Validation of the VOF method}
	\label{subsec4.1}
	To quantitatively demonstrate the accuracy of the interface tracking method~(OS-MSTACS), a classical volume fraction advection test~\cite{liovic2006,saincher2022} case involving time-reversing shearing velocity field is simulated in a 3D environment. The test case is initialized with a sphere of radius $0.15$ centered at ($0.5, 0.75, 0.25$) in a cuboid of size $1.0\times1.0\times2.25$. The patch of sphere is subjected to a shearing velocity field given below as:
	\begin{linenomath}
		\begin{align}
			\begin{split}
				u &= \sin(2\pi y)\sin^2(\pi x)\cos\left(\frac{\pi t}{T}\right),  \\
				v &= -\sin(2\pi x)\sin^2(\pi y)\cos\left(\frac{\pi t}{T}\right), \\
				w &= \left( 1-2 \sqrt{(x-0.5)^2+(y-0.5)^2} \right) ^2 \cos\left(\frac{\pi t}{T}\right),
			\end{split}
			\label{eqn:shearing_velo}
		\end{align}
	\end{linenomath}
	where the cosine term implies time reversal: the sign of the velocity components change at $t=T/2$. The total simulation time $T$ is taken as $9$ and the time step corresponding to Courant number $Cou_{adv}=0.1$ is used. A uniform grid of $128 \times 128 \times 288$ is used to simulate the problem.
	
\begin{figure}[H]
		\centering
		{
			\includegraphics[height=7.5cm]{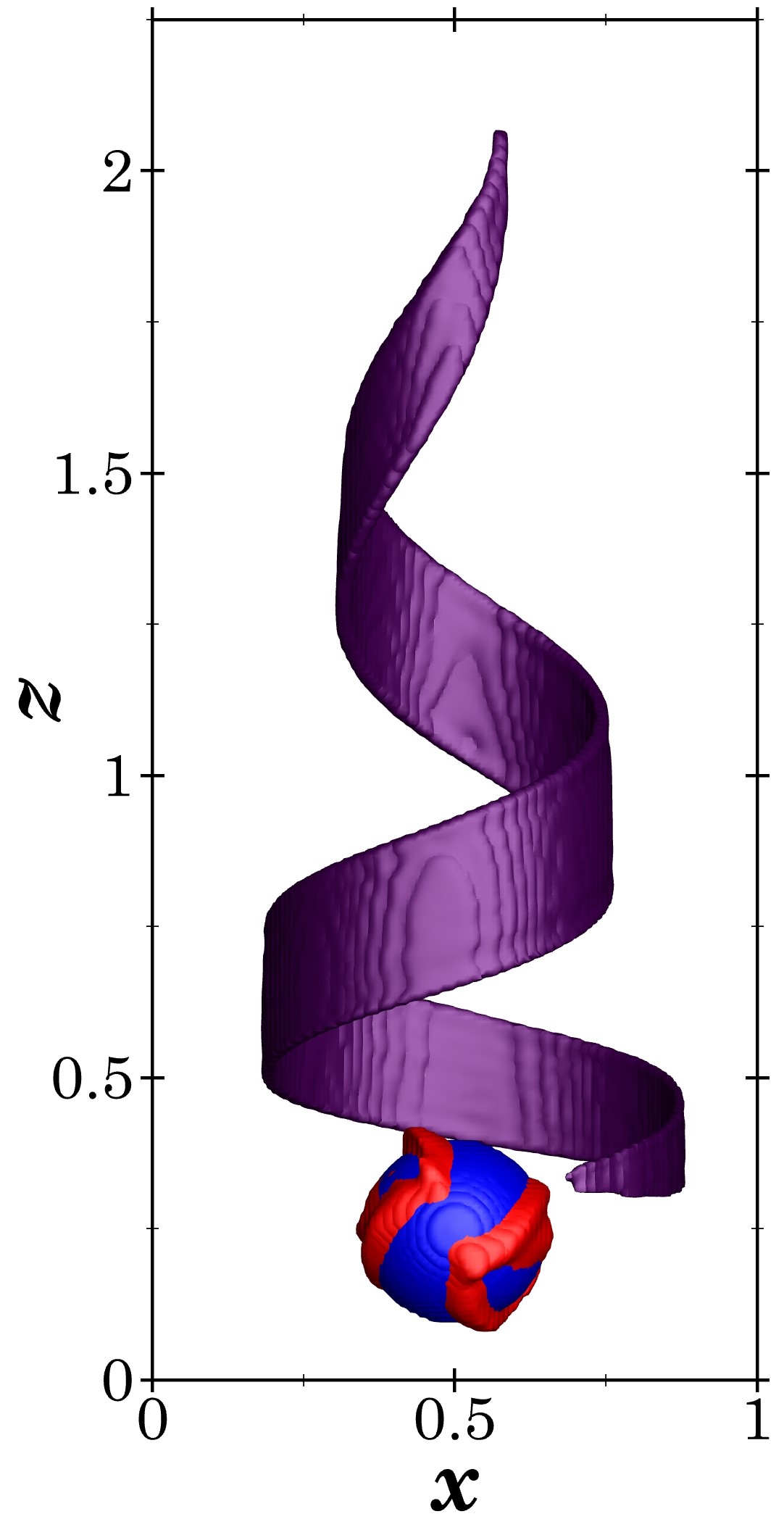}
		}
		\caption{Interface shape ($C=0.5$ iso-contour) of the volume fraction advection test case: $t=0$ in blue, $t=T/2$ (maximum interface deformation) in purple and $t=T$ in red.} 
		\label{fig_adv_test}
	\end{figure}

	The interface evolution predicted by OS-MSTACS is depicted in figure \ref{fig_adv_test}. The figure shows a smooth interface topology of the spiral (at $t=T/2$) which is better than the one obtained by~\cite{saincher2022} with OS-CICSAM. This is attributed to a lower numerical diffusion offered by MSTACS scheme compared to CICSAM~\cite{anghan2021}.  After $t=T/2$, the deformed sphere reverts to its initial position and the numerical error for the same is quantified by computing the $L_1$-norm and the global volume error (expressed in $\%$) of the volume fraction which are defined as:
	
	\begin{linenomath}
		\begin{align*}
			E_{L_1} &= \frac{\sum_{n=1}^{N} |C^T_n\forall_n-C^0_n\forall_n|}{\sum_{n=1}^{N}C^0_n\forall_n}, \\
			E_G &= \frac{|\sum_{n=1}^{N} C^T_n\forall_n-\sum_{n=1}^{N}C^0_n\forall_n|}{\sum_{n=1}^{N}C^0_n\forall_n} \times 100\%
		\end{align*}
	\end{linenomath}
	where $N$ denotes the total number of cells, $\forall_n$ denotes the cell volume; superscripts $0$ and $T$ refer to the initial and final time volume fractions, respectively.
	\begin{table}[H]
		\centering
		\caption{Comparison of volume errors for MSTACS, OS-CICSAM and OS-MSTACS.}
		\begin{tabular}{lccc}
			\hline
			\textbf{Volume error} & \textbf{MSTACS}~\cite{saincher2022} & \textbf{OS-CICSAM}~\cite{saincher2022} & \textbf{OS-MSTACS} \\
			\hline
			$E_{L_1}$ & $2.307 \times 10^{-1}$ & $2.075 \times 10^{-1}$ & $1.988 \times 10^{-1}$ \\
			$E_G$ & $2.537 \times 10^{-4}$ & $8.175 \times 10^{-4}$ & $6.568 \times 10^{-8}$ \\
			\hline
		\end{tabular}
		\label{table:os_mstacs}
	\end{table}
	The present simulation resulted in the $L_1$ error of $E_{L_1}=1.988 \times 10^{-1}$ and global volume error of $E_G=6.568 \times 10^{-8}$ which is better than the OS-CICSAM method and the unsplit MSTACS, as can be seen from table \ref{table:os_mstacs}. This is due to the superior formulation of MSTACS scheme compared to CICSAM. We can conclude from these results that the MSTACS with operator split can accurately capture the evolution of the interface.

One of the contributions of the present work is the formulation of OS-MSTACS. Hence, it is of interest to compare the performance of MSTACS and OS-MSTACS. We repeated the volume fraction advection test for $Cou_{adv}$ of 0.1, 0.5 and 1.0 and compared $E_{L_1}$, $E_G$ and the computational time; these data are reported in table~\ref{table:os_mstacs_timeComparison}. Saincher and Sriram~\cite{saincher2022} provided a comparison of different schemes, in which they solved the system of linear algebraic equations with tolerance set to $10^{-13}$. In contrast, in this work, we set the tolerance for unsplit MSTACS to $10^{-6}$ in order to compare the accuracy for approximately the same computational cost. From table~\ref{table:os_mstacs_timeComparison}, it can be seen that for all Courant numbers, OS-MSTACS provides more accurate results despite requiring reduced computational time. The conclusion that OS-MSTACS is more accurate remains the same even if the tolerance for the unsplit MSTACS is set to $10^{-13}$, as can be seen from table~\ref{table:os_mstacs}. We used an Intel Core i9-9920X CPU  with a clock speed of 3.50GHz for these comparison studies.

	\begin{table}[h!]
		\centering
		\caption{Comparison of simulation time and volume errors for OS-MSTACS and unsplit MSTACS with convergence criterion $10^{-6}$. }
		\begin{tabular}{c|ccc|ccc}
                \hline
                & \multicolumn{3}{c|}{\textbf{MSTACS}} & \multicolumn{3}{c}{\textbf{OS-MSTACS}}\\
                \hline
                \hline
			$\boldsymbol{Cou_{adv}}$ & $\boldsymbol{ E_{L_1} }$ & $\boldsymbol{ E_G }$ & \textbf{Time (s)} & $\boldsymbol{ E_{L_1} }$ & $\boldsymbol{ E_G }$ & \textbf{Time (s)}  \\
			\hline
			0.1    & $2.82 \times 10^{-1}$   & $2.17 \times 10^{0}$   & $101562.38$ & $ 1.98 \times 10^{-1}$   & $6.51 \times 10^{-8}$    & $76713.58$ \\
			0.5    & $5.32 \times 10^{-1}$    & $4.95 \times 10^{0}$  & $23612.99$ & $2.44 \times 10^{-1}$     & $1.43 \times 10^{-6}$   & $14909.49$  \\
			1.0    & $9.42 \times 10^{-1}$   & $1.14 \times 10^{1}$   & $14006.22$ & $1.05 \times 10^{0}$   & $1.93 \times 10^{-1}$    & $7447.58$ \\
			\hline
			\hline
		\end{tabular} 
		\label{table:os_mstacs_timeComparison}
	\end{table}

\subsection{2D static droplet}
\label{appendix_staticDroplet}
To ensure correct implementation of the surface tension model, we examine the 2D static droplet test case. We place a droplet of diameter $D=1m$ at the centre of a domain of $2m \times 2m$. We set the density of the two inviscid fluids as $\rho_1=1000 kg/m^3$ and $\rho_2=1 kg/m^3$, the surface tension coefficient, $\sigma =1 N/m$ and the speed of sound, $c_s=5 m/s$ (corresponding to Ma=0.2 based on the velocity scale, $U_{\sigma}=\sqrt{\sigma/\rho_2 D}$~\cite{popinet2009}). The problem setup is the same as that of Yang and Aoki~\cite{yang2021}, including the initial condition for pressure. On all boundaries, we enforce the free-slip condition for velocity and zero normal gradient for pressure. A uniform grid of $128 \times 128$  is used and the time step $\Delta t =1.25 \times 10^{-3}s$. We examine the results at $t=0.625s$, as discussed by Yang and Aoki~\cite{yang2021}.

 \begin{figure}[H]
		\centering
		\subfigure[]
		{
			\includegraphics[trim = 0mm 30mm 0mm 0mm, clip, width=7.5cm]{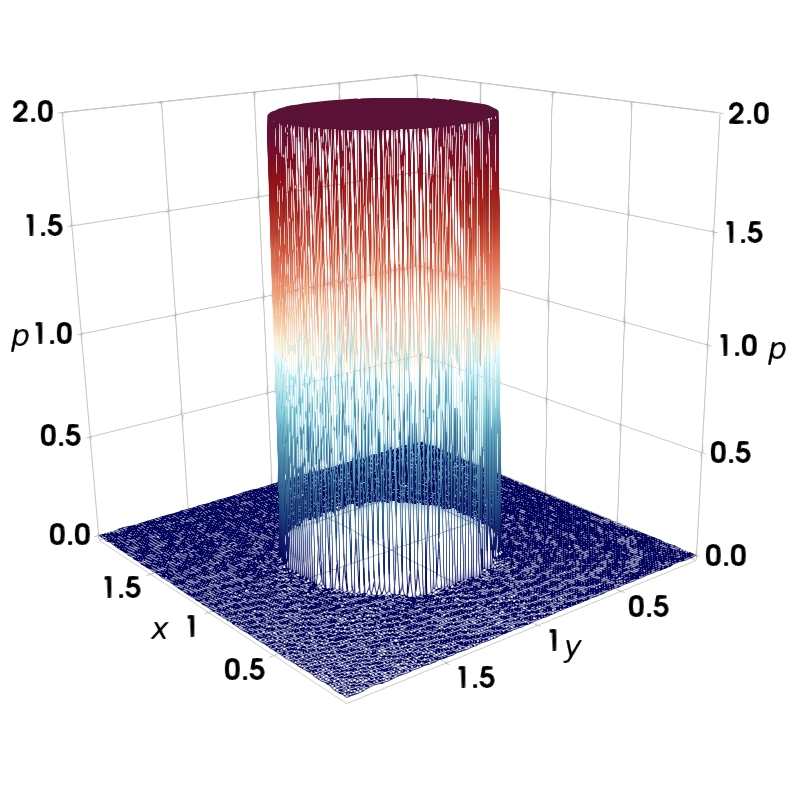}
			\label{fig_staticDrop_pressure1}
		}
		\subfigure[]
		{
			\includegraphics[height=5.5cm]{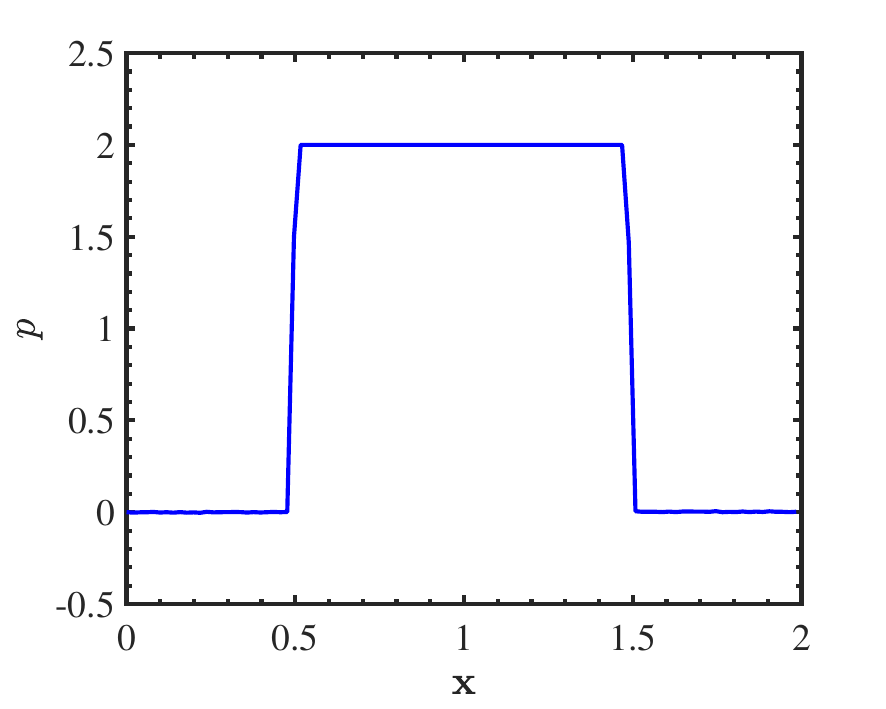}
			\label{fig_staticDrop_pressure2}
		}
		\caption{Results of the static droplet test case at $t=0.625s$. (a) Surface plot of pressure in the domain (b) Line plot of pressure through the centreline along the x-axis.}
		\label{fig_staticDrop_pressure}
	\end{figure}
In the static environment, the suspended droplet under equilibrium achieves a balance between the pressure and surface tension forces. The theoretical pressure jump across the interface is given as $\Delta p = 2\sigma/D$, and the obtained pressure jump is in good agreement with the same. This is directly evident from figure~\ref{fig_staticDrop_pressure}, which shows the pressure in the computational domain.

\begin{figure}[H]
		\centering
		\subfigure[]
		{
			\includegraphics[height=6.5cm]{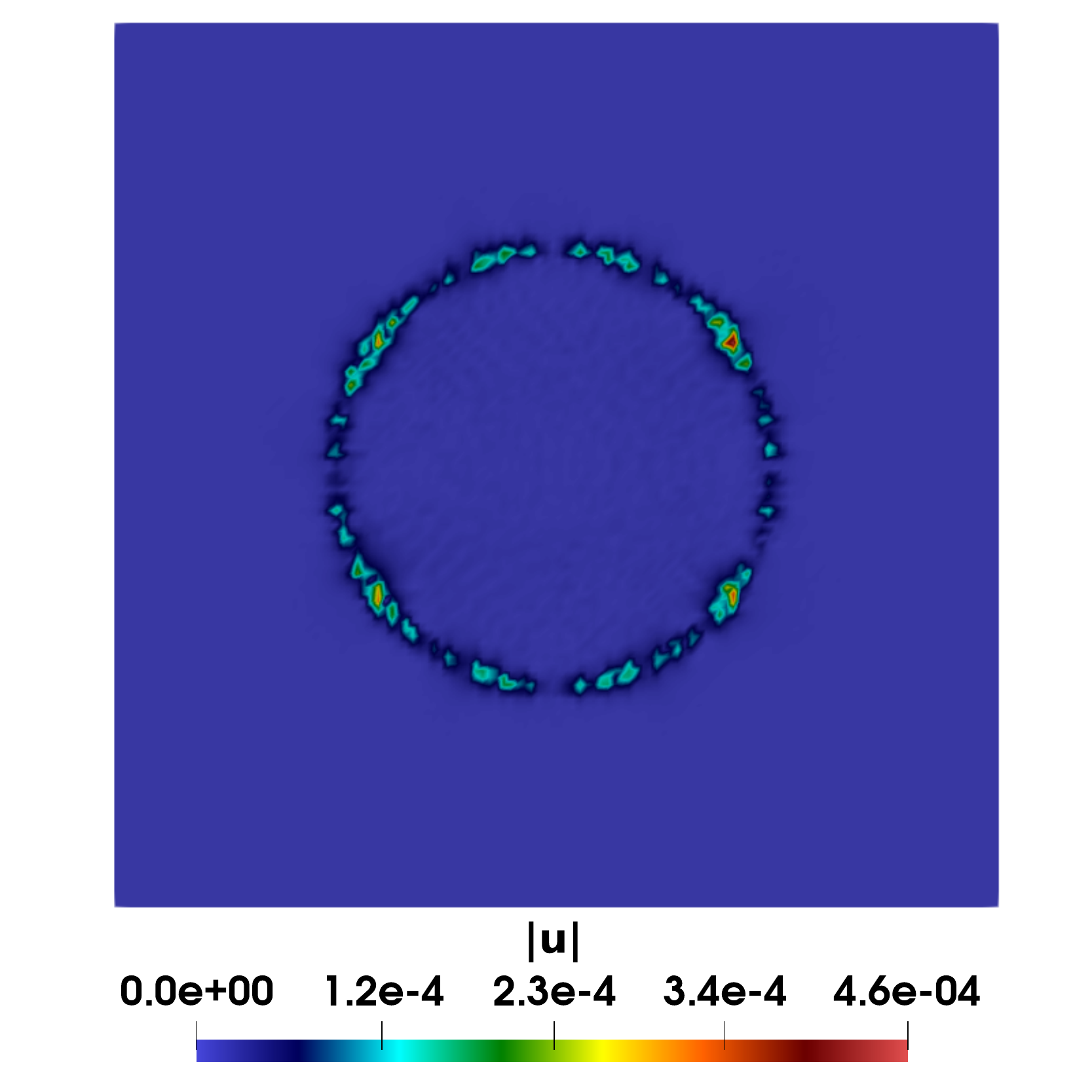}
			\label{fig_spurious_inviscid_gpe}
		}
		\subfigure[]
		{
			\includegraphics[height=6.5cm]{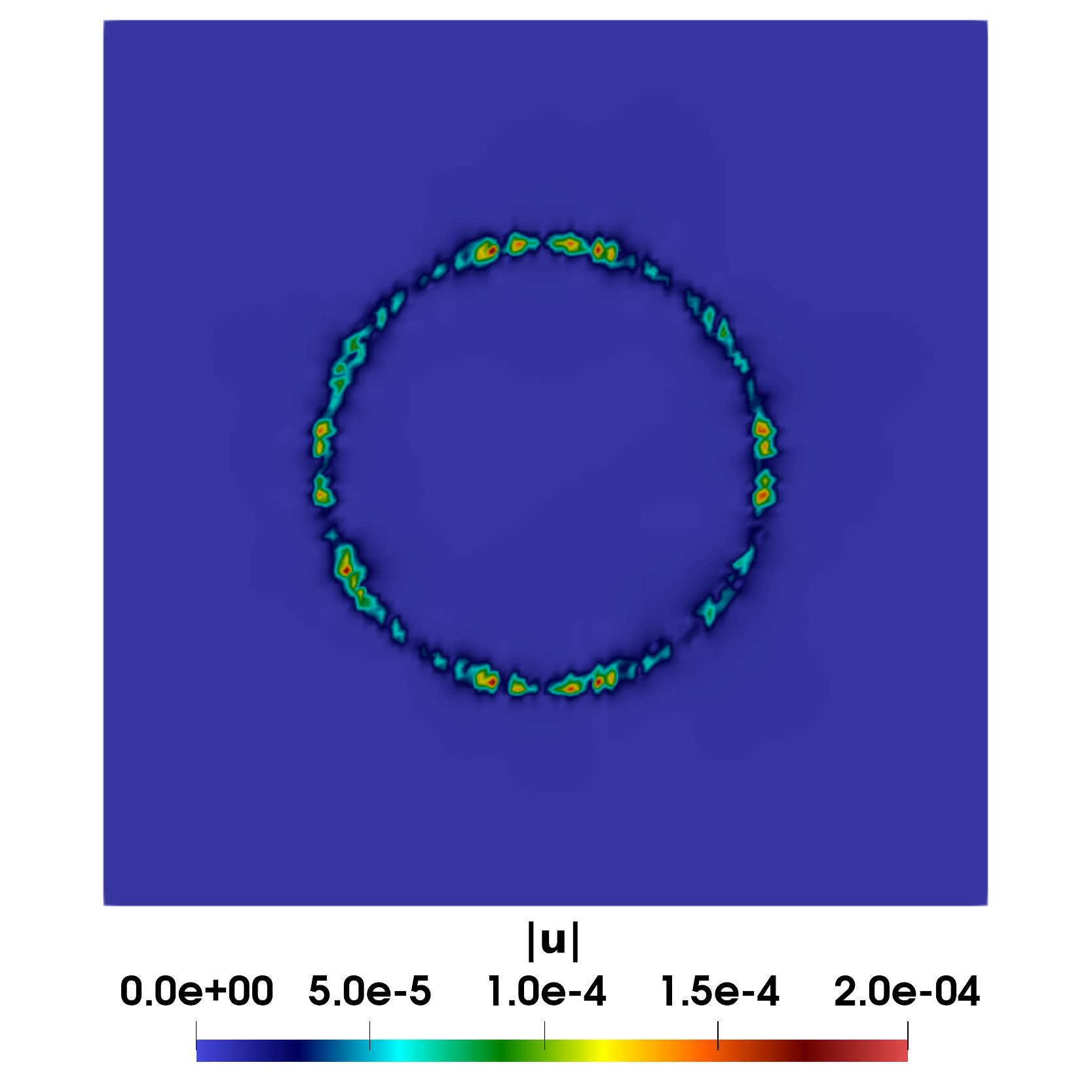}
			\label{fig_spurious_inviscid_ins}
		}
		 \caption{Magnitude of spurious currents for the static droplet test case from (a) GPE solver (b) INS solver, at $t=0.625s$.}
		\label{fig_spurious_inviscid}
	\end{figure}

Next, we examine the presence of parasitic (spurious) currents, which are indications of the discretisation error associated with the surface tension model used. Figure~\ref{fig_spurious_inviscid} compares the magnitude of these spurious currents obtained from GPE and INS solvers. We observe that these are of the same order, and the difference is negligible. We emphasize that the obtained $\bf |u|_{\max}$ is lower than that reported by another weakly compressible solver~\cite{yang2021} for which $|\textbf{u}|_{\max}  = 8\times 10^{-3}$.

\begin{figure}[H]
		\centering
		\subfigure[]
		{
			\includegraphics[height=5cm]{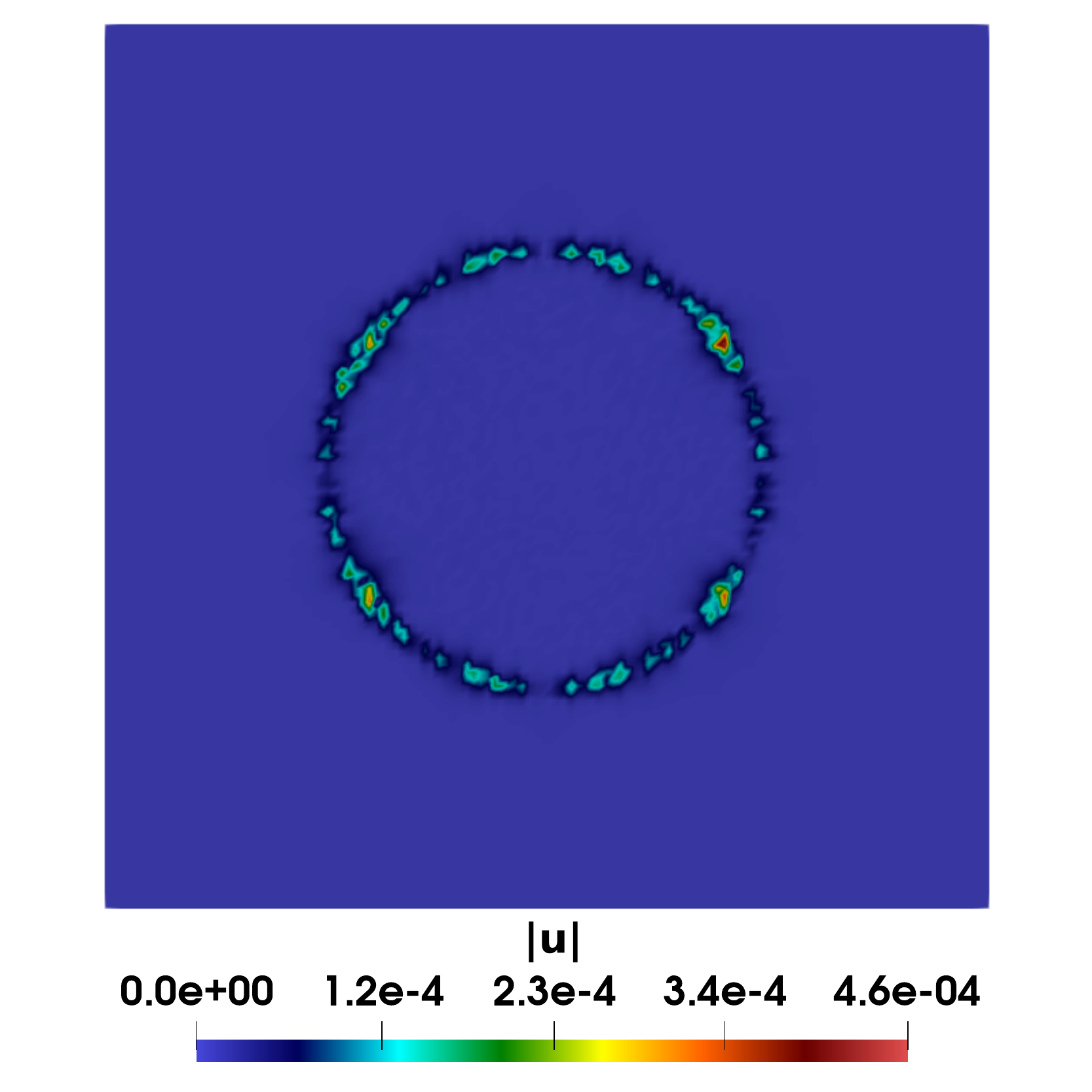}
			\label{fig_staticDrop_MaStudy1}
		}
		\subfigure[]
		{
			\includegraphics[height=5cm]{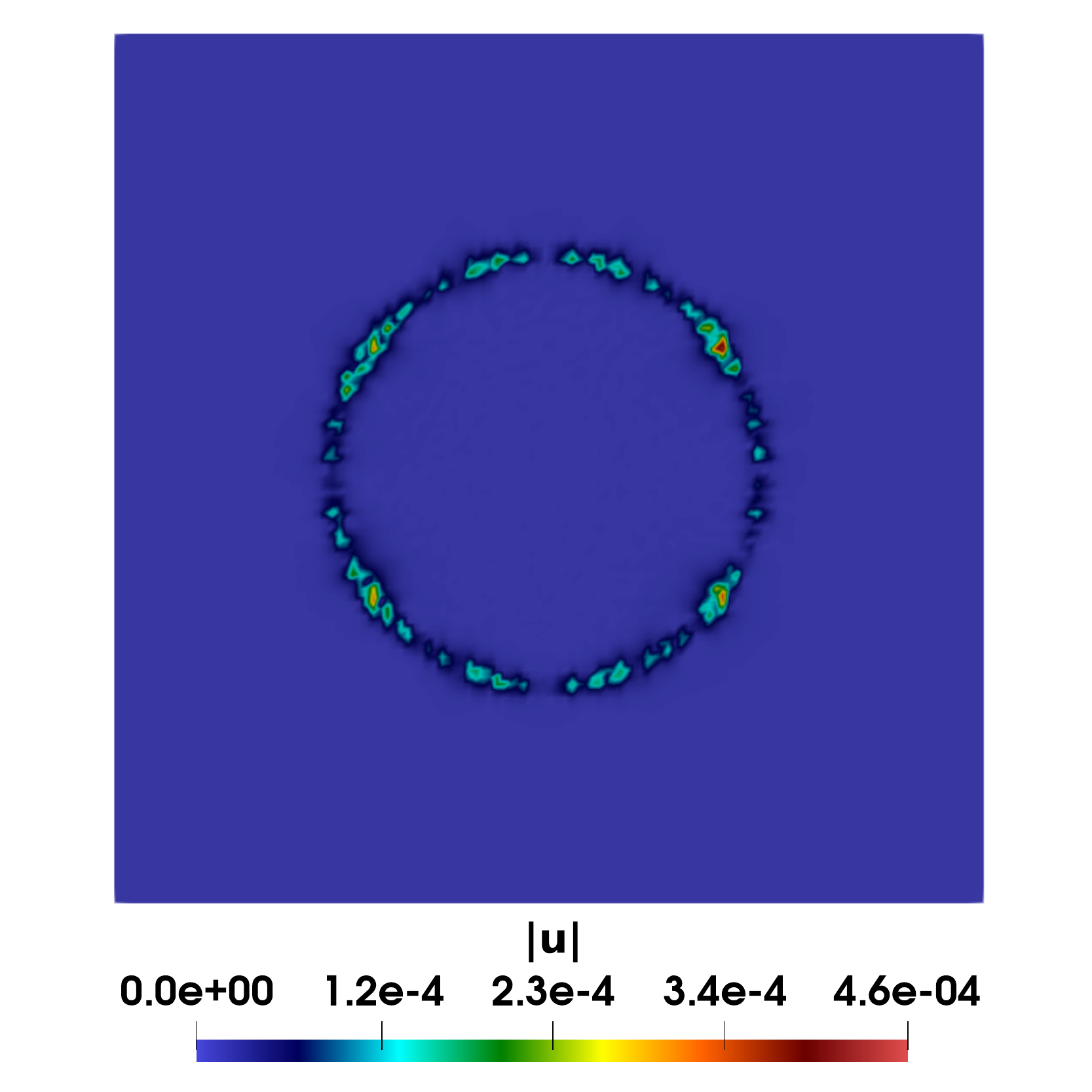}
			\label{fig_staticDrop_MaStudy2}
		}
            \subfigure[]
		{
			\includegraphics[height=5cm]{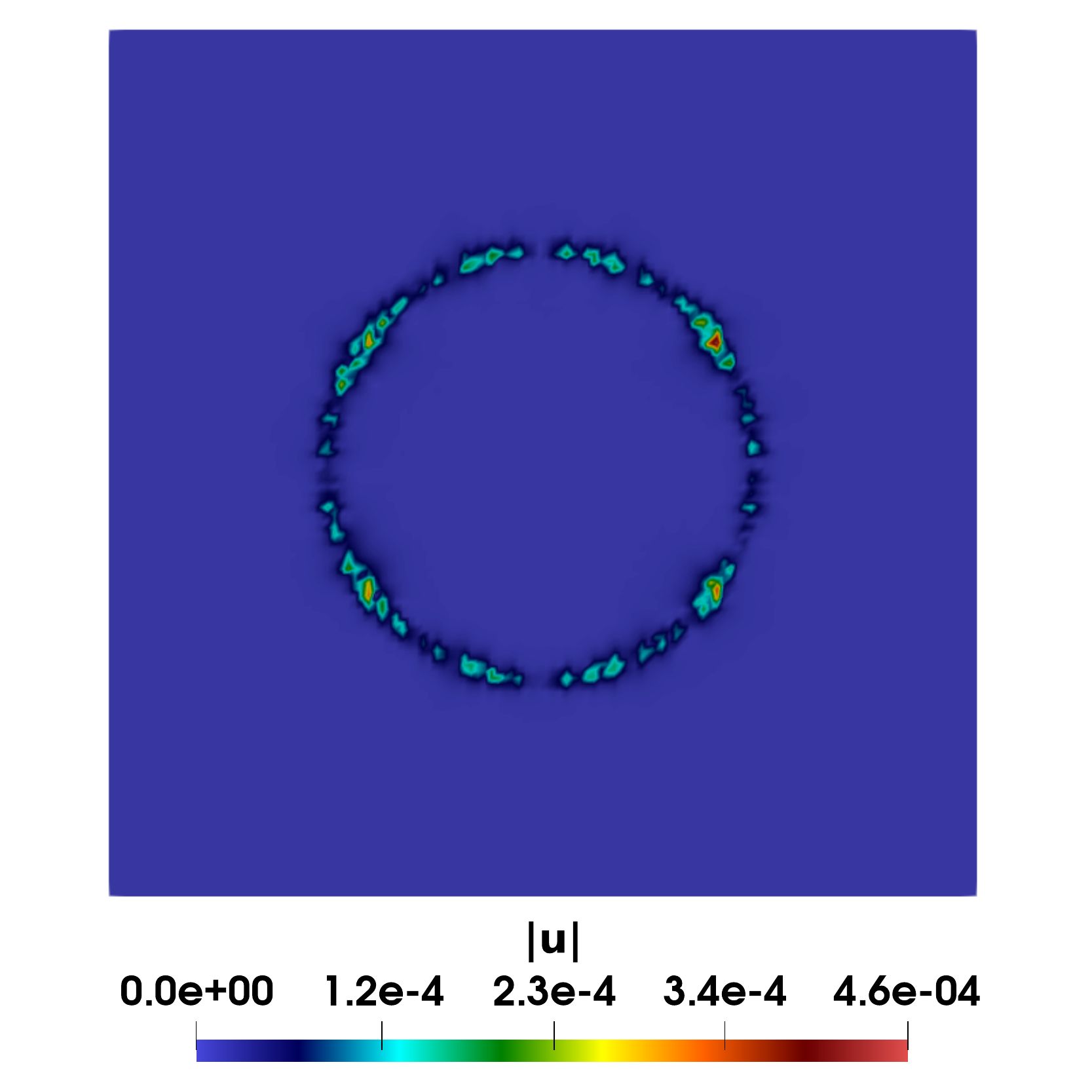}
			\label{fig_staticDrop_MaStudy3}
		}
		\caption{Effect of Mach number on spurious currents associated with static droplet test case. The contours of velocity magnitude ($|\textbf{u}|$) is plotted for (a) $\text{Ma}=0.1$ (b) $\text{Ma}=0.05$ and (c) $\text{Ma}=0.01$, at $t=0.625s$.}
		\label{fig_staticDrop_MaStudy}
	\end{figure}

We further examine the effect of \text{Ma} on the spurious currents. In order to do this, we plot $|\textbf{u}|$ for different \text{Ma} as shown in figure~\ref{fig_staticDrop_MaStudy}. We use $c_s=10,\ 20$ and $ 100$ corresponding to $\text{Ma}=0.1,\ 0.05$ and $0.01$. Similar to the findings of~\cite{kajzer2018_conference}, we observe that \text{Ma} doesn't influence the magnitude of spurious currents. Although not visible clearly, the acoustic waves become less pronounced as we lower the \text{Ma}, an observation we discuss in detail in the subsequent test cases.

\begin{table}[H]
    \centering
    \caption{Effect of Laplace number on spurious currents for the static droplet test case quantified by the maximum value of velocity.}
    \begin{tabular}{|l|c|}
        \hline
        \hline
        \textbf{La} & $\boldsymbol{ | u |_{\max} }$ \\
        \hline 
        $\infty$  & $4.6 \times 10^{-4}$  \\
        12000  & $1.2 \times 10^{-2}$  \\
        1200 & $2.5 \times 10^{-2}$ \\
        120 & $4.0 \times 10^{-2}$ \\
        \hline
        \hline
    \end{tabular}
    \label{table:La_staticDrop}
\end{table}
Finally, we also examine the influence of Laplace number, defined as $\text{La} = \rho D \sigma / {\mu}^2 $, on the spurious currents. The values of $\bf |u|_{\max}$ are presented in table~\ref{table:La_staticDrop} for different values of $\text{La}$ ranging from 120 to infinity. We observe that, unlike other INS-based works~\cite{popinet2009}, where the $\bf |u|_{\max}$ decreases with decreasing La, we obtain an increase in $\bf |u|_{\max}$ when La is reduced. However, this result is consistent with those reported for other similar weakly compressible formulations~\cite{yang2021,kajzer2018_conference}. From these observations, we can conclude that the implemented model for the surface tension is accurate enough compared to other weakly compressible solvers.

	\subsection{2D Rayleigh-Taylor Instability}
	\label{subsec4.2}
	When a system comprising a heavier fluid gently placed atop a lighter fluid is slightly perturbed externally, an interfacial instability called Rayleigh-Taylor instability is manifested. Due to gravity, the heavier fluid penetrates the lighter one. This results in the evolution of the complex topology of the interface, and the prediction of the interface shape is a standard test case for the two-phase flows. The test case is adopted from Garoosi et al.~\cite{garoosi2022}; the parameters are listed in table \ref{table:RT_parameters}.
	
\begin{figure}[H]
	 	\centering
	 	\includegraphics[width=7.5cm]{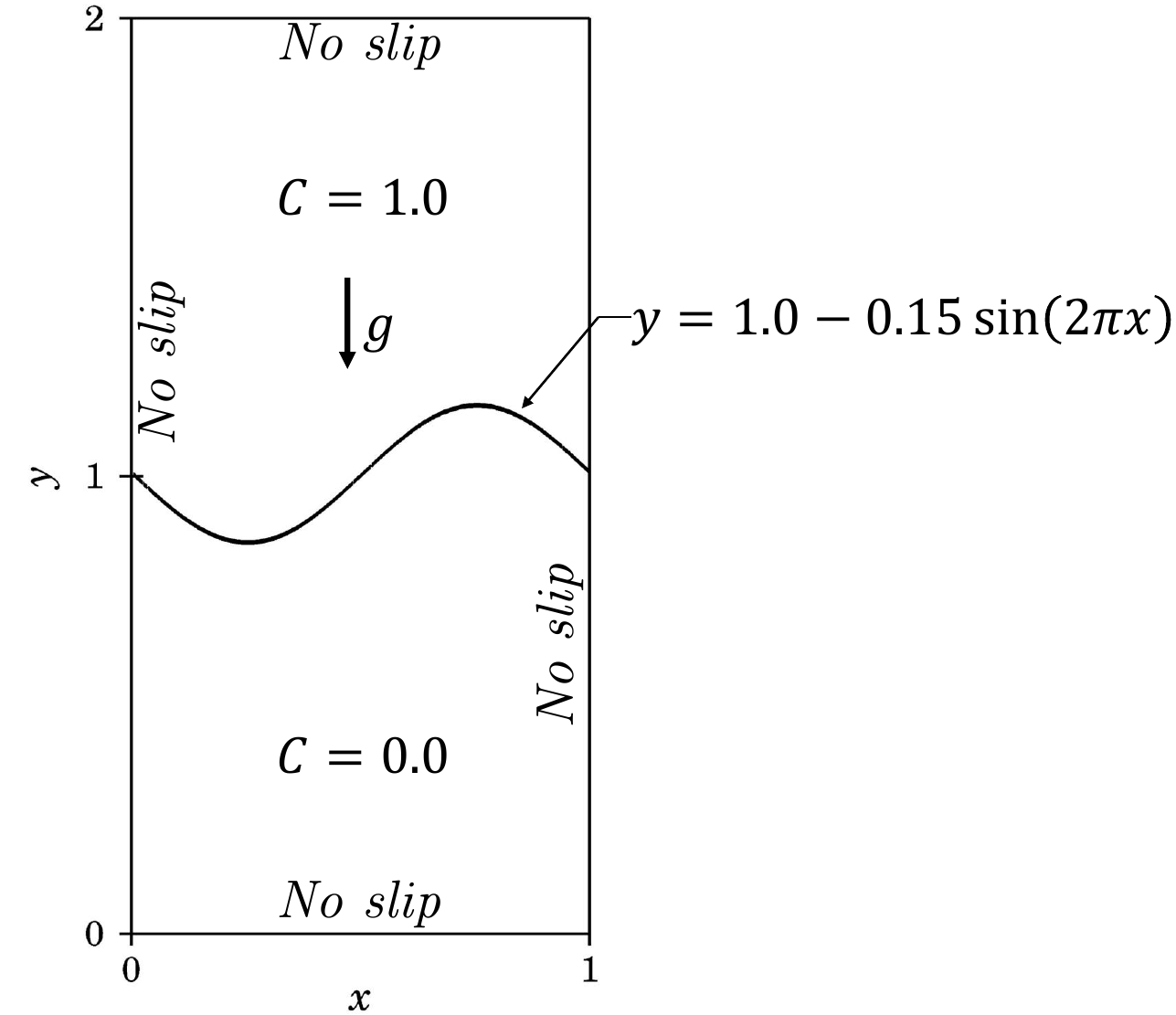}
	 	\caption{Initial flow configuration for Rayleigh-Taylor instability including the initial perturbation and boundary conditions.}
	 	\label{fig_initial_RT}
	 \end{figure}

	The initial configuration for simulating the RT instability is shown in figure \ref{fig_initial_RT} wherein the heavier fluid ($C=1.0$) is placed atop the lighter fluid ($C=0.0$) separated by an interface ($0<C<1$). As shown in the figure, the interface is sinusoidally perturbed at the initial time instant to trigger the interfacial instability. All the boundaries are equipped with no-slip boundary condition, and the configuration is initialized with zero velocities and pressure. The advection scheme used for the test case is the second-order central difference scheme for discretizing the advected velocity component, and the simulation is carried out for 5 non-dimensional time units ($t^*=t\sqrt{g/L}$).
	
	\begin{table}[H]
		\centering
		\caption{Parameters for Rayleigh-Taylor instability adopted from Garoosi et al.~\cite{garoosi2022}.}
		\begin{tabular}{lccc}
			\hline
			\hline
			\textbf{Name} & \textbf{Definition} & \textbf{Value} \\
			\hline 
			Length & $L\text{ }(m)$ & $1.0$ \\
			Height & $W\text{ }(m)$ & $2.0$ \\
			Heavier fluid density & $\rho_1\text{ }(kg/m^3)$ & $1.8$ \\
			Heavier fluid viscosity & $\mu_1\text{ }(kg/ms)$&$0.018$ \\
			Lighter fluid density & $\rho_2\text{ }(kg/m^3)$&$1.0$ \\
			Lighter fluid viscosity & $\mu_2\text{ }(kg/ms)$&$0.01$ \\
			Reynolds number & $\text{Re}=\rho_1L\sqrt{gL}/\mu_1$&$420$ \\
			Froude number & $\text{Fr}=(\sqrt{gL})^2/(gL)$&$1.0$ \\
			Grid resolution & $\text{Nx}\times \text{Ny}$ & $200 \times 400$ \\
			\hline
			\hline
		\end{tabular}
		\label{table:RT_parameters}
	\end{table}
	 	 
\begin{figure}[H]
		\centering
		\subfigure[]
		{
			\includegraphics[height=6.5cm]{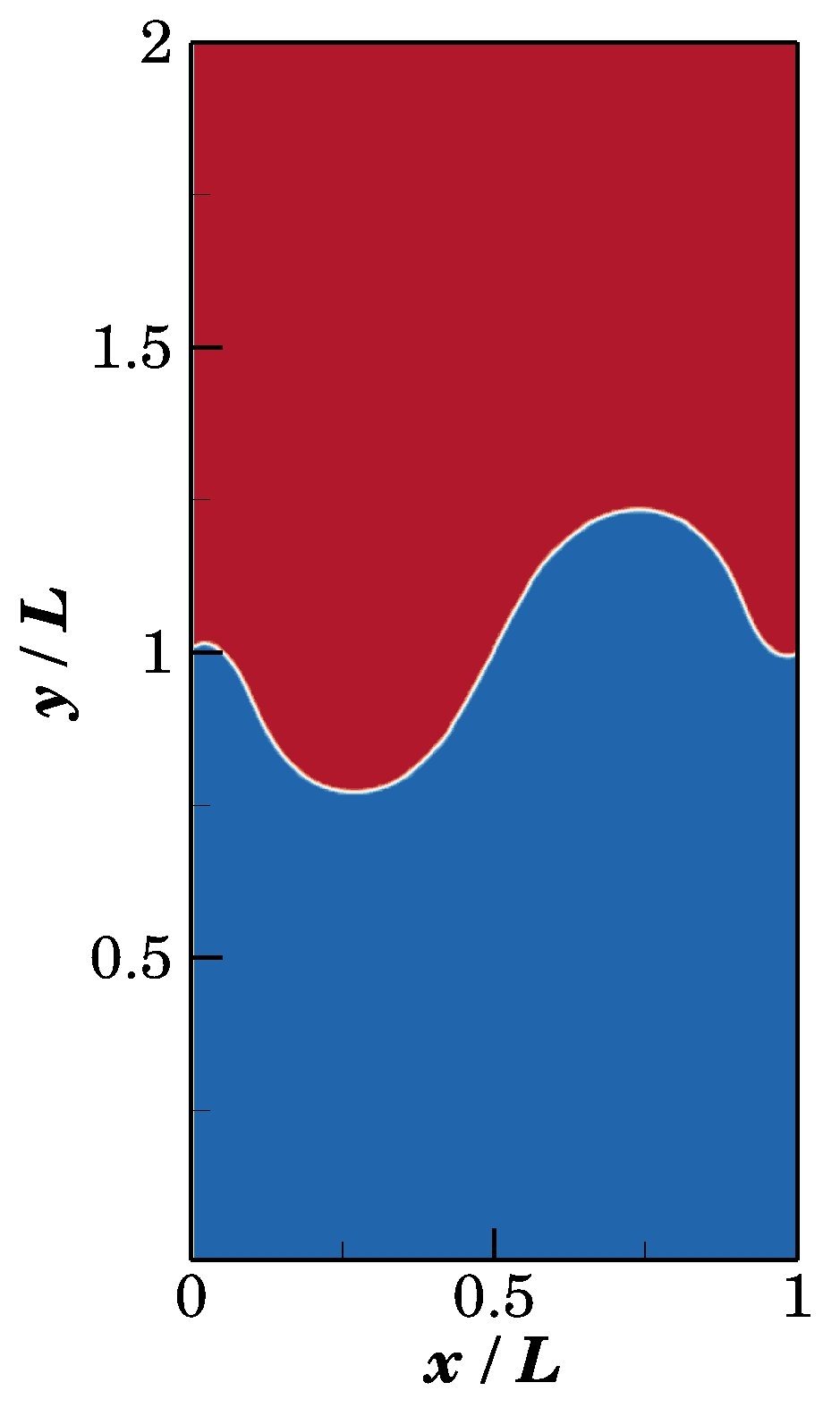}
			\label{fig_2D_RT_evolve_a}
		}
		\subfigure[]
		{
			\includegraphics[height=6.5cm]{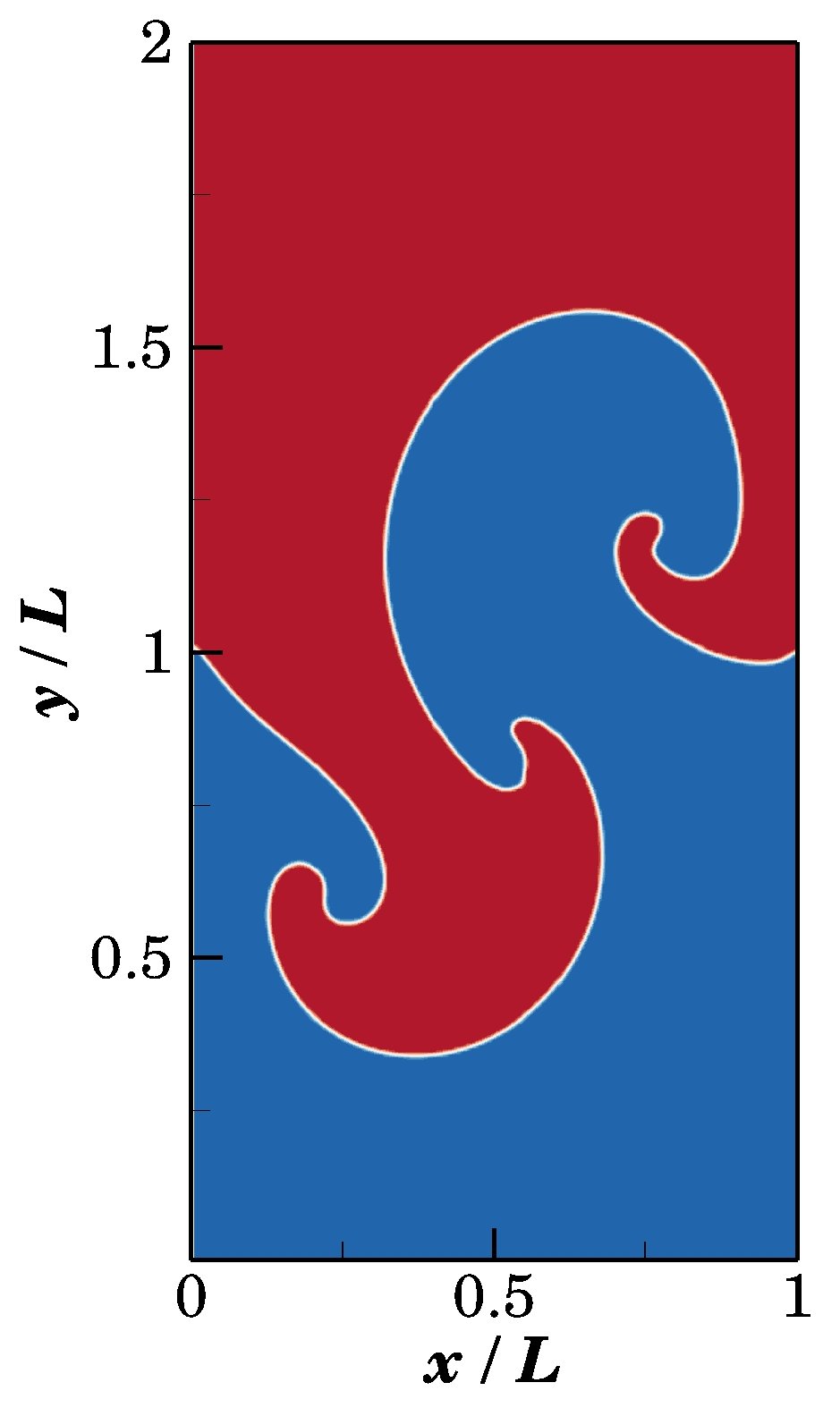}
			\label{fig_2D_RT_evolve_b}
		}
		\subfigure[]
		{
			\includegraphics[height=6.5cm]{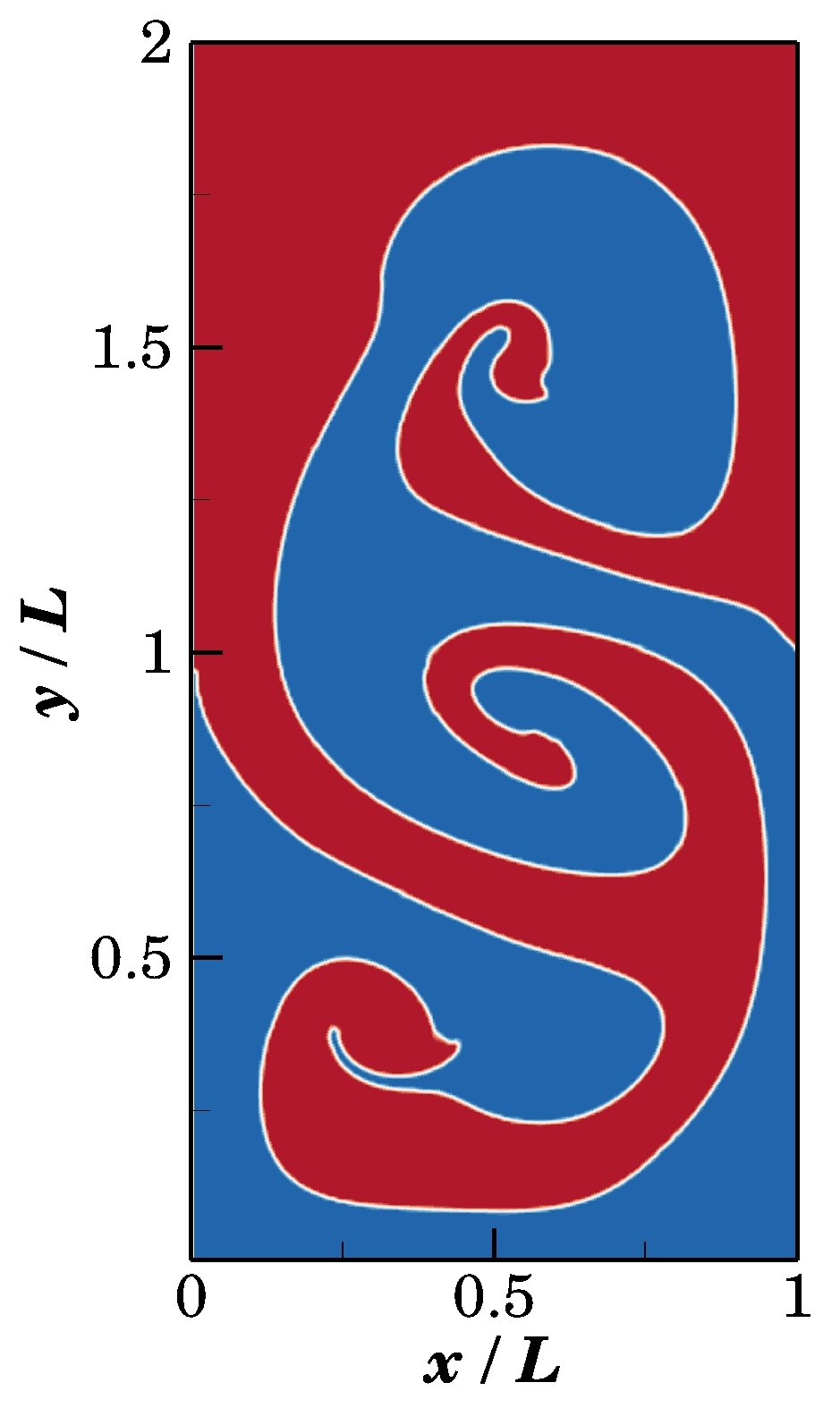}
			\label{fig_2D_RT_evolve_c}
		}
		\subfigure[]
		{
			\includegraphics[height=6.0cm]{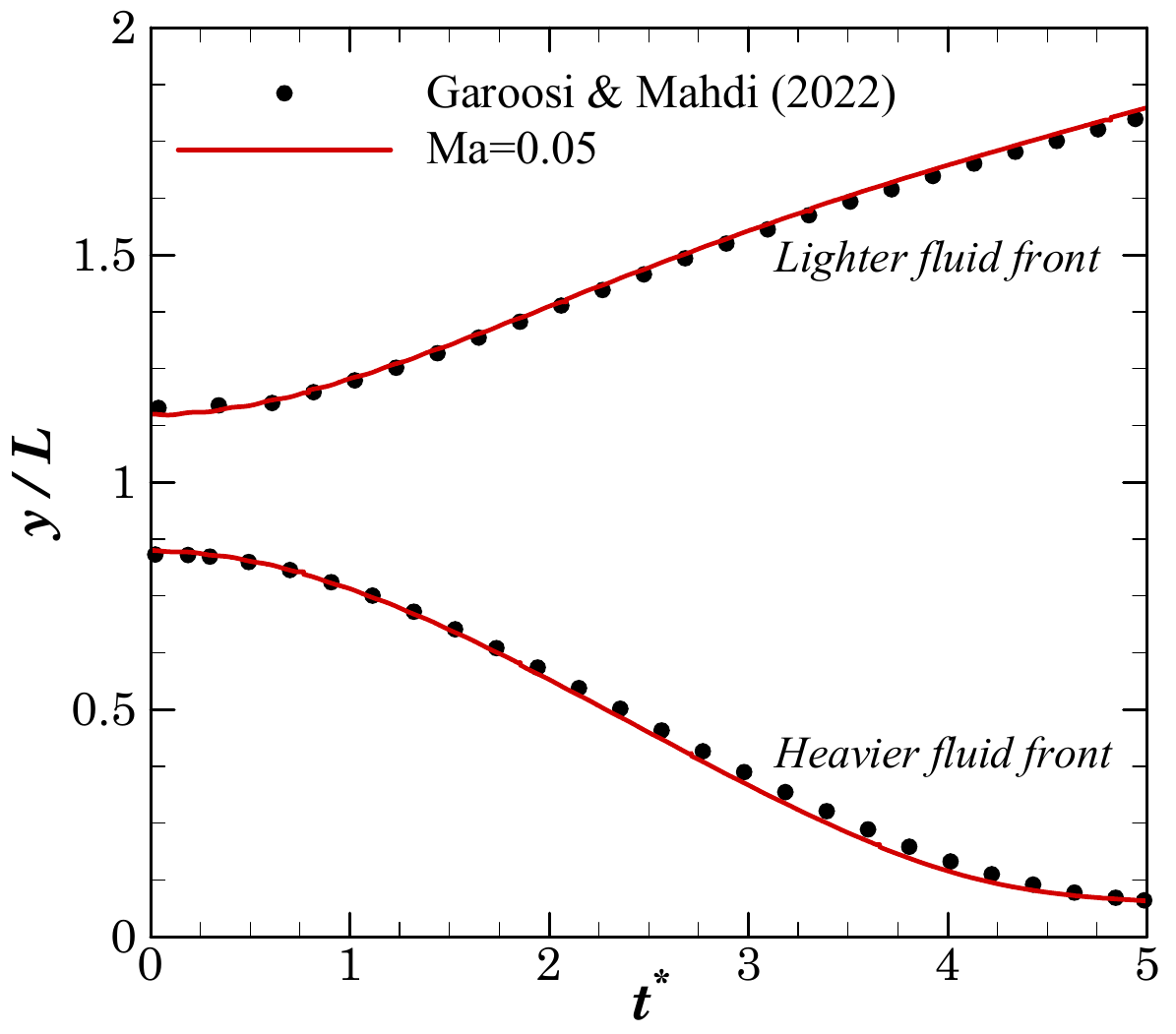}
			\label{fig_2D_RT_evolve_d}
		}
		\caption{Temporal evolution of the interface for RT instability visualized at (a) $t^*=1$  (b) $t^*=3$ and (c) $t^*=5$, (d) position of the maximum and minimum fluid fronts with time.}
		\label{fig_2D_RT_evolve}
	\end{figure}

	The temporal evolution of the interface is qualitatively assessed at $t^*=1,3,5$ with a time step of $10^{-4}$ and $\text{Ma}=0.05$ for the simulation. As shown in figure \ref{fig_2D_RT_evolve_a}, due to the existence of gravity and density gradient between the two immiscible fluids, the heavier fluid accelerates into the lighter fluid volume. Simultaneously, owing to the buoyancy force, the lighter fluid flows upward into the heavier fluid. As a result, the amplitude of the sinuous perturbation starts evolving into mushroom-shaped plumes, as seen in figure \ref{fig_2D_RT_evolve_b}. As time progresses, the instability increases and the resulting complex interface topology can be observed in figure \ref{fig_2D_RT_evolve_c}. Qualitatively, our flow field matches well with those reported in the literature.
	
	The maximum (minimum) position of the lighter fluid (heavier fluid) is tracked with time and compared with Garoosi et al.~\cite{garoosi2022}. It is evident from figure \ref{fig_2D_RT_evolve_d} that the evolution of the interface position with time is in excellent agreement with the existing literature.
	
	\begin{figure}[H]
		\centering
		\subfigure[]
		{
			\includegraphics[height=6cm]{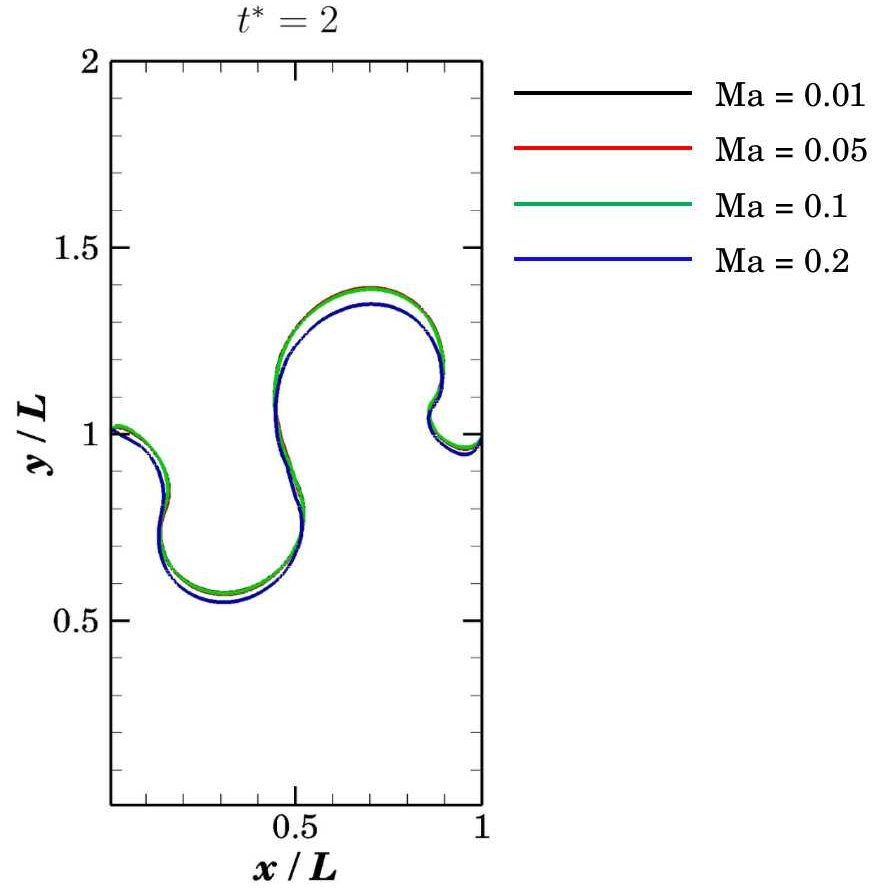}
			\label{fig_2D_RT_Ma_a}
		}
		\subfigure[]
		{
			\includegraphics[height=6cm]{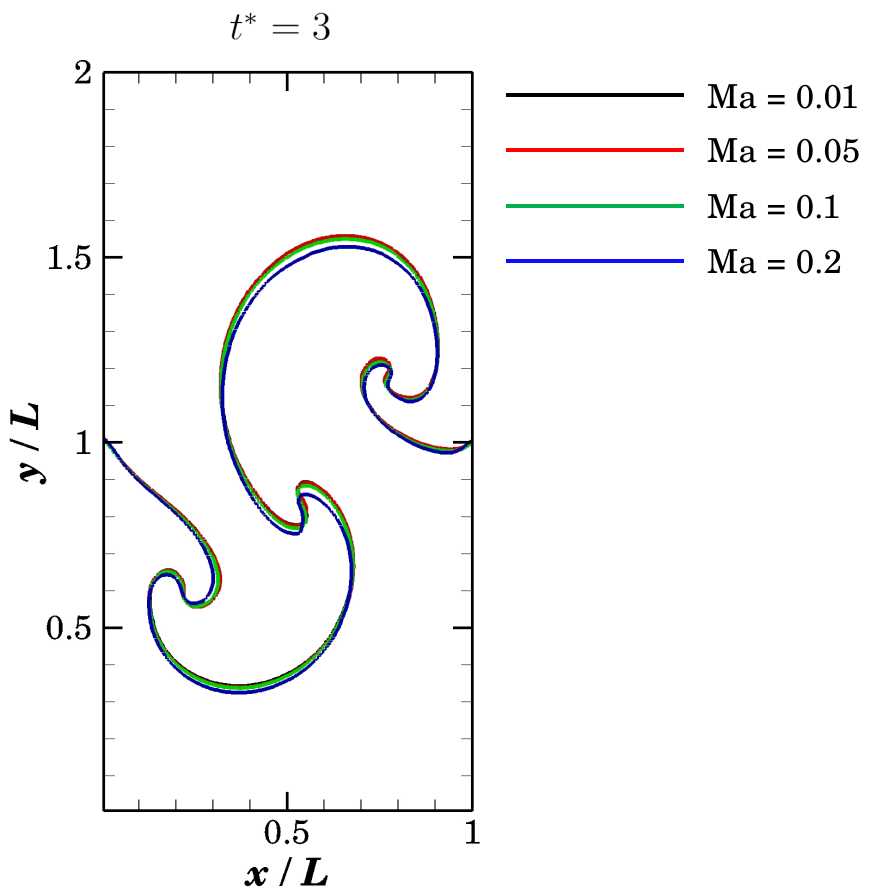}
			\label{fig_2D_RT_Ma_b}
		}
		\subfigure[]
		{
			\includegraphics[height=6cm]{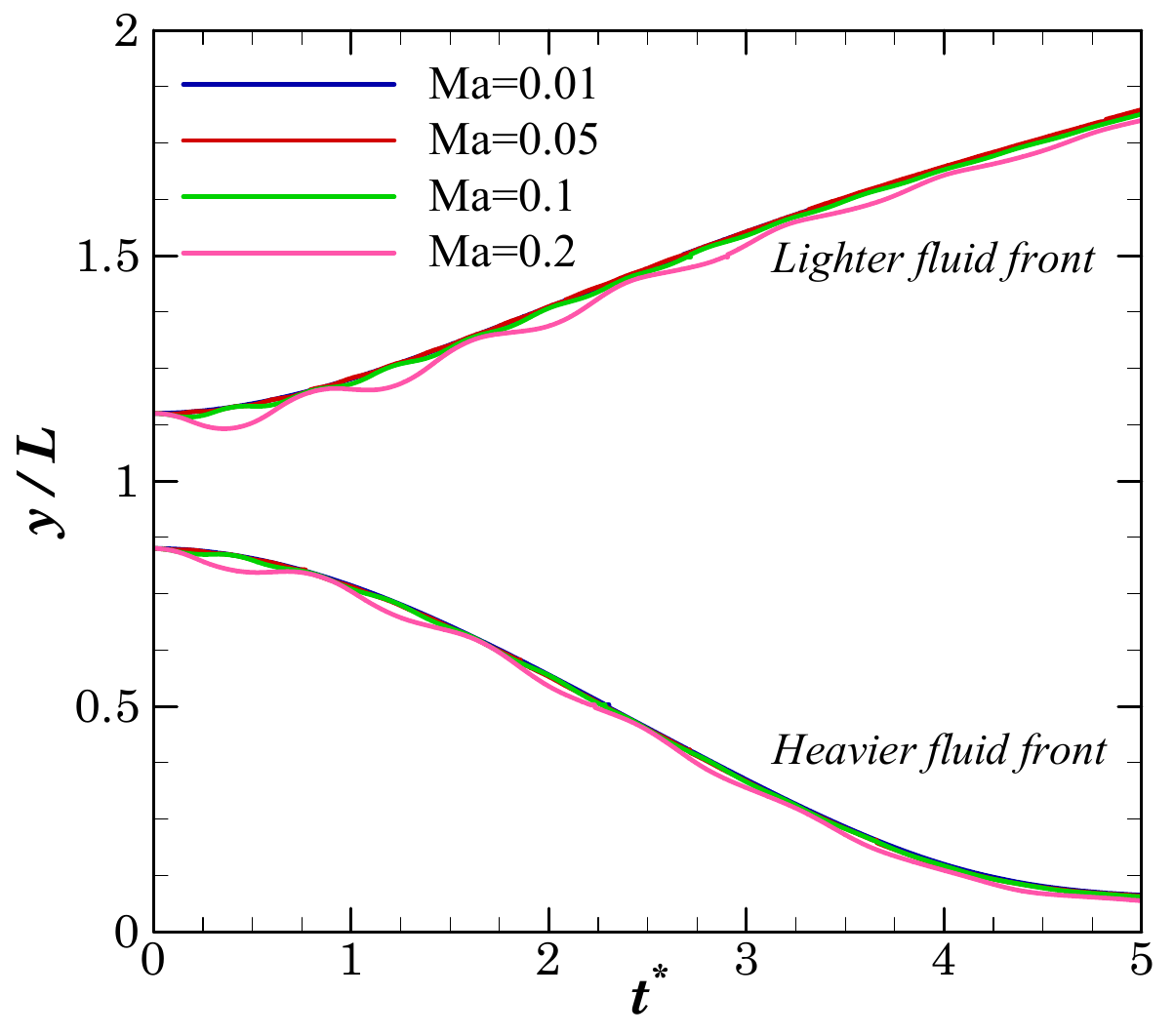}
			\label{fig_2D_RT_Ma_c}
		}
		\caption{Effect of Mach number on RT instability. (a)~and (b)~denote the interface shapes for $t^*=2$ and $t^*=3$, respectively, (c)~position of the maximum and minimum fluid fronts with time.}
		\label{fig_2D_RT_Ma}
	\end{figure}
	
	Toutant~\cite{toutant2018} studied the effect of Mach number on the accuracy of single-phase flow simulations. The Mach number appearing in equation \eqref{eqn:gpe_nondim} is an artificial parameter in GPE. The choice of Ma is crucial to dictate the solver's accuracy and computational efficiency; hence, we study the effect of the Mach number for this problem. Figure \ref{fig_2D_RT_Ma} depicts the effect of Mach number for the proposed method, and in this study, the values considered are $0.01$, $0.05$, $0.1$ and $0.2$. It is apparent from figures \ref{fig_2D_RT_Ma_a} and \ref{fig_2D_RT_Ma_b} that for the case of $\text{Ma}=0.01 \text{ and Ma}=0.05$, the discrepancies in the interface topology are negligible, which is also evidenced quantitatively from figure \ref{fig_2D_RT_Ma_c}. However, for the case of $\text{Ma}=0.1$, minute oscillations can be seen in the interface evolution plot for a finite simulation time. Furthermore, although $\text{Ma}=0.2$ provides a spatially smooth interface shape, unphysical oscillations are predominantly observed in the time evolution plot, especially during the initial time instants. The oscillations are damped as time progresses. Clausen reports a similar fluctuating nature of the plot for the case of single-phase flows using EDAC as the pressure transport equation~\cite{clausen2013}.
	
	\subsection{Broken Dam flow}
	\label{subsec4.3}
\begin{figure}[H]
	\centering
	{
		\includegraphics[width=12cm]{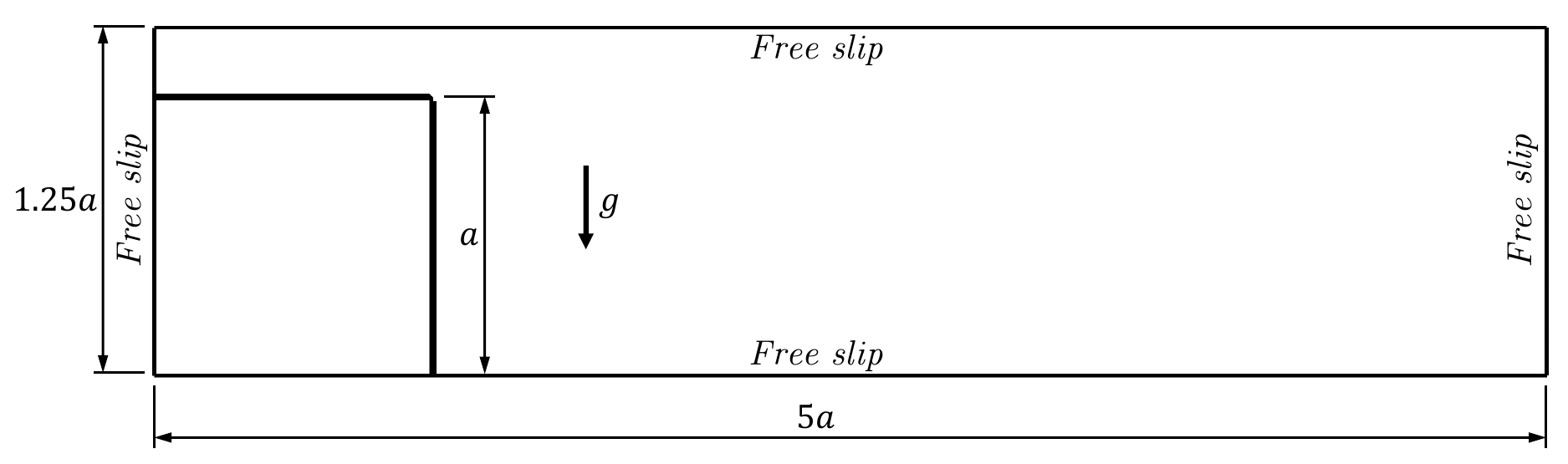}
	}
	\caption{Initial flow configuration for the broken dam flow test case.}
	\label{fig_initial_BD}
	\end{figure}

	The previous test case involves immiscible fluids with a minor difference in density and viscosity. To investigate the accuracy of our method for problems involving a large density ratio, we simulate the broken dam flow, a classical benchmark test for multiphase flows. A rectangular liquid column collapses in this simulation and percolates over the bottom wall. The test case was first experimentally investigated by Martin and Moyce~\cite{martin1952} and numerically tested by many authors. In the present work, we adopt the numerical parameters of Kelecy and Pletcher~\cite{kelecy1997}, also studied by Ling et al.~\cite{ling2019}. The initial configuration for the simulation is shown in figure~\ref{fig_initial_BD}. Zero velocity and hydrostatic pressure are used as the initial conditions. The parameters considered for the simulation are listed in table \ref{table:BD_parameters}. Due to the violent nature of the flow, a total variable diminishing scheme with a Van Leer limiter~\cite{vanLeer1974} is used for discretizing the advected velocity component. The time step chosen for this problem is $10^{-5}$. Analogous to lattice-Boltzmann simulations~\cite{huang2020,wang2015} we set the artificial speed of sound, $c_s = \Delta x/(\sqrt{3} \Delta t)$ which corresponds to $\text{Ma}=0.018$.
	
	\begin{table}[H]
		\centering
		\caption{Parameters for the broken dam flow test case.}
		\begin{tabular}{lcc}
			\hline
			\hline
			\textbf{Name} & \textbf{Definition} & \textbf{Value}\\
			\hline 
			Characteristic length & $a\text{ }(m)$ & $0.05715$ \\
			Length & $L\text{ }(m)$ & $5a$\\
			Height & $W\text{ }(m)$ & $1.25a$\\
			Heavier fluid density & $\rho_1\text{ }(kg/m^3)$ & $1000$\\
			Heavier fluid viscosity & $\mu_1\text{ }(kg/ms)$&$0.001$\\
			Lighter fluid density & $\rho_2\text{ }(kg/m^3)$&$1.25$\\
			Lighter fluid viscosity & $\mu_2\text{ }(kg/ms)$&$1.8\times 10^{-5}$\\
			Surface tension coefficient & $\sigma\text{ }(N/m)$ & $0.0755$\\
			Reynolds number & $\text{Re}=\rho_1a\sqrt{ga}/\mu_1$ & $42792$\\ 
			Grid resolution & $\text{Nx}\times \text{Ny}$ & $400 \times 100$\\
			\hline
			\hline
		\end{tabular}
		\label{table:BD_parameters}
	\end{table}

\begin{figure}[H]
		\centering
		\subfigure[]
		{
			\includegraphics[width=6.75cm]{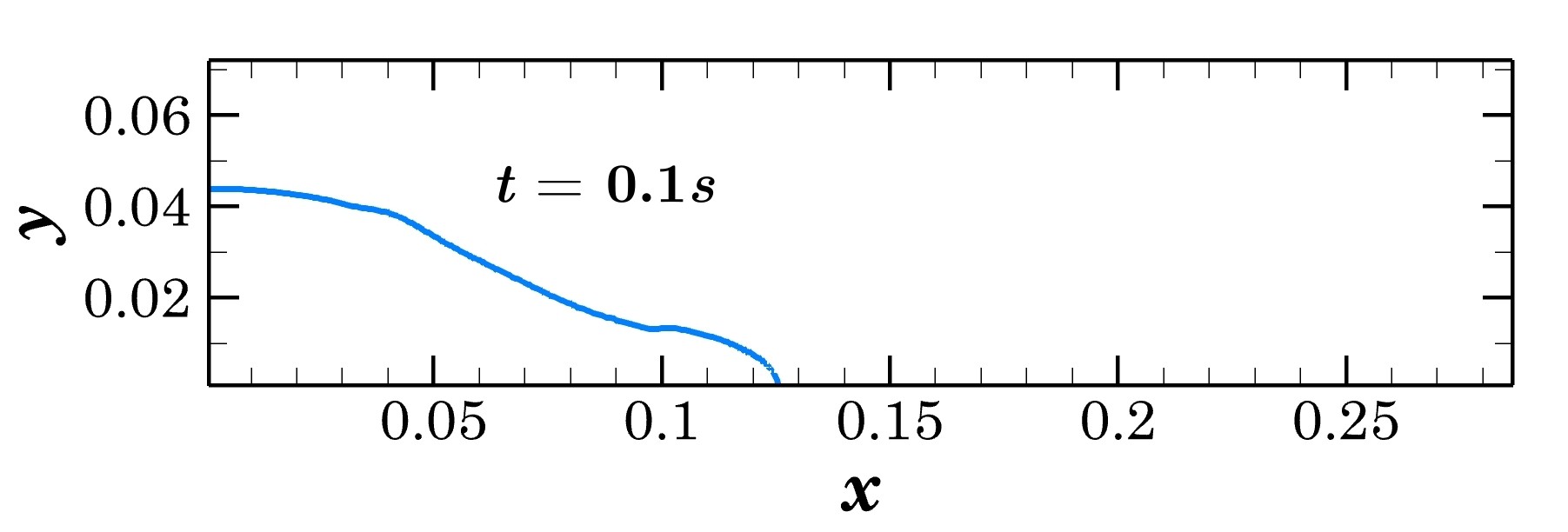}
			\label{fig_BD_evolve_a}
		}
		\subfigure[]
		{
			\includegraphics[width=6.75cm]{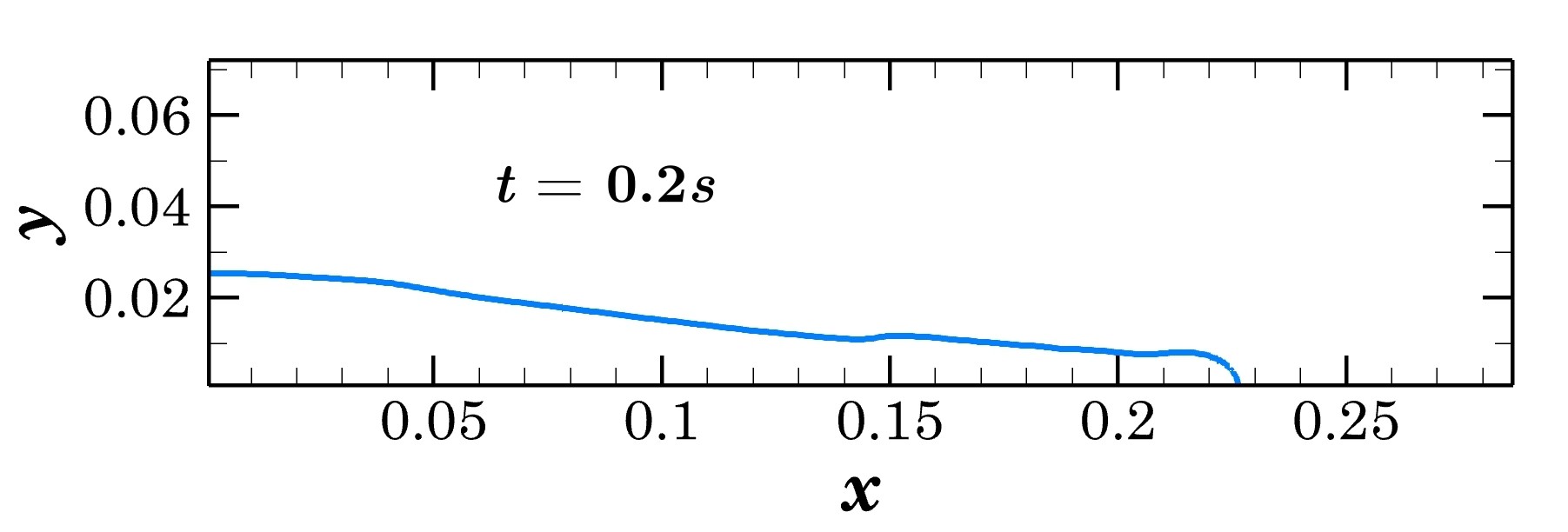}
			\label{fig_BD_evolve_b}
		}\\
		\subfigure[]
		{
			\includegraphics[width=6.75cm]{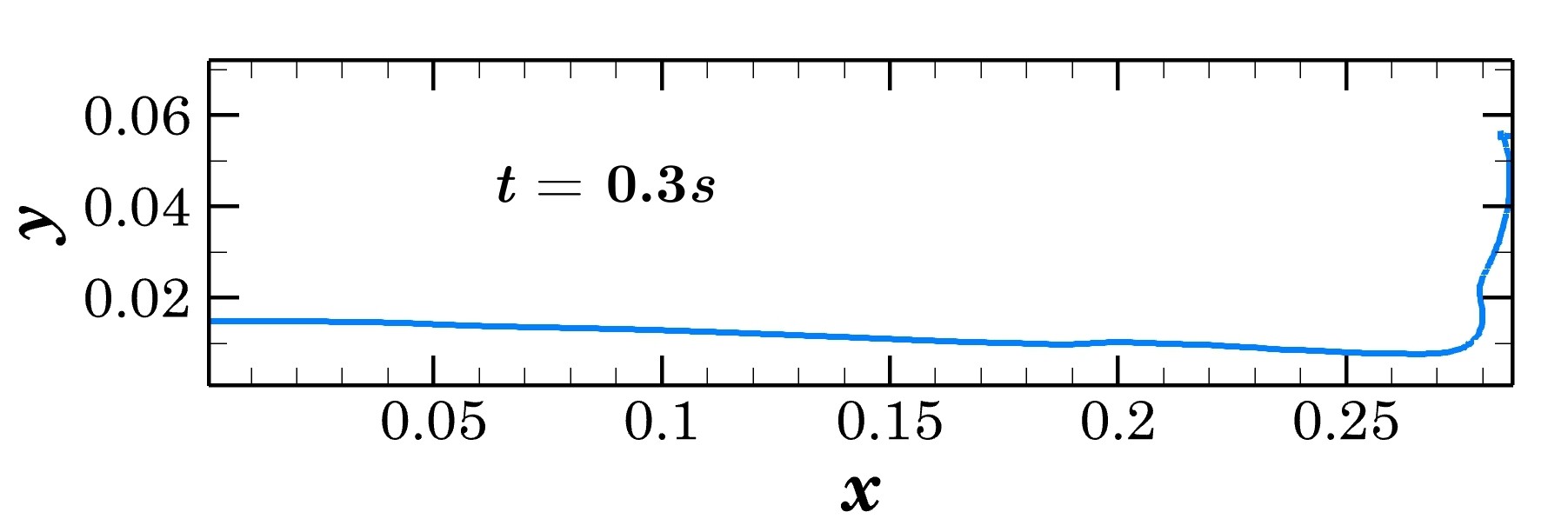}
			\label{fig_BD_evolve_c}
		}
		\subfigure[]
		{
			\includegraphics[width=6.75cm]{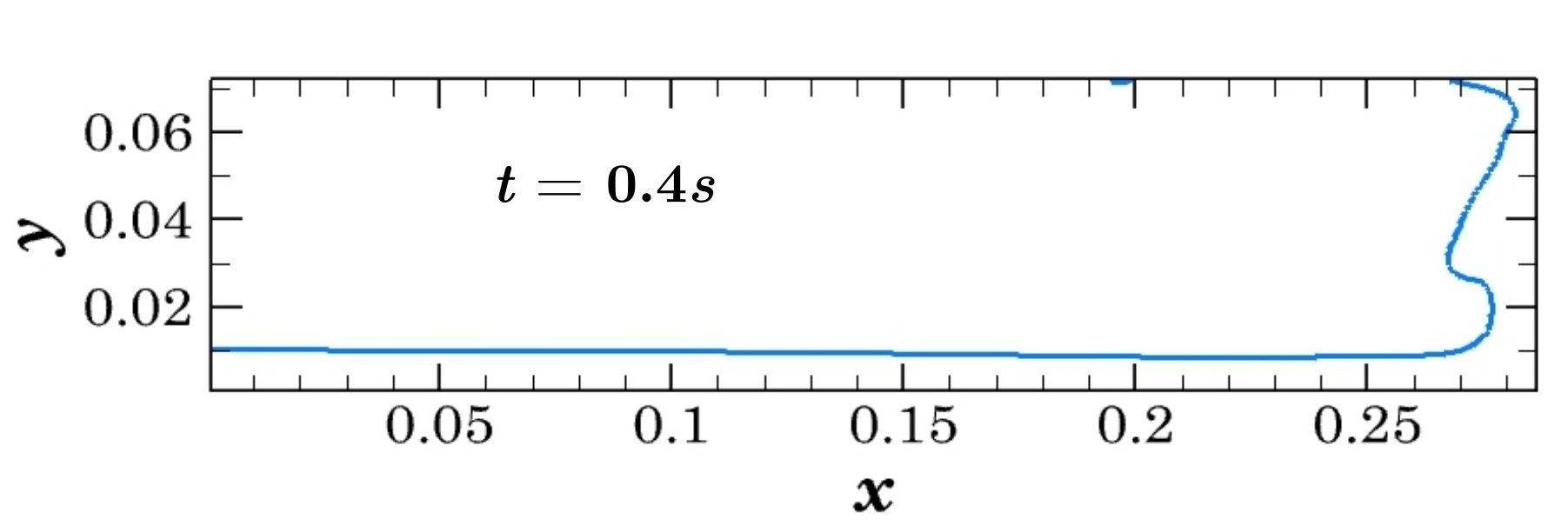}
			\label{fig_BD_evolve_d}
		}\\
		\subfigure[]
		{
			\includegraphics[width=6.75cm]{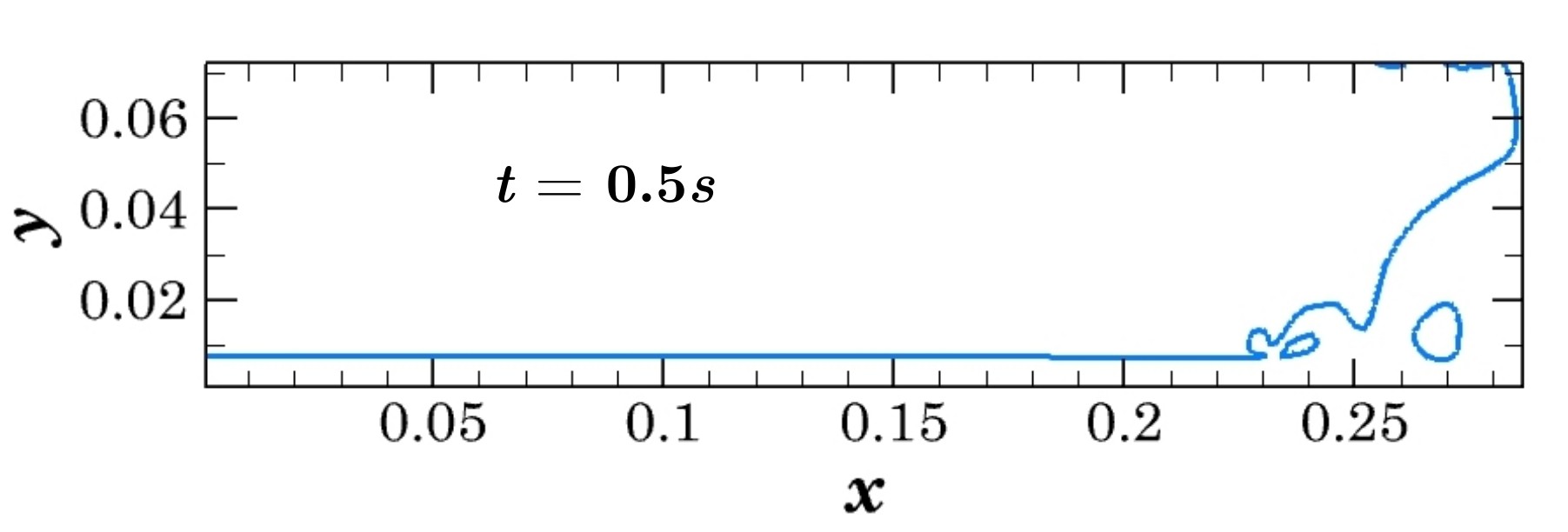}
			\label{fig_BD_evolve_e}
		}\\
		\subfigure[]
		{
			\includegraphics[width=6cm]{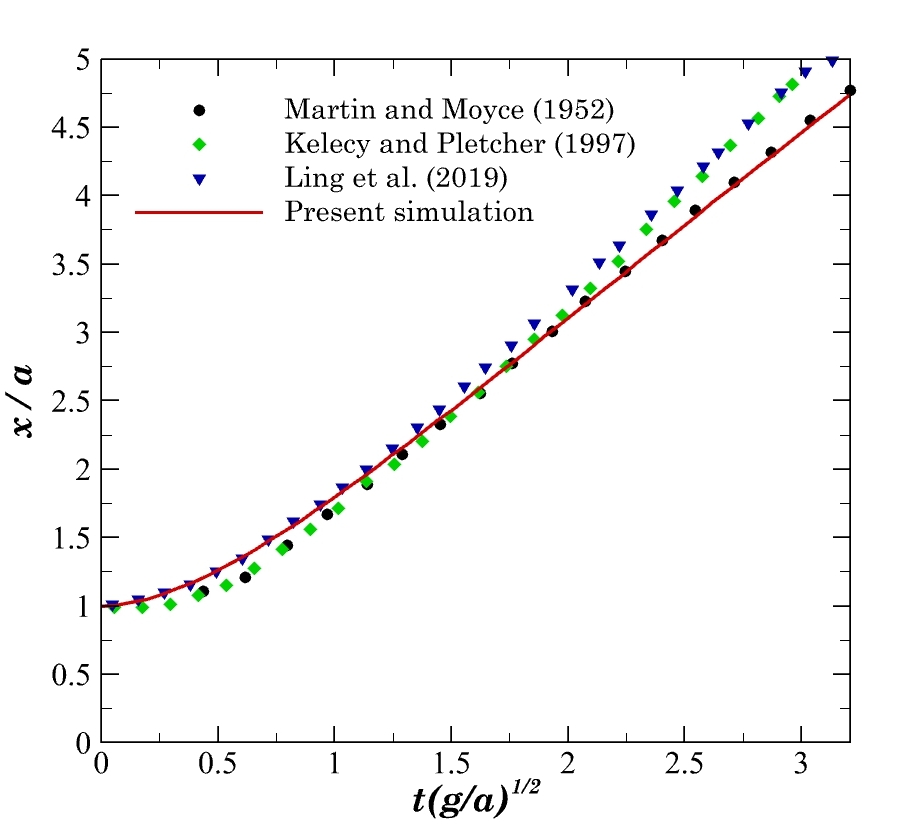}
			\label{fig_BD_evolve_f}
		}
		\subfigure[]
		{
			\includegraphics[width=6cm]{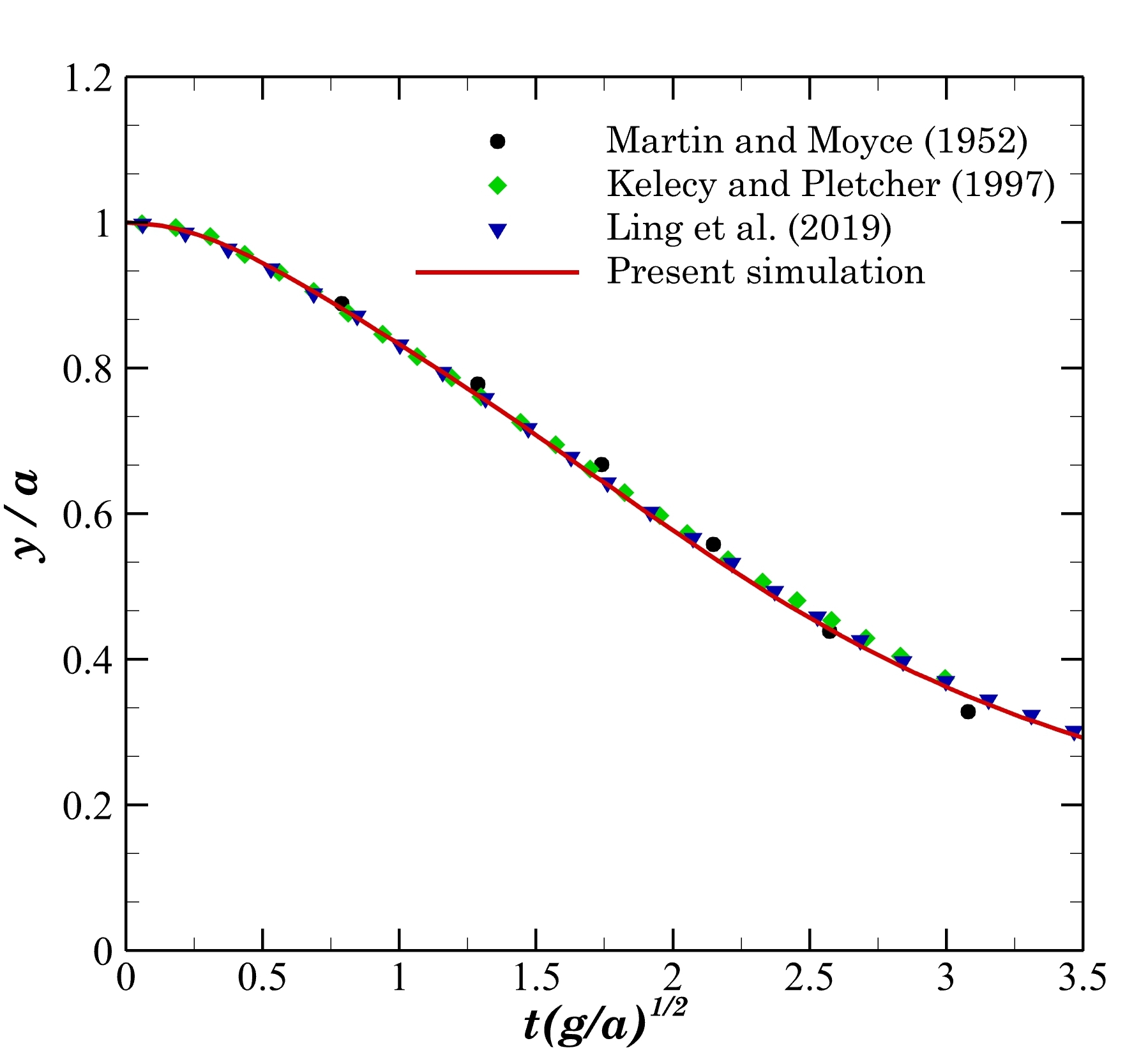}
			\label{fig_BD_evolve_g}
		}
		\caption{Temporal evolution of the interface for the broken dam flow visualized at (a)~$t=0.1s$  (b)~$t=0.2s$ and (c)~$t=0.3s$ (d)~$t=0.4s$ (e)~$t=0.5s$, (f) Propagation of the interface front along the bottom wall with time, and (g) Reduction of the liquid height along the left wall with time.}
		\label{fig_BD_evolve}
	\end{figure}

	Figure \ref{fig_BD_evolve} depicts the temporal evolution of the interface for the considered test case.  Initially,  the liquid column collapses and flows over the bottom wall (figures \ref{fig_BD_evolve_a} and \ref{fig_BD_evolve_b}). The liquid front reaches the right wall after time $t=0.2s$, crawls up the wall (figures \ref{fig_BD_evolve_c} and \ref{fig_BD_evolve_d}) and traps an air bubble near the corner, as shown in figure \ref{fig_BD_evolve_e}.
	
	The propagation of the liquid front along the bottom wall and the reduction in the liquid height along the left wall are standard quantities to investigate the accuracy of this simulation.  The time variation of these quantities is presented in figures~\ref{fig_BD_evolve_f} and \ref{fig_BD_evolve_g}, respectively.  It can be clearly seen that both the parameters describing the interface tracking are in excellent agreement with the existing literature~\cite{kelecy1997,ling2019,martin1952}. 
 
	The qualitative and quantitative agreement of the present results with those reported in the literature strongly suggests that our GPE-based solver can accurately resolve the unsteady flow field of two-phase flows, even in the case of a large density ratio without any \textit{ad hoc} modifications in the algorithm.

 	\begin{figure}[H]
		\centering
		\subfigure[]
		{
			\includegraphics[width=7.5cm]{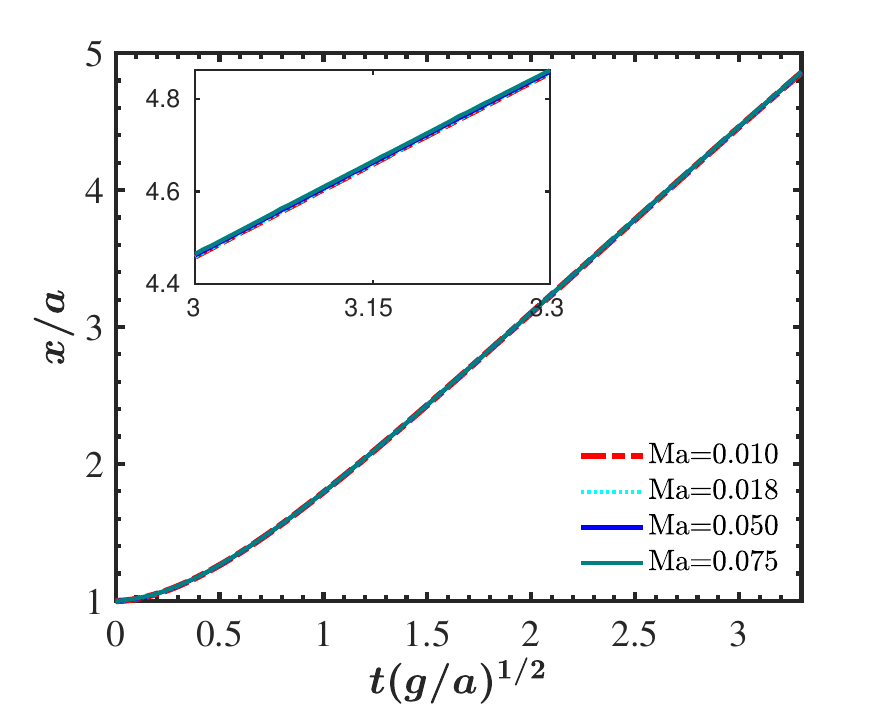}
			\label{fig_BD_Ma_xTrack}
		}
		\subfigure[]
		{
			\includegraphics[width=7.5cm]{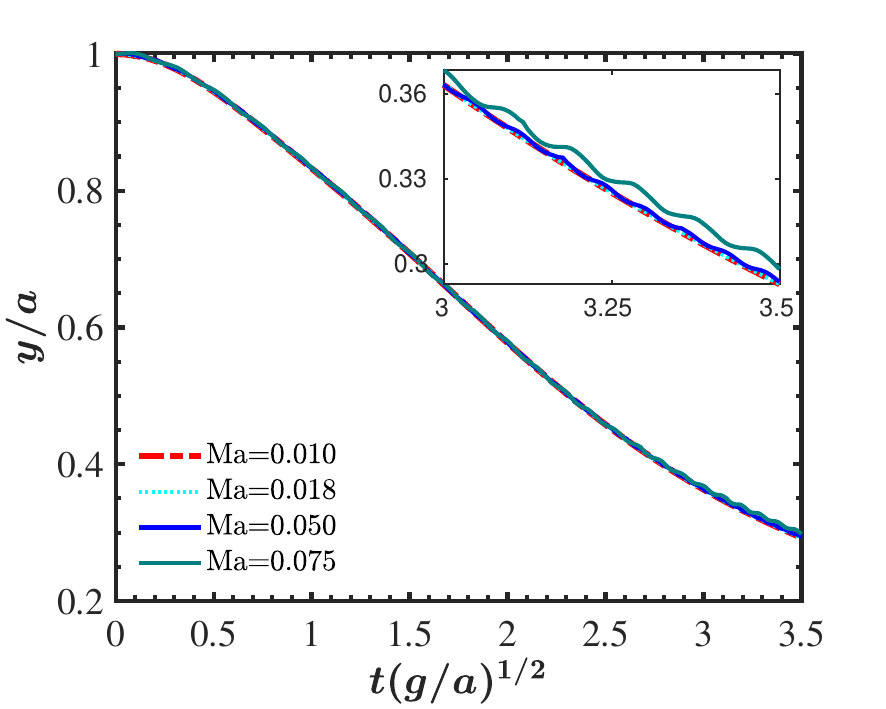}
			\label{fig_BD_Ma_yTrack}
		}
		\caption{Effect of Mach number on the broken dam flow problem (a) Propagation of the interface front along the bottom wall with time, and (b) Reduction of the liquid height along the left wall with time.}
		\label{fig_damBreak_MaStudy}
	\end{figure}

Similar to the one performed for the 2D Rayleigh Taylor instability problem, we study the effect of $\text{Ma}$ for this high-density ratio test case. The $\text{Ma}$ values considered are 0.01, 0.018, 0.05 and 0.075; the simulations with $\text{Ma}>0.075$ did not converge. However, we would like to emphasize that this is not a limitation of the present approach. In theory, weakly compressible approaches model incompressible flows when $\text{Ma}\ll 1$. Figure \ref{fig_damBreak_MaStudy} depicts the effect of Mach number on the propagation of the interface front along the bottom wall and the reduction of the liquid height along the left wall with time. While figure~\ref{fig_BD_Ma_xTrack}, shows no noticeable difference, figure~\ref{fig_BD_Ma_yTrack} indicates nonphysical oscillations at $\text{Ma}=0.075$. We observe that the oscillations are clearly visible at $\text{Ma}=0.075$, less pronounced when $\text{Ma}=0.05$, and negligible for the lower Ma cases. This indicates that acoustic sound waves are quickly damped when $\text{Ma}$ is small, imparting more stable and accurate results. This observation is similar to the results reported in the Rayleigh Taylor instability test case.

	\subsection{Bubble rise}
	\label{subsec4.4}
	In many practical applications of two-phase flows, the influence of the surface tension is high enough to manifest complex topological changes. It is of utmost importance that the numerical method is capable of producing accurate results for such cases. Hence, we test our solver against the problem of a rising gas bubble in a liquid-filled domain. This bubble rise test case was proposed as a rigorous benchmark case for two-phase flow solvers~\cite{hysing2009}. Two test cases from Hysing et al.~\cite{hysing2009} are considered. The first test case involves moderate bubble deformation, while the second test case features significantly complex topological changes due to high density and viscosity ratios. The later test case offers a considerable challenge to the method. 
	
	The test simulation is initialized with a patch of a circle of radius $r_0=0.25$ centred at ($0.5,0.5$) in a $1\times 2$ rectangular domain. The initial velocities and pressure in the domain are zero, with gravity acting in the negative $y$-direction. The domain is equipped with no-slip boundaries in the $y$-direction and free-slip boundaries in the $x$-direction. The parameters considered for the simulation are listed in table \ref{table:BR_parameters}. The third-order QUICK scheme is used to discretize the advected velocity component in the advection term. Note that unlike the test cases mentioned in the previous subsections, the gravitational constant for the current problem is taken as $0.98$, as used in Hysing et al.~\cite{hysing2009}. The time step chosen for case 1 is $10^{-4}$, whereas for case 2 (challenging test case) is $10^{-5}$. Similar to the broken dam problem, $c_s = \Delta x/(\sqrt{3} \Delta t)$.
	
	\begin{table}[H]
		\centering
		\caption{Parameters defining the bubble rise problem.}
		\begin{tabular}{lccc}
			\hline
			\hline
			\textbf{Name} & \textbf{Definition} &
			\multicolumn{2}{c}{\textbf{Value}} \\
			& & \textbf{Case 1} & \textbf{Case 2} \\ 
			\hline 
			Surrounding fluid density & $\rho_1$ & $1000$ & $1000$\\
			Surrounding fluid viscosity & $\mu_1$&$10$& $10$\\
			Bubble density & $\rho_2$&$100$& $1$\\
			Bubble viscosity & $\mu_2$&$1$& $0.1$\\
			Surface tension coefficient & $\sigma$ & $24.5$ & $1.96$\\
			Reynolds number & $\text{Re}=\rho_1r_0\sqrt{2gr_0}/\mu_1$&$420$& $300$\\
			E$\ddot{o}$tv$\ddot{o}$s number & $\text{Eo}=4\rho_1g{r_0}^2/\sigma$ & $10$ & $125$ \\
			Grid resolution & $\text{Nx}\times \text{Ny}$ & $80 \times 160$ & $80 \times 160$ \\
			\hline
			\hline
		\end{tabular}
		\label{table:BR_parameters}
	\end{table}
	
\begin{figure}[]
		\centering
		\subfigure[]
		{
			\includegraphics[height=5cm]{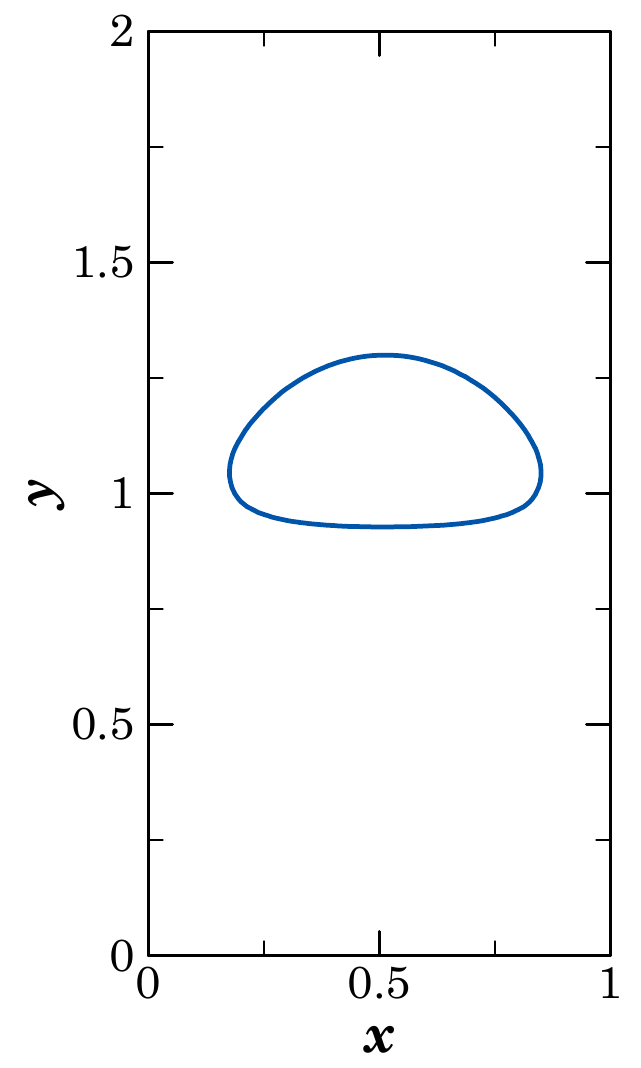}
			\label{fig_BR_evolve_a}
		}
		\subfigure[]
		{
			\includegraphics[height=5cm]{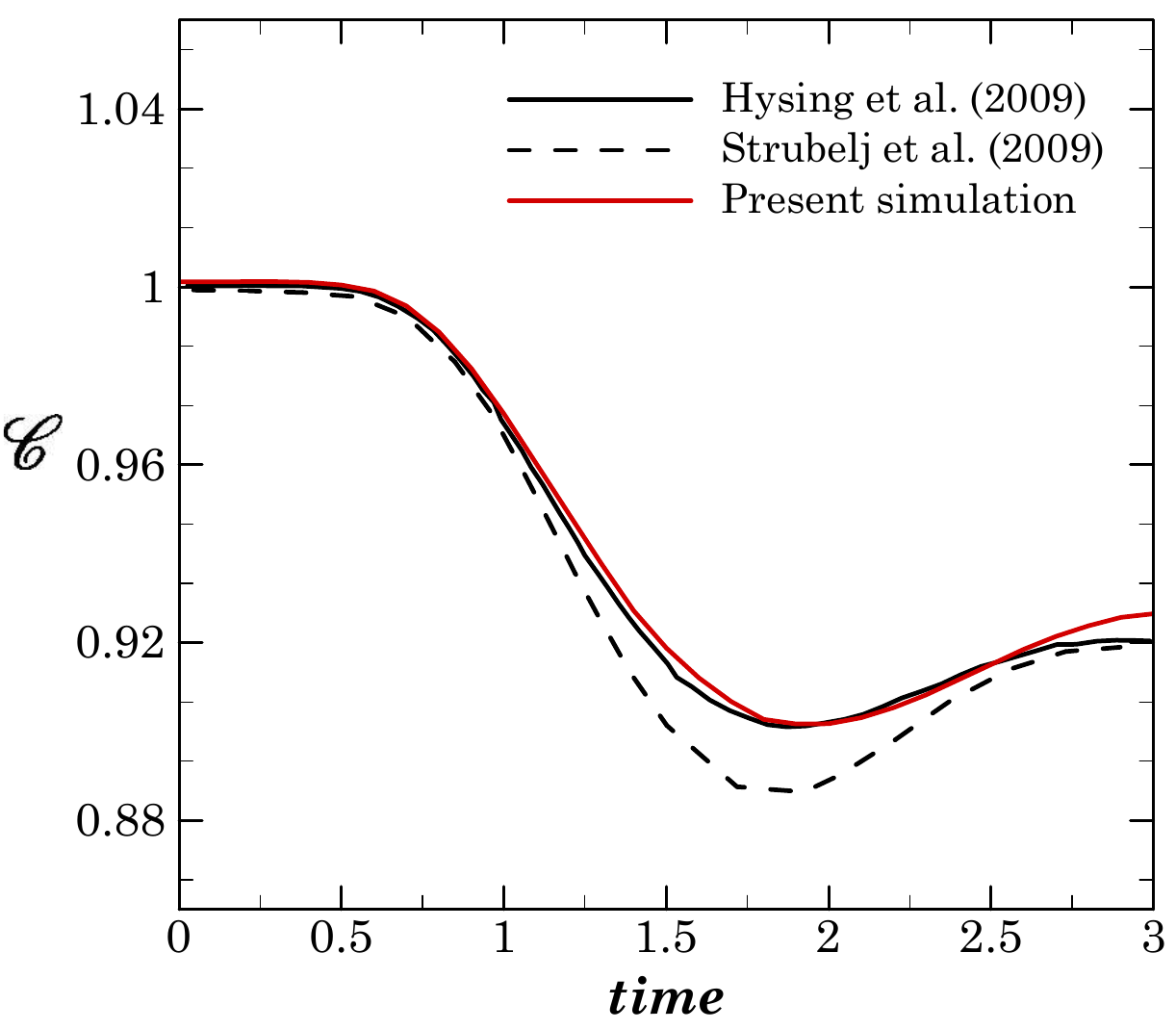}
			\label{fig_BR_evolve_b}
		}
		\subfigure[]
		{
			\includegraphics[height=5cm]{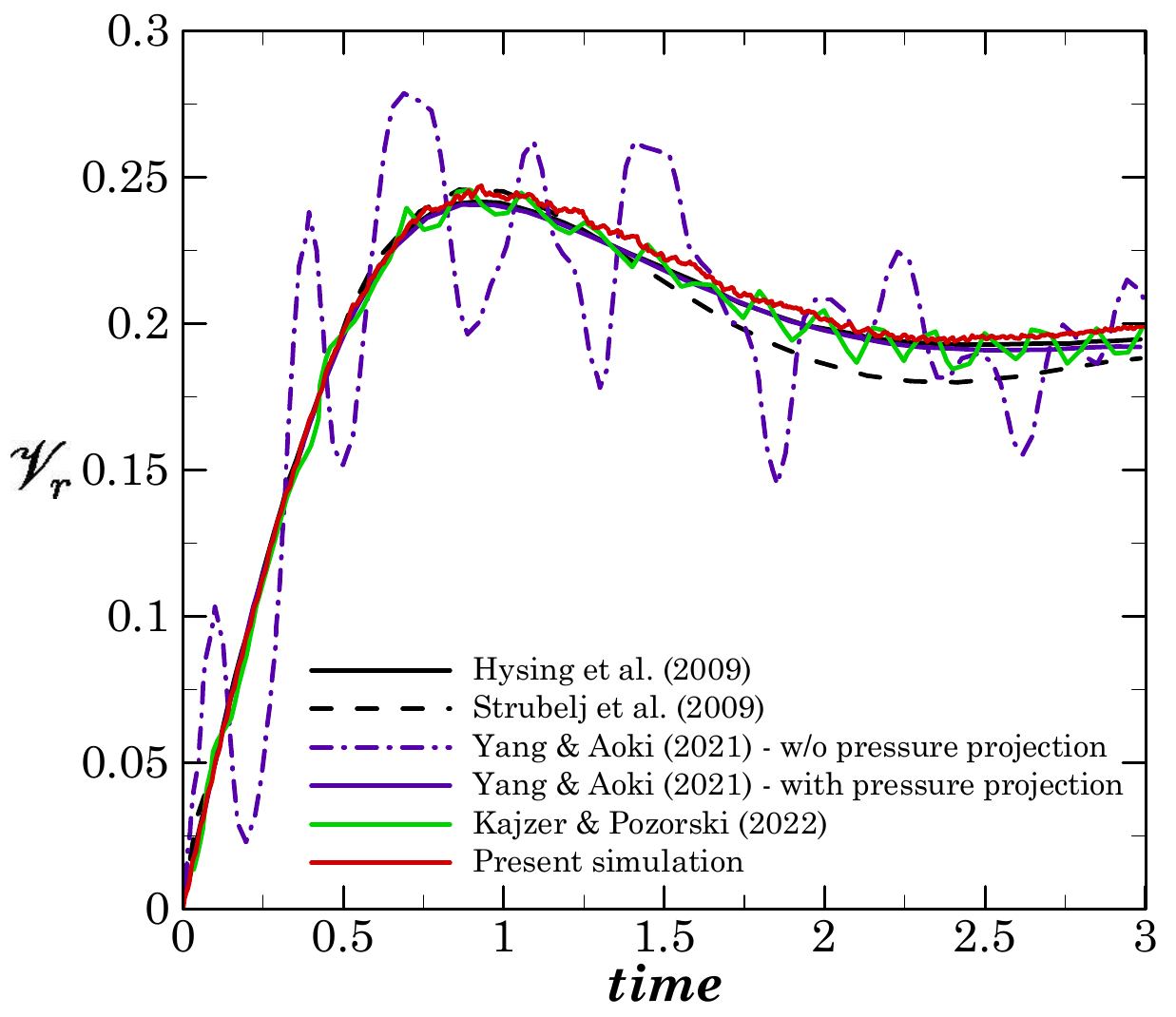}
			\label{fig_BR_evolve_c}
		}\\
		\subfigure[]
		{
			\includegraphics[height=5cm]{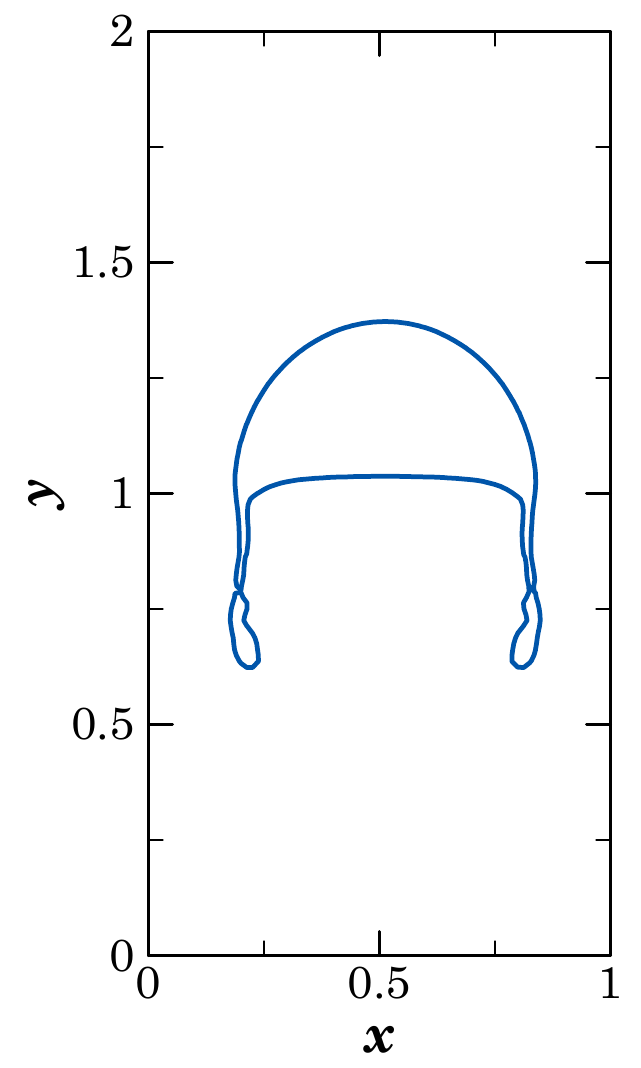}
			\label{fig_BR_evolve_d}
		}
		\subfigure[]
		{
			\includegraphics[height=5cm]{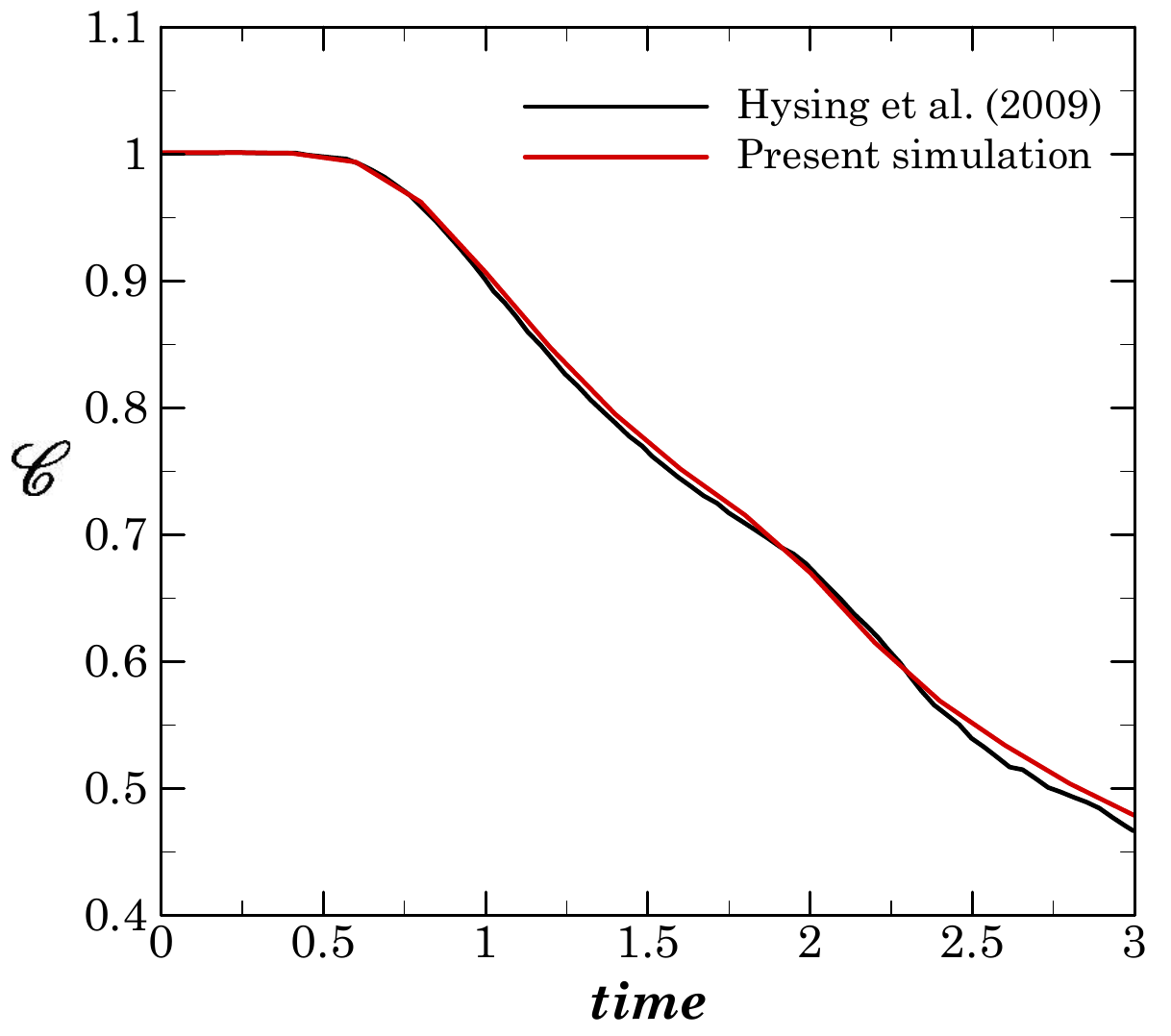}
			\label{fig_BR_evolve_e}
		}
		\subfigure[]
		{
			\includegraphics[height=5cm]{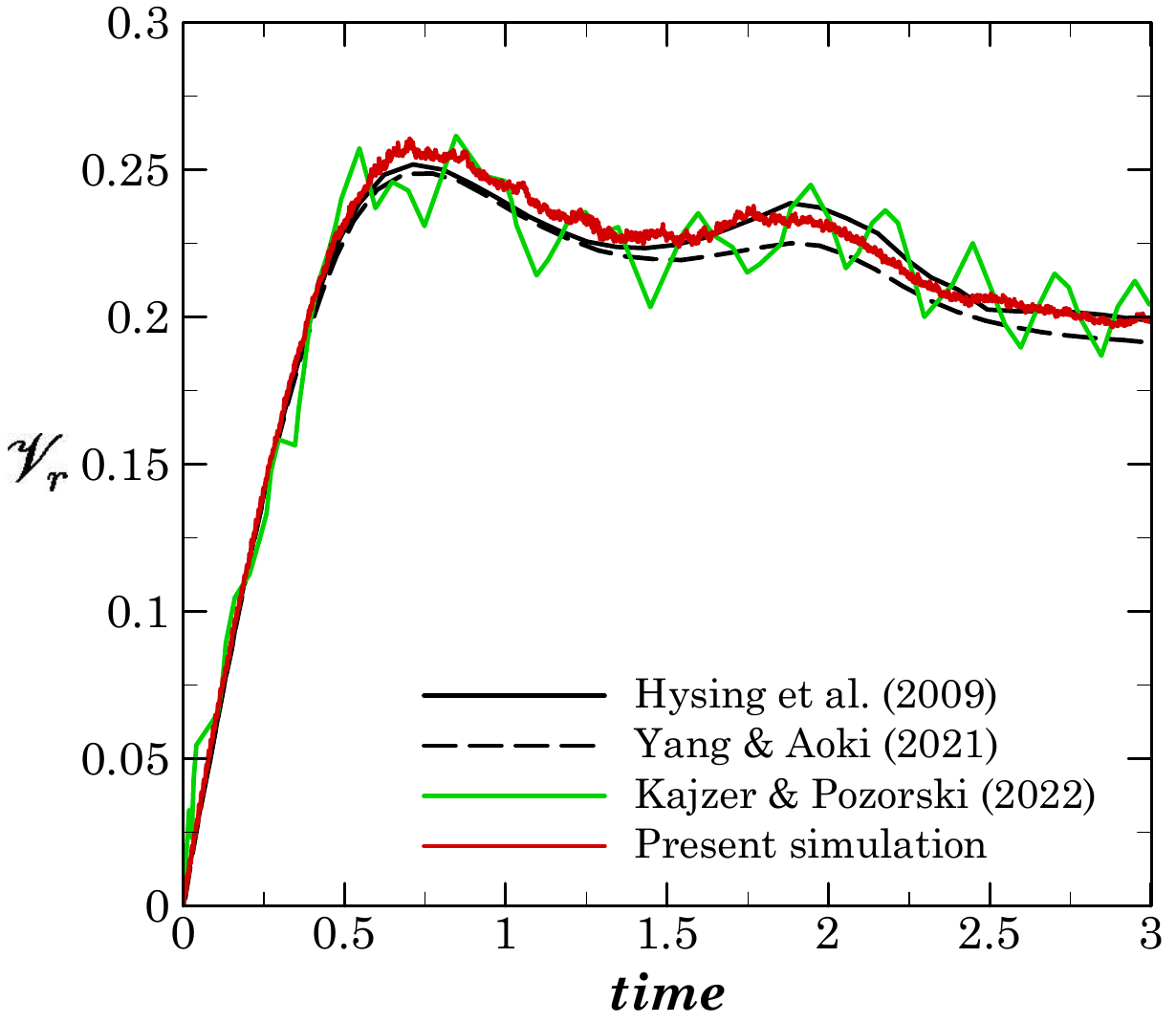}
			\label{fig_BR_evolve_f}
		}
		\caption{Results of the bubble rise problem for Case 1 (top row) and Case 2 (bottom row). Final shape of the bubble at $t=3.0$ (left column), and circularity ($\mathscr{C}$) (centre column) and rise velocity $\mathscr{V}_r$ (right column) of the bubble.} 
		\label{fig_BR_evolve}
	\end{figure}

	The standard quantification parameters to describe the temporal evolution of the bubble are the circularity~($\mathscr{C}$) and the rise velocity~($\mathscr{V}_r$). $\mathscr{C}$ is the ratio of the perimeter of the area-equivalent circle to the instantaneous perimeter of the bubble. $\mathscr{V}_r$ refers to the mean velocity with which the bubble is rising. Mathematically, 
	\begin{equation}
		\mathscr{C} = \frac{2\pi r_0}{P_b},
		\label{eqn:circularity}
	\end{equation}
	\begin{equation}
		\mathscr{V}_r = \frac{\sum_{n=1}^{N}|v|\forall_n}{\sum_{n=1}^{N}\forall_n},
		\label{eqn:rise_velo}
	\end{equation}
	where $\forall_n$ refers to the cell volume, $v$ refers to the velocity opposing the gravity and $P_b$ denotes the instantaneous perimeter of the bubble, which is computed using the `integrate variables' filter available in the open source tool, \textit{Paraview}~\cite{ayachit2015paraview}. The first test case involves a high surface tension coefficient value and low density and viscosity ratios. As a result of prevailing surface tension effects, the bubble deforms moderately: the initially circular drop takes an ellipsoidal shape at the final time instant, as evidenced by figure \ref{fig_BR_evolve_a}. As can be seen from figures \ref{fig_BR_evolve_b} and \ref{fig_BR_evolve_c}, the quantification parameters are in excellent agreement with Hysing et al.~\cite{hysing2009} and better than the simulation results of {\v{S}}trubelj et al.~\cite{strubelj2009} and the EDAC based solver of Kajzer and Pozorski~\cite{kajzer2022}. We want to point out that in the study of Yang and Aoki~\cite{yang2021}, an evolving pressure equation is solved for a finite number of iterations to dampen the acoustic waves for case 1 of the bubble rise problem. The smooth curve obtained by Yang and Aoki, as observed in figure \ref{fig_BR_evolve_c} ({solid purple line}), is a result of $10$ iterations of the evolving pressure equation. However, in the present study, no iterative technique is used in the algorithm, and the current results exhibit much less oscillatory behaviour than Yang and Aoki's counterpart ({purple dashed-dot line}). 
	
	In the second test case, the bubble undergoes a significant deformation due to the combined effect of low surface tension and a large jump in material properties across the interface, as depicted in figure \ref{fig_BR_evolve_d}. The quantification parameters are compared with those in the existing literature. As can be evidenced from figures \ref{fig_BR_evolve_e} and \ref{fig_BR_evolve_f}, the present study results are in excellent agreement with Hysing et al.~\cite{hysing2009}. We stress that the oscillations in the plot are negligible when compared to the other relevant weakly compressible approaches~\cite{kajzer2022,yang2021}.
	
\begin{figure}[]
		\centering
		\subfigure[]
		{
			\includegraphics[height=6cm]{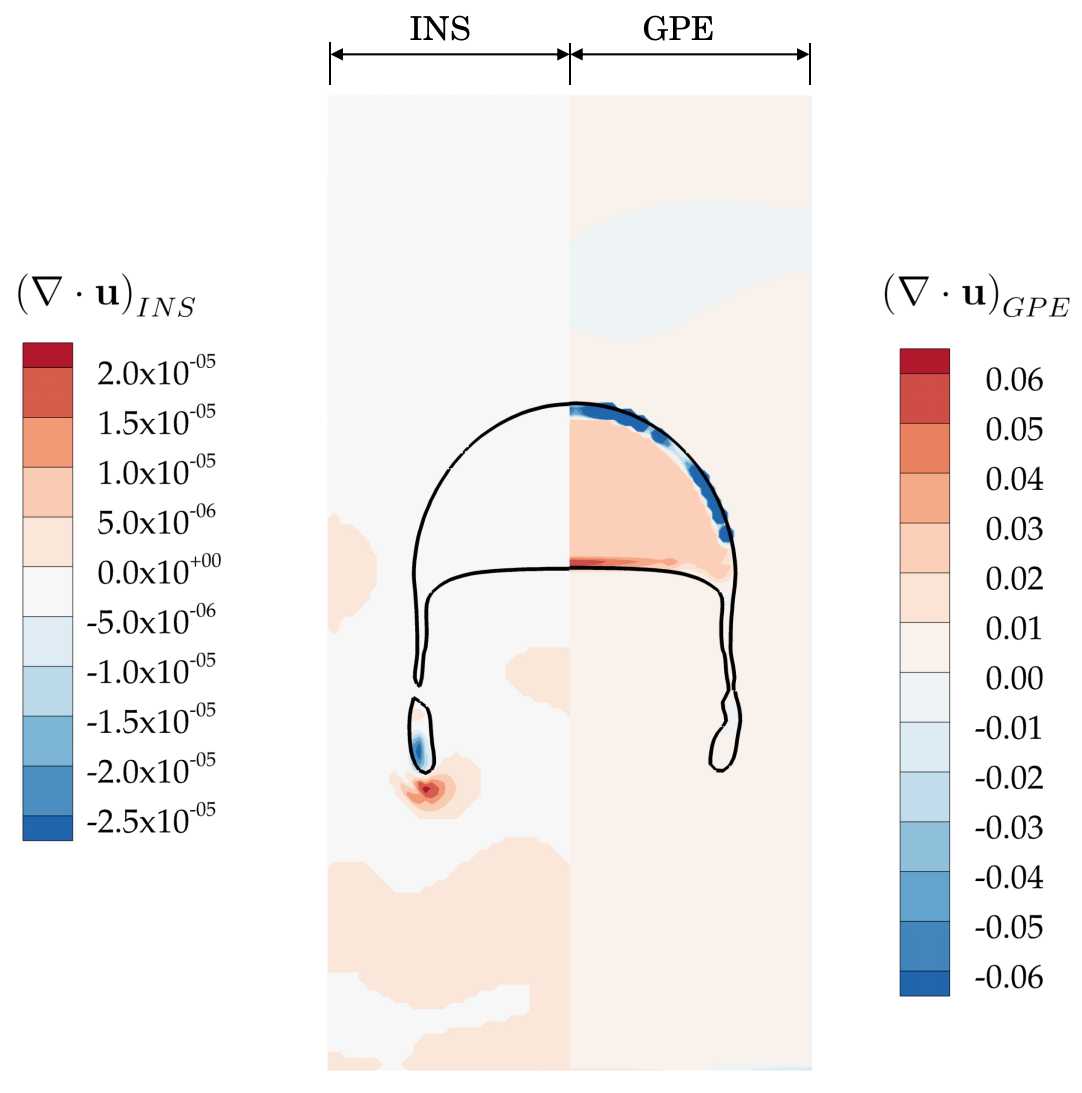}
			\label{fig_BR_vel_div_a}
		}
		\subfigure[]
		{
			\includegraphics[height=6cm]{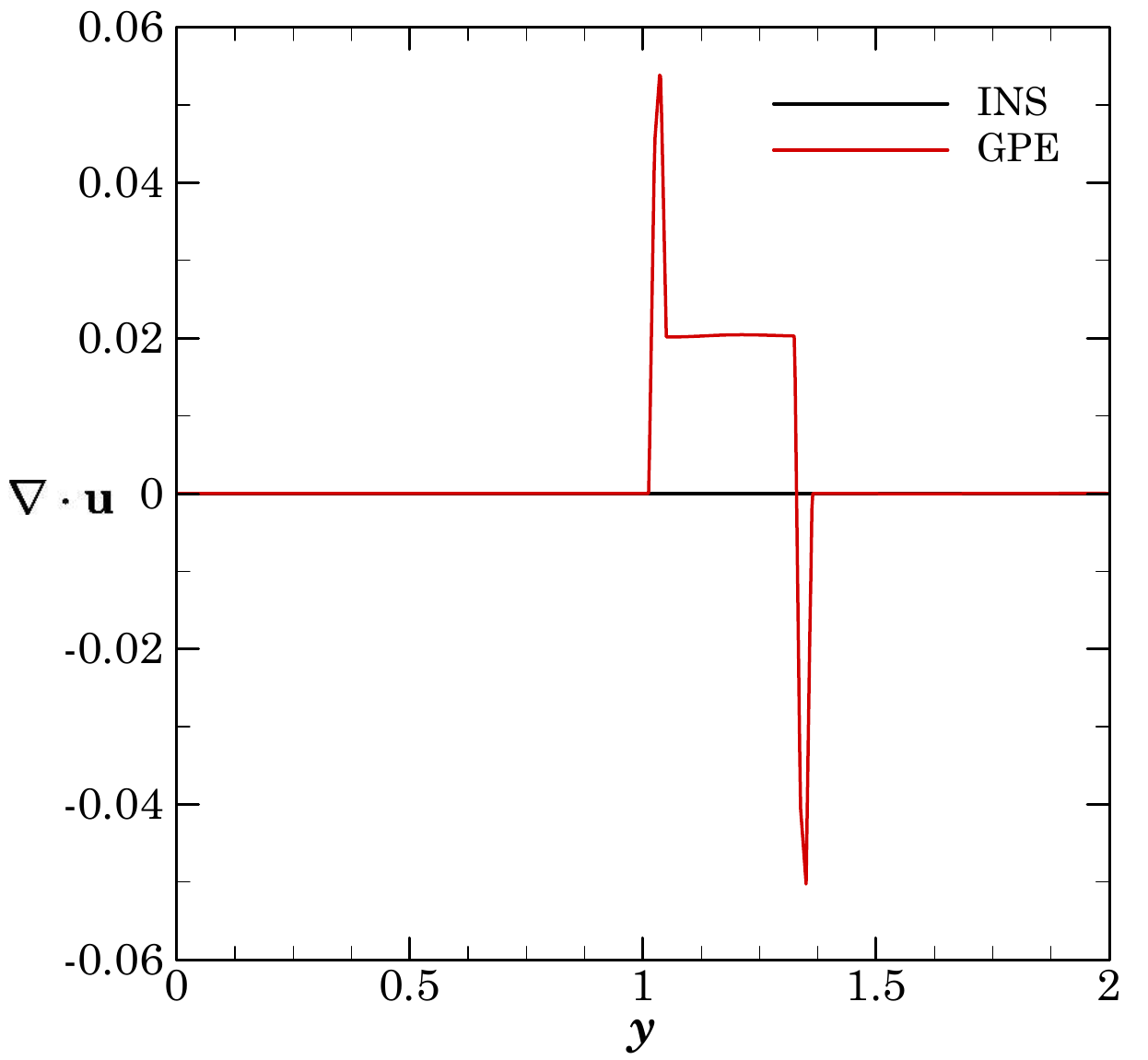}
			\label{fig_BR_vel_div_b}
		}
		\caption{Comparison of the velocity divergence between the INS and the GPE-based solver for Case 2 of the bubble rise: (a) filled contour, with solid black lines representing interface, and (b) along the vertical centreline at $x=0.5$.}
		\label{fig_BR_vel_div}
	\end{figure}

	Figure \ref{fig_BR_vel_div_a} shows the velocity divergence ($\mathbf{\nabla} \cdot \mathbf{u}$) field for the challenging test case of bubble rise at $time=3.0$. It is computed using the GPE-based solver and the traditional INS involving the pressure Poisson equation. The same is quantified in figure \ref{fig_BR_vel_div_b}, which shows the velocity divergence along the vertical centreline of the domain at $x=0.5$. The discretization details in the staggered framework for the pressure Poisson-based algorithm are similar to the GPE-based solver, except that the algorithm features the conventional predictor-corrector approach~\cite{ferziger2002}. Figure \ref{fig_BR_vel_div_b} shows that the GPE-based solver's velocity divergence is significantly higher than the INS. Despite such a large error in ($\mathbf{\nabla} \cdot \mathbf{u}$), our solver can accurately capture the qualitative and quantitative features of the challenging test case. We reiterate that the oscillations present in our results are significantly lower than the other weakly compressible two-phase flow solvers~\cite{kajzer2022,yang2021}. 

\begin{figure}[]
		\centering
		\subfigure[]
		{
			\includegraphics[height=6.5cm]{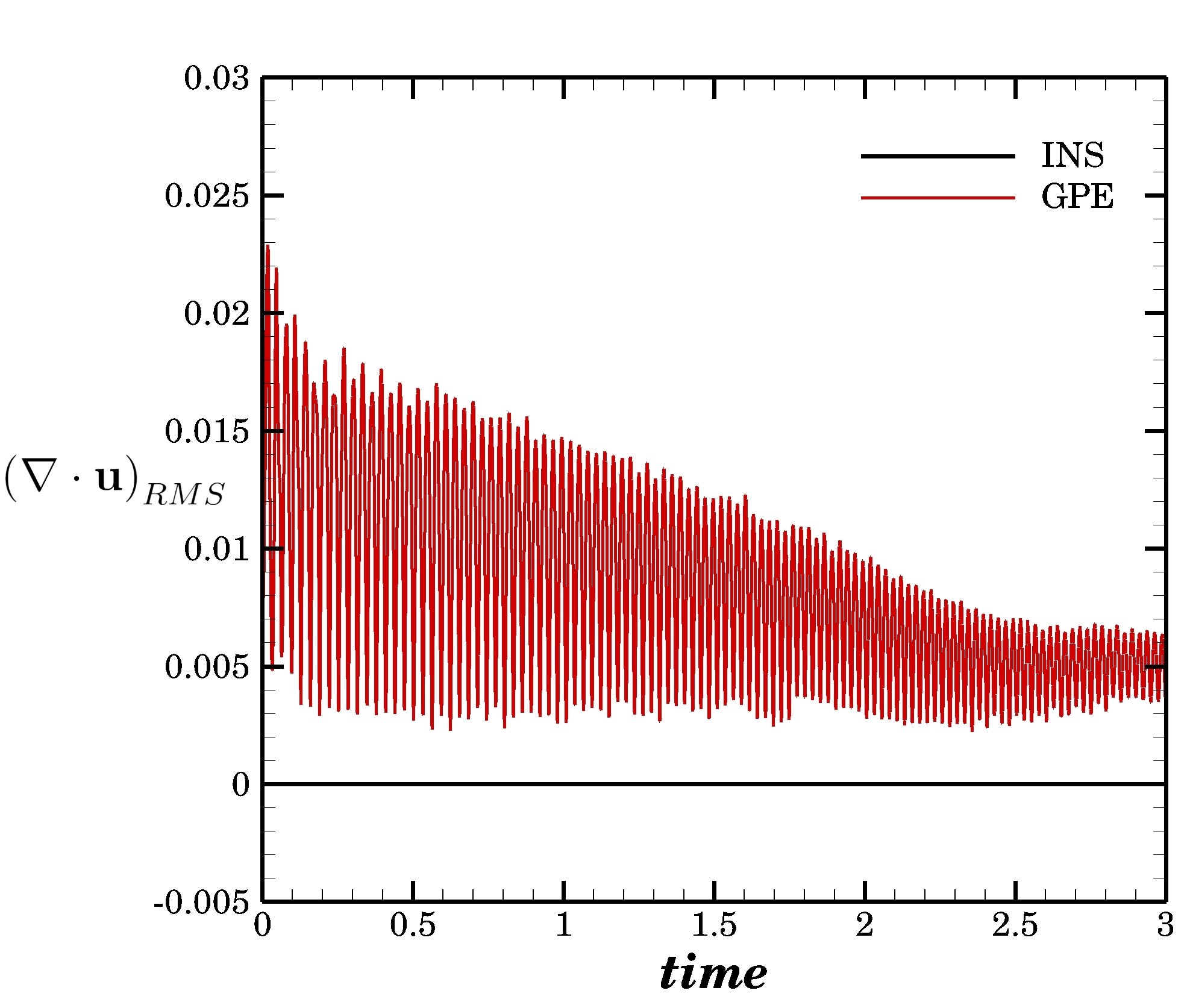}
			\label{rms_velDiv_hysing1}
		}
		\subfigure[]
		{
			\includegraphics[height=6.5cm]{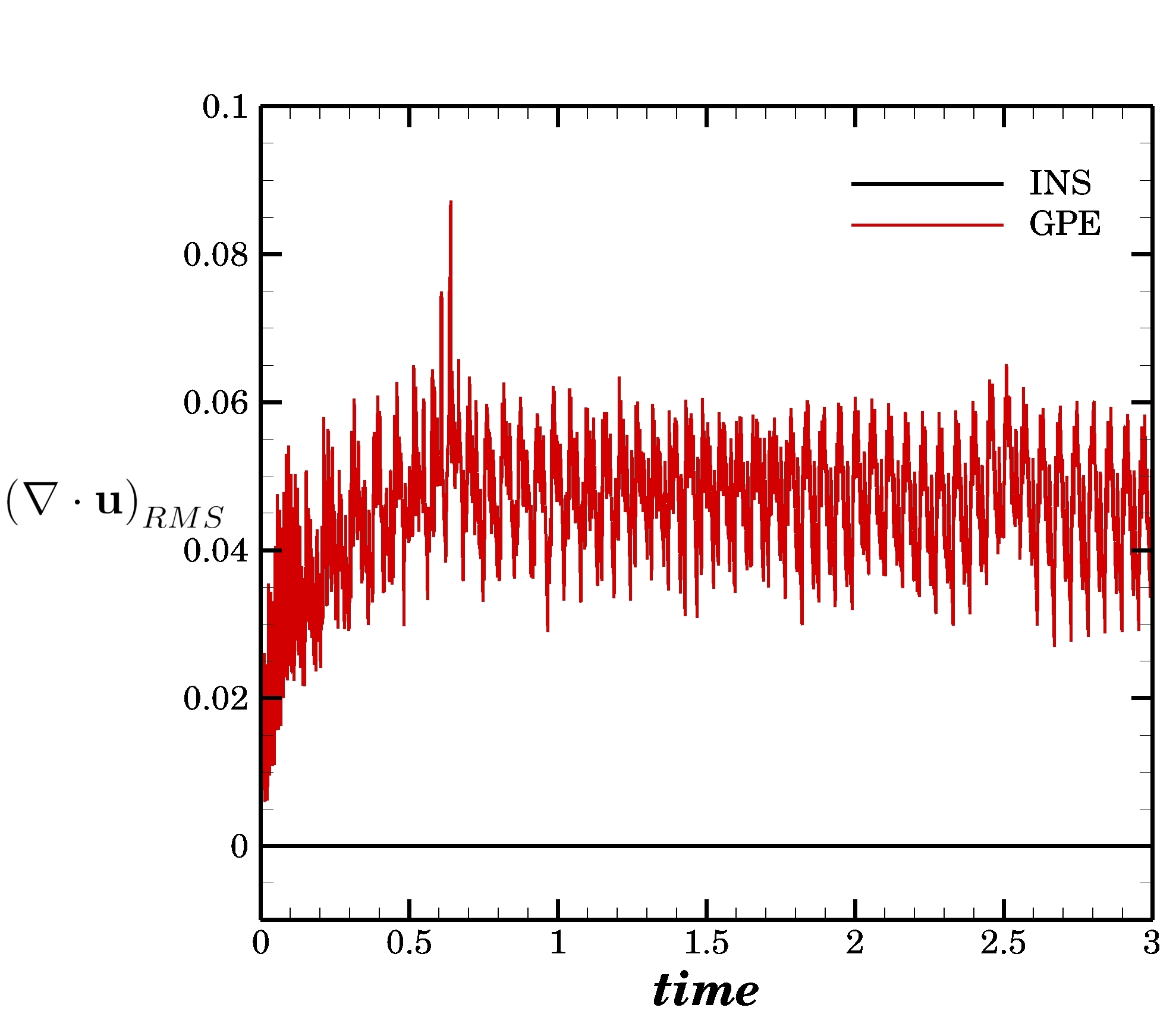}
			\label{rms_velDiv_hysing2}
		}
		\caption{Evolution of the RMS of velocity divergence with time for the INS and the GPE-based solver for (a)~case 1 and (b)~case 2 of the bubble rise problem.}
		\label{fig_BR_vel_div_rms}
	\end{figure}

While the error in $\mathbf{\nabla} \cdot \mathbf{u}$ at a particular time presented in figure~\ref{fig_BR_vel_div} reveals an important information, a more relevant quantity to consider is the root-mean-square (RMS) value of the velocity divergence, denoted as $\left(\nabla \cdot \mathbf{u}\right)_{RMS}$, as the time evolves. For both variants of the bubble rise problem, the variation of $\left(\nabla\cdot\mathbf{u}\right)_{RMS}$ with time is illustrated in figure~\ref{fig_BR_vel_div_rms}, for both GPE and INS solvers. As expected, the error produced by the INS solver is very small (of the order of $10^{-5}$). However, for the GPE solver, we see that the error is of the order of $10^{-2}$. In comparison, Toutant~\cite{toutant2018} reported the RMS of velocity divergence error of the order of $10^{-3}$ for simulations of single-phase flows using GPE. This larger error for the two-phase GPE solver is due to the peak occurring in the interface region, as shown in figure~\ref{fig_BR_vel_div}. However, key takeaways from figure~\ref{fig_BR_vel_div_rms} are as follows: (i)~although the density ratio is increased by two orders of magnitude from case 1 to 2, the corresponding $\left(\nabla\cdot\mathbf{u}\right)_{RMS}$ increased only marginally, and (ii)~the velocity divergence error does not get accumulated over time. The later observation indicates that the proposed method works for large density ratio problems also.

	\subsection{3D Rayleigh-Taylor Instability}
	\label{subsec4.5}
\begin{figure}[]
		\centering
		\includegraphics[width=12cm]{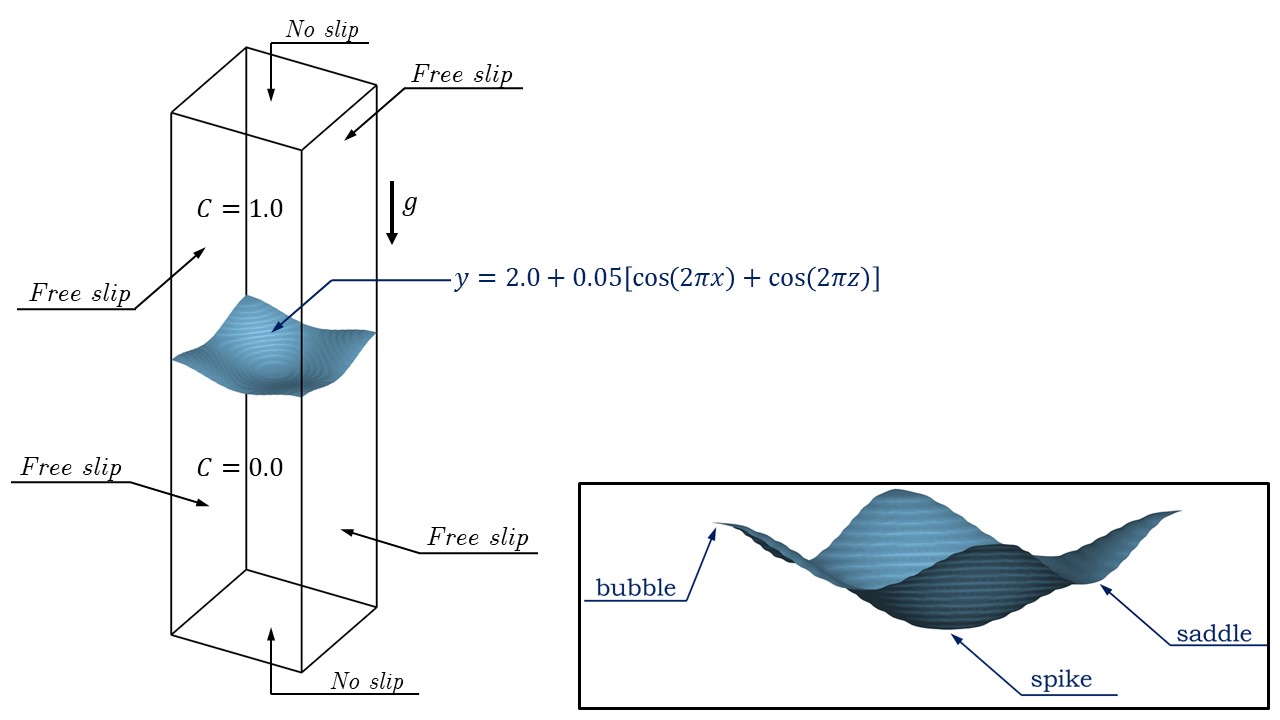}
		\caption{Initial flow configuration for 3D Rayleigh-Taylor instability including the initial perturbation and boundary conditions.}
		\label{fig_initial_3D_RT}
	\end{figure}

	All four previous test cases are two-dimensional. To show the proposed method's applicability to three-dimensional flows, we simulate the 3D Rayleigh-Taylor instability. The test case is adopted from Saito et al.~\cite{saito2017} and the parameters are listed in table \ref{table:3D_RT_parameters}. The simulation is set up with the heavier fluid placed on top of the lighter fluid in a cuboidal domain of $1.0 \times 4.0 \times 1.0$. The interface between the two fluids is initially perturbed sinusoidally, as shown in figure \ref{fig_initial_3D_RT}, to trigger the instability. The entire domain is employed with the free-slip condition on the lateral sides in conjunction with the no-slip condition at the top and bottom boundary. The advection scheme used for the test case is the second-order central difference scheme for discretizing the advected velocity component in the advection term.

	\begin{table}[H]
	\centering
	\caption{Parameters for 3D Rayleigh-Taylor instability adopted from Saito et al.~\cite{saito2017}.}
	\begin{tabular}{lccc}
		\hline
		\hline
		\textbf{Name} & \textbf{Definition} & \textbf{Value} \\
		\hline 
		Length & $L\text{ }(m)$ & $1.0$ \\
		Height & $H\text{ }(m)$ & $4.0$ \\
		Width  & $W\text{ }(m)$ & $1.0$ \\
		Heavier fluid density & $\rho_1\text{ }(kg/m^3)$ & $3.0$ \\
		Heavier fluid viscosity & $\mu_1\text{ }(kg/ms)$&$3.0$ \\
		Lighter fluid density & $\rho_2\text{ }(kg/m^3)$&$1.0$ \\
		Lighter fluid viscosity & $\mu_2\text{ }(kg/ms)$&$1.0$ \\
		Reynolds number & $\text{Re}=\rho_1L\sqrt{gL}/\mu_1$&$512$ \\
		Froude number & $\text{Fr}=(\sqrt{gL})^2/(gL)$&$1.0$ \\
		Grid resolution & $\text{Nx}\times \text{Ny} \times \text{Nz}$ & $64 \times 256\times 64$ \\
		\hline
		\hline
	\end{tabular}
	\label{table:3D_RT_parameters}
	\end{table}

\begin{figure}[]
		\centering
		\subfigure[]
		{
			\includegraphics[height=8cm]{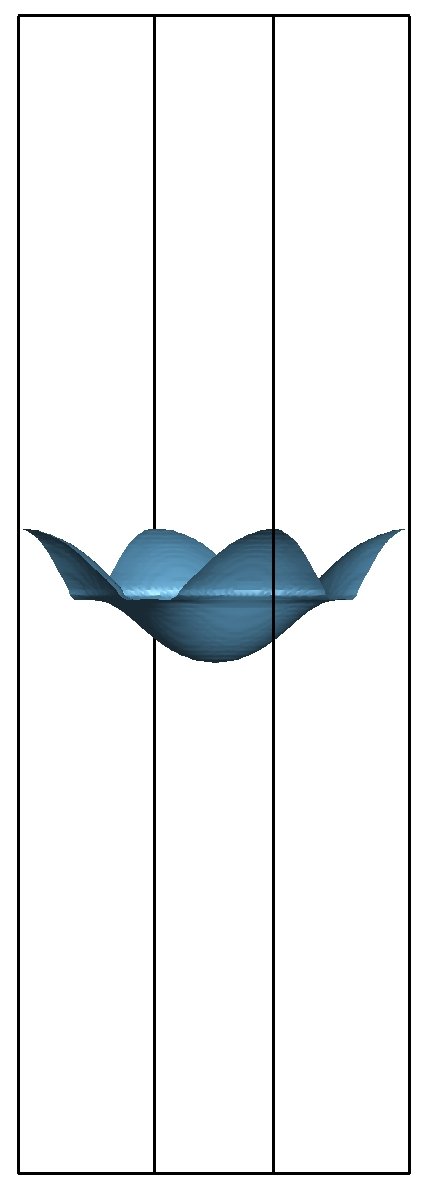}
			\label{fig_3D_RT_evolve_a}
		}
		\subfigure[]
		{
			\includegraphics[height=8cm]{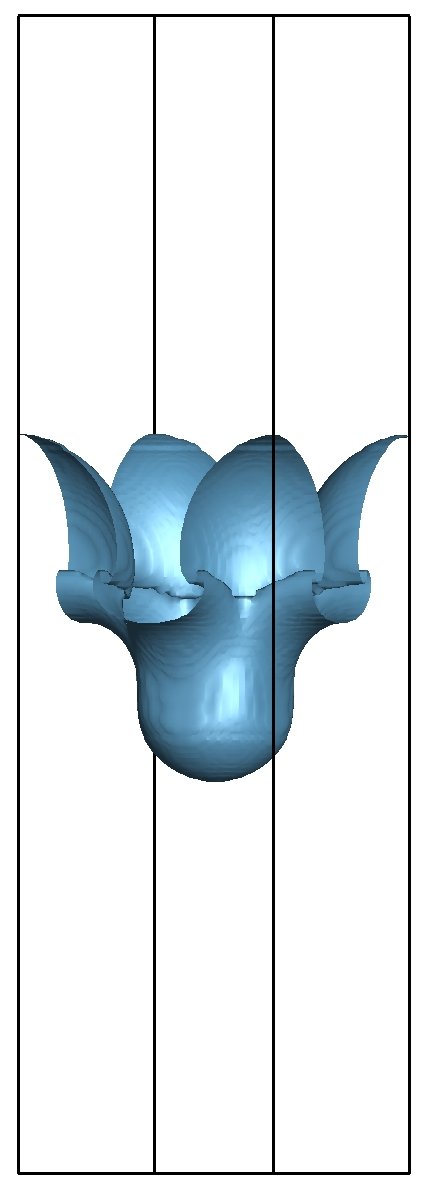}
			\label{fig_3D_RT_evolve_b}
		}
		\subfigure[]
		{
			\includegraphics[height=8cm]{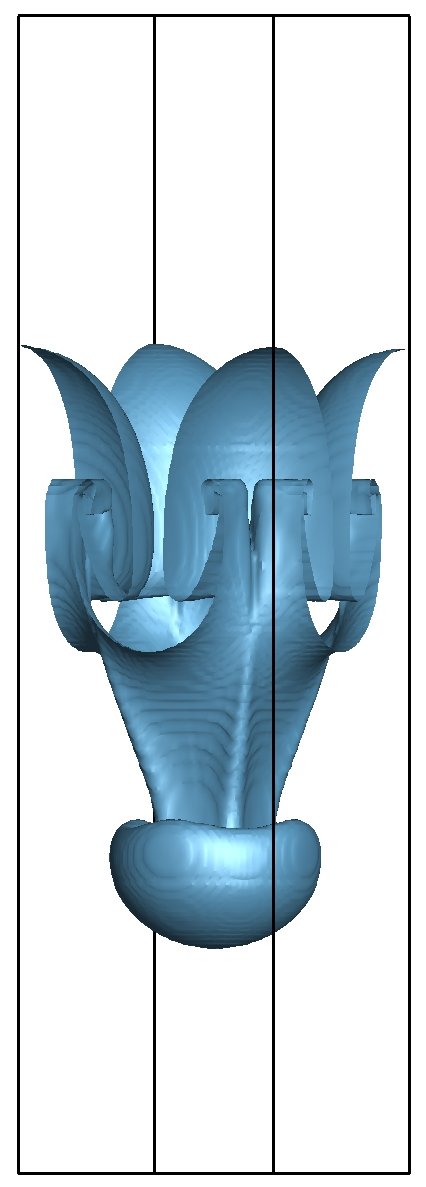}
			\label{fig_3D_RT_evolve_c}
		}
		\subfigure[]
		{
			\includegraphics[height=8cm]{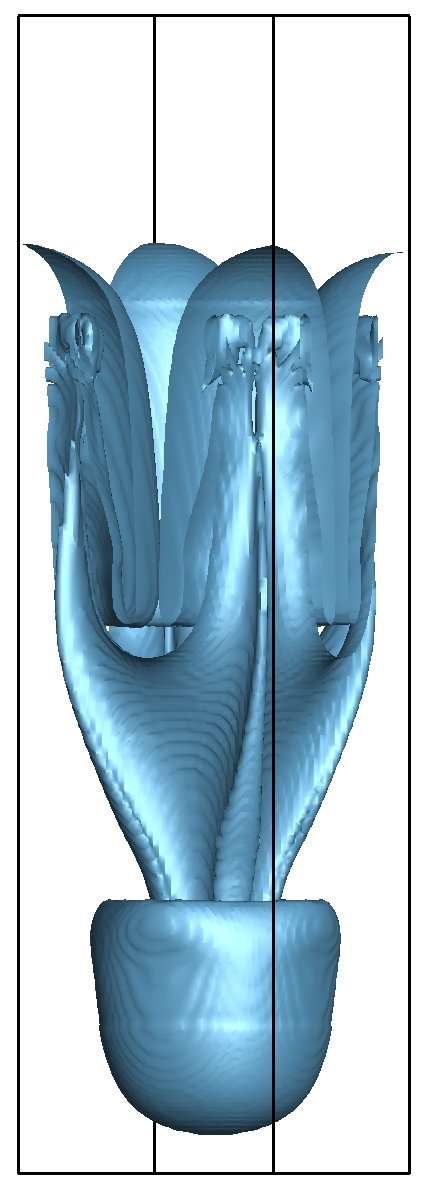}
			\label{fig_3D_RT_evolve_d}
		}\\
		\subfigure[]
		{
			\includegraphics[height=9cm]{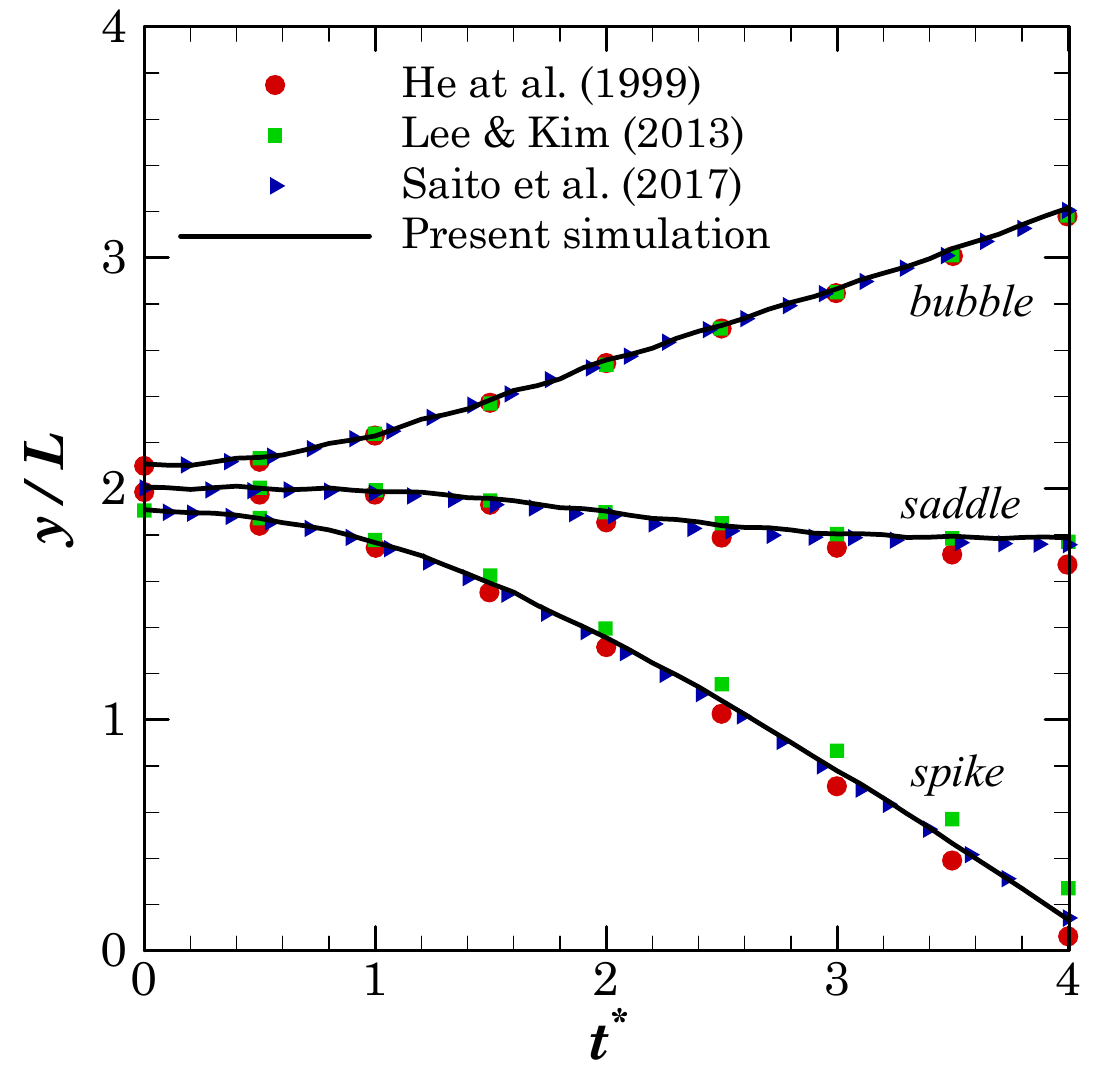}
			\label{fig_3D_RT_evolve_e}
		}
		\caption{Temporal evolution of the interface for RT instability visualized at (a) $t^*=1$  (b) $t^*=2$  (c) $t^*=3$ and (d) $t^*=4$, and (e) evolution of bubble, saddle and spike with time.}
		\label{fig_3D_RT_evolve}
	\end{figure}

	The test case of RT instability was simulated for $4$ non-dimensional time units ($t^*=t\sqrt{g/L}$), and the evolution of the interface is qualitatively assessed at $t^*=1,2,3,4$ with $\text{Ma}=0.05$ and time step of $10^{-4}$. Similar to the 2D test case, the acceleration of heavier fluid into the lighter fluid results in complex topological features. The process can be evidenced in figures \ref{fig_3D_RT_evolve_a}-\ref{fig_3D_RT_evolve_d}. The position of the bubble, saddle and spike (as shown in figure \ref{fig_initial_3D_RT}) are tracked with time and compared with the existing literature~\cite{he1999,saito2017,lee2013}. It is evident from figure \ref{fig_3D_RT_evolve_e} that the evolution of the interface shape and position with time is in excellent agreement with the existing literature.
	
	\begin{figure}[]
		\centering
		{
			\includegraphics[height=8cm]{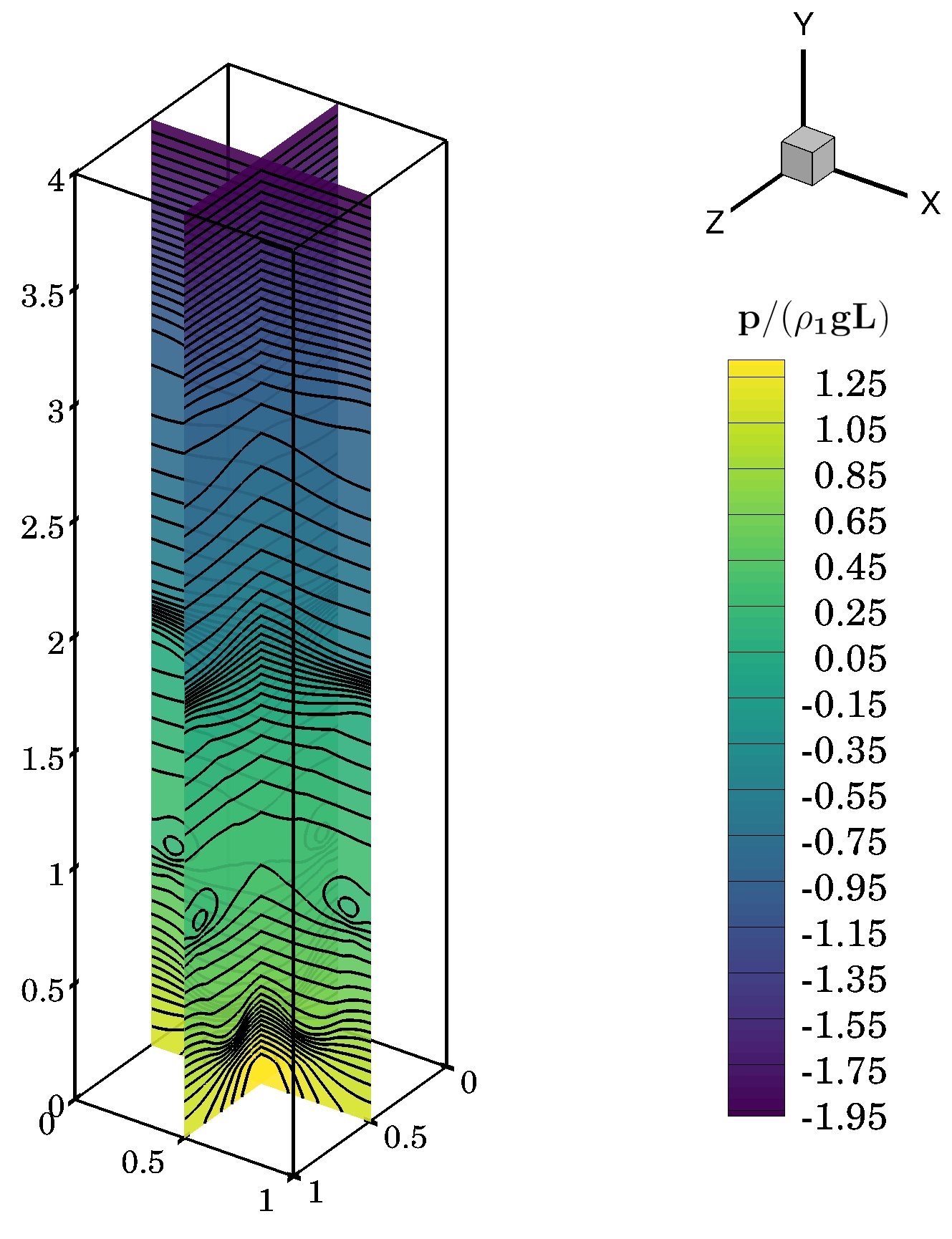}
		}
		\caption{Pressure contours at $t^*=4$ for the RT instability.}
		\label{fig_3D_RT_pressure}
	\end{figure}
	
	As mentioned in the introduction, the only existing GPE-based method to solve two-phase flows is by Huang~\cite{huang2020}, who used two special procedures for a stable computation: the addition of a second viscosity term to eliminate checkerboard oscillations and a certain averaging of pressure for stabilizing 3D flows. However, we emphasize that neither of these procedures is needed in our method for accuracy and stability. Figure \ref{fig_3D_RT_pressure} shows the pressure contour for the 3D Rayleigh-Taylor problem at the final time instant. In the figure, the contour lines depict smooth variation in pressure instead of a checkerboard pattern reported by Huang~\cite{huang2020}. Moreover, the pressure gradient term is discretized in the usual manner and we observed no stability issues. Huang~\cite{huang2020} reported that the simulation stopped abruptly without the special pressure averaging. Thus, we stress that our solver offers simplicity and inherent stability for the accurate simulation of 3D flows. 

	\section{Conclusion}
	\label{sec5}
	The pressure Poisson equation in the incompressible Navier-Stokes solver is replaced with the general pressure equation to simulate the two-phase flow problems. The method is tested against various two-phase configurations, and the results are validated with the existing literature. The validation study reveals that our algorithm accurately solves 2D and 3D problems. Our quantitative results show that the unphysical oscillations, due to pressure acoustics, are significantly smaller than the existing weakly compressible methods. In addition, the existing methods need at least one of the following for numerical stability: different treatments for the interface region, the addition of second viscosity or special averaging techniques for pressure. However, our algorithm does not require such modifications, making it convenient to implement and stable even for the simulations involving high density and viscosity ratios and surface tension effects. Furthermore, thanks to the GPE and OS framework for VOF, the algorithm is \textit{fully-explicit}. Hence, it can be extended to carry out high-performance simulations over multi-node CPUs or GPUs to solve more complex problems. In the world of perpetually developing high-performance resources, we need efficient solvers that can exploit computing power without compromising the accuracy of the problem. Thus, a performance study of our solver in terms of scalability can be pivotal in satisfying the aforementioned need, which serves as the future scope of our work. 

	\nolinenumbers
	
	\section*{Acknowledgements}
We acknowledge the financial support provided by IIT Goa in the form of a startup grant. Simulations are run on computational resources setup from the DST-SERB Ramanujan fellowship~(SB/S2/RJN-037/2018).
	
	\section*{Data Availability}
	The data supporting this study's findings are available from the corresponding author upon reasonable request.
	
\appendix

\section{}
\label{appendix_gpeDreivation}
In this section, we explain the form of the GPE used in the paper. The equation derived by Toutant (equation (31) of~\cite{toutant2017}) is written as,
\begin{equation}
    \frac{\partial p}{\partial t} + \rho {c_s}^2 \frac{\partial u_i}{\partial x_i} = \frac{\kappa}{\rho C_v} \frac{\partial^2 p}{\partial x_i \partial x_i} 
\end{equation}
where $\rho$ is the density, $\mu$ is the dynamic viscosity, $\kappa$ is the thermal conductivity, and $C_v$ is the isochoric heat capacity. We rewrite the above equation using the definitions of heat capacity ratio, $\gamma = C_p/C_v$, thermal diffusivity, $\alpha = \kappa / (\rho C_p)$ and Prandtl number, $\text{Pr} = \mu / (\rho \alpha)$, to obtain,
\begin{equation}
    \frac{\partial p}{\partial t} + \rho {c_s}^2 \frac{\partial u_i}{\partial x_i} = \frac{ \mu \gamma}{\rho \text{Pr}} \frac{\partial^2 p}{\partial x_i \partial x_i}.
\end{equation}

We simplify the above equation by setting $\gamma = \text{Pr}$ to obtain the following
\begin{equation}
    \frac{\partial p}{\partial t} + \rho {c_s}^2 \frac{\partial u_i}{\partial x_i} = \frac{ \mu}{\rho} \frac{\partial^2 p}{\partial x_i \partial x_i}.
\end{equation}
Equating $\gamma$ and $\text{Pr}$ is a common assumption employed in all GPE and EDAC based solvers for simulating single-phase~\cite{toutant2018, dupuy2020, shi2020,pan2022, clausen2013,kajzer2018} and two-phase flows~\cite{yang2021, kajzer2018_conference, kajzer2020}. Clausen~\cite{clausen2013} directly used this assumption while deriving the EDAC equation whereas, Toutant~\cite{toutant2017} retained $\kappa$ and $C_v$ in his formulation of GPE. However, while showcasing the capability of GPE to simulate benchmark test cases~\cite{toutant2018}, he set $\gamma=\text{Pr}$, and this is later followed by all the others working with GPE. Finally, expressing the diffusion term similar to that of the momentum equation, we wrote the above equation in vector notation to get equation (2) presented in the paper.

\bibliographystyle{ieeetr}
\bibliography{ref}

\begin{thebibliography}{10}

\bibitem{hirt1981}
C.~W. Hirt and B.~D. Nichols, ``Volume of fluid ({VOF}) method for the dynamics
  of free boundaries,'' {\em Journal of computational physics}, vol.~39, no.~1,
  pp.~201--225, 1981.

\bibitem{saincher2015}
S.~Saincher and J.~Banerjee, ``A redistribution-based volume-preserving
  {PLIC-VOF} technique,'' {\em Numerical Heat Transfer, Part B: Fundamentals},
  vol.~67, no.~4, pp.~338--362, 2015.

\bibitem{bodhanwalla2022}
H.~Bodhanwalla, C.~Anghan, and J.~Banerjee, ``The effect of one-sided
  confinement on nappe oscillations in free falling liquid sheet,'' {\em
  Physics of Fluids}, vol.~34, no.~12, p.~124107, 2022.

\bibitem{sussman1994}
M.~Sussman, P.~Smereka, and S.~Osher, ``A level set approach for computing
  solutions to incompressible two-phase flow,'' {\em Journal of Computational
  {P}hysics}, vol.~114, no.~1, pp.~146--159, 1994.

\bibitem{gada2011}
V.~H. Gada and A.~Sharma, ``On a novel dual-grid level-set method for two-phase
  flow simulation,'' {\em Numerical Heat Transfer, Part B: Fundamentals},
  vol.~59, no.~1, pp.~26--57, 2011.

\bibitem{gada2012}
V.~H. Gada and A.~Sharma, ``Analytical and level-set method based numerical
  study on oil--water smooth/wavy stratified-flow in an inclined
  plane-channel,'' {\em International Journal of Multiphase Flow}, vol.~38,
  no.~1, pp.~99--117, 2012.

\bibitem{allen1979}
S.~M. Allen and J.~W. Cahn, ``A microscopic theory for antiphase boundary
  motion and its application to antiphase domain coarsening,'' {\em Acta
  {M}etallurgica}, vol.~27, no.~6, pp.~1085--1095, 1979.

\bibitem{mirjalili2020}
S.~Mirjalili, C.~B. Ivey, and A.~Mani, ``A conservative diffuse interface
  method for two-phase flows with provable boundedness properties,'' {\em
  Journal of Computational Physics}, vol.~401, p.~109006, 2020.

\bibitem{dadvand2021}
A.~Dadvand, M.~Bagheri, N.~Samkhaniani, H.~Marschall, and M.~W{\"o}rner,
  ``Advected phase-field method for bounded solution of the {C}ahn-{H}illiard
  {N}avier-{S}tokes equations,'' {\em Physics of Fluids}, vol.~33, no.~5,
  p.~053311, 2021.

\bibitem{tryggvason2011}
G.~Tryggvason, R.~Scardovelli, and S.~Zaleski, {\em Direct numerical
  simulations of gas--liquid multiphase flows}.
\newblock Cambridge university press, 2011.

\bibitem{ferziger2002}
J.~H. Ferziger, M.~Peri{\'c}, and R.~L. Street, {\em Computational methods for
  fluid dynamics}, vol.~3.
\newblock Springer, 2002.

\bibitem{gunstensen1991}
A.~K. Gunstensen, D.~H. Rothman, S.~Zaleski, and G.~Zanetti, ``Lattice
  {B}oltzmann model of immiscible fluids,'' {\em Physical Review A}, vol.~43,
  no.~8, p.~4320, 1991.

\bibitem{he1999}
X.~He, S.~Chen, and R.~Zhang, ``A lattice {B}oltzmann scheme for incompressible
  multiphase flow and its application in simulation of {R}ayleigh-{T}aylor
  instability,'' {\em Journal of {C}omputational {P}hysics}, vol.~152, no.~2,
  pp.~642--663, 1999.

\bibitem{inamuro2004}
T.~Inamuro, T.~Ogata, S.~Tajima, and N.~Konishi, ``A lattice {B}oltzmann method
  for incompressible two-phase flows with large density differences,'' {\em
  Journal of Computational {P}hysics}, vol.~198, no.~2, pp.~628--644, 2004.

\bibitem{wang2015}
Y.~Wang, C.~Shu, H.~Huang, and C.~J. Teo, ``Multiphase lattice {B}oltzmann flux
  solver for incompressible multiphase flows with large density ratio,'' {\em
  Journal of Computational Physics}, vol.~280, pp.~404--423, 2015.

\bibitem{chorin1997}
A.~J. Chorin, ``A numerical method for solving incompressible viscous flow
  problems,'' {\em Journal of {C}omputational {P}hysics}, vol.~135, no.~2,
  pp.~118--125, 1997.

\bibitem{soh1988}
W.~Soh and J.~W. Goodrich, ``Unsteady solution of incompressible
  {N}avier-{S}tokes equations,'' {\em Journal of {C}omputational {P}hysics},
  vol.~79, no.~1, pp.~113--134, 1988.

\bibitem{malan2002}
A.~Malan, R.~Lewis, and P.~Nithiarasu, ``An improved unsteady, unstructured,
  artificial compressibility, finite volume scheme for viscous incompressible
  flows: Part {I}. theory and implementation,'' {\em International Journal for
  Numerical Methods in Engineering}, vol.~54, no.~5, pp.~695--714, 2002.

\bibitem{kelecy1997}
F.~Kelecy and R.~Pletcher, ``The development of a free surface capturing
  approach for multidimensional free surface flows in closed containers,'' {\em
  Journal of {C}omputational {P}hysics}, vol.~138, no.~2, pp.~939--980, 1997.

\bibitem{shah2011}
A.~Shah and L.~Yuan, ``Numerical solution of a phase field model for
  incompressible two-phase flows based on artificial compressibility,'' {\em
  Computers \& Fluids}, vol.~42, no.~1, pp.~54--61, 2011.

\bibitem{ansumali2005}
S.~Ansumali, I.~V. Karlin, and H.~S. {\"O}ttinger, ``Thermodynamic theory of
  incompressible hydrodynamics,'' {\em Physical Review Letters}, vol.~94,
  no.~8, p.~080602, 2005.

\bibitem{karlin2006}
I.~V. Karlin, A.~G. Tomboulides, C.~E. Frouzakis, and S.~Ansumali,
  ``Kinetically reduced local {N}avier-{S}tokes equations: An alternative
  approach to hydrodynamics,'' {\em Physical Review E}, vol.~74, no.~3,
  p.~035702, 2006.

\bibitem{hashimoto2013}
T.~Hashimoto, I.~Tanno, Y.~Tanaka, K.~Morinishi, and N.~Satofuka, ``Simulation
  of doubly periodic shear layers using kinetically reduced local
  {N}avier-{S}tokes equations on a {GPU},'' {\em Computers \& Fluids}, vol.~88,
  pp.~715--718, 2013.

\bibitem{hashimoto2015}
T.~Hashimoto, I.~Tanno, T.~Yasuda, Y.~Tanaka, K.~Morinishi, and N.~Satofuka,
  ``Higher order numerical simulation of unsteady viscous incompressible flows
  using kinetically reduced local {N}avier-{S}tokes equations on a {GPU},''
  {\em Computers \& Fluids}, vol.~110, pp.~108--113, 2015.

\bibitem{hashimoto2018}
T.~Hashimoto, T.~Yasuda, I.~Tanno, Y.~Tanaka, K.~Morinishi, and N.~Satofuka,
  ``Multi-{GPU} parallel computation of unsteady incompressible flows using
  kinetically reduced local {N}avier-{S}tokes equations,'' {\em Computers \&
  Fluids}, vol.~167, pp.~215--220, 2018.

\bibitem{clausen2013}
J.~R. Clausen, ``Entropically damped form of artificial compressibility for
  explicit simulation of incompressible flow,'' {\em Physical Review E},
  vol.~87, no.~1, p.~013309, 2013.

\bibitem{delorme2017}
Y.~T. Delorme, K.~Puri, J.~Nordstrom, V.~Linders, S.~Dong, and S.~H. Frankel,
  ``A simple and efficient incompressible {N}avier-{S}tokes solver for unsteady
  complex geometry flows on truncated domains,'' {\em Computers \& Fluids},
  vol.~150, pp.~84--94, 2017.

\bibitem{kajzer2018}
A.~Kajzer and J.~Pozorski, ``Application of the entropically damped artificial
  compressibility model to direct numerical simulation of turbulent channel
  flow,'' {\em Computers \& Mathematics with Applications}, vol.~76, no.~5,
  pp.~997--1013, 2018.

\bibitem{kajzer2020}
A.~Kajzer and J.~Pozorski, ``A weakly compressible, diffuse-interface model for
  two-phase flows,'' {\em Flow, Turbulence and Combustion}, vol.~105,
  pp.~299--333, 2020.

\bibitem{kajzer2022}
A.~Kajzer and J.~Pozorski, ``A weakly compressible, diffuse interface model of
  two-phase flows: Numerical development and validation,'' {\em Computers \&
  Mathematics with Applications}, vol.~106, pp.~74--91, 2022.

\bibitem{toutant2017}
A.~Toutant, ``General and exact pressure evolution equation,'' {\em Physics
  Letters A}, vol.~381, no.~44, pp.~3739--3742, 2017.

\bibitem{toutant2018}
A.~Toutant, ``Numerical simulations of unsteady viscous incompressible flows
  using general pressure equation,'' {\em Journal of Computational Physics},
  vol.~374, pp.~822--842, 2018.

\bibitem{dupuy2020}
D.~Dupuy, A.~Toutant, and F.~Bataille, ``Analysis of artificial pressure
  equations in numerical simulations of a turbulent channel flow,'' {\em
  Journal of Computational Physics}, vol.~411, p.~109407, 2020.

\bibitem{shi2020}
X.~Shi and C.-A. Lin, ``Simulations of wall bounded turbulent flows using
  general pressure equation,'' {\em Flow, Turbulence and Combustion}, vol.~105,
  no.~1, pp.~67--82, 2020.

\bibitem{huang2020}
J.-J. Huang, ``Numerical simulation of two-phase incompressible viscous flows
  using general pressure equation,'' {\em arXiv preprint arXiv:2011.00814},
  2020.

\bibitem{yang2021}
K.~Yang and T.~Aoki, ``Weakly compressible {N}avier-{S}tokes solver based on
  evolving pressure projection method for two-phase flow simulations,'' {\em
  Journal of Computational Physics}, vol.~431, p.~110113, 2021.

\bibitem{ubbink1999}
O.~Ubbink and R.~Issa, ``A method for capturing sharp fluid interfaces on
  arbitrary meshes,'' {\em Journal of {C}omputational {P}hysics}, vol.~153,
  no.~1, pp.~26--50, 1999.

\bibitem{arote2020}
A.~Arote, M.~Bade, and J.~Banerjee, ``An improved compressive volume of fluid
  scheme for capturing sharp interfaces using hybridization,'' {\em Numerical
  Heat Transfer, Part B: Fundamentals}, vol.~79, no.~1, pp.~29--53, 2020.

\bibitem{anghan2021}
C.~Anghan, M.~H. Bade, and J.~Banerjee, ``A modified switching technique for
  advection and capturing of surfaces,'' {\em Applied Mathematical Modelling},
  vol.~92, pp.~349--379, 2021.

\bibitem{xiao2005}
F.~Xiao, Y.~Honma, and T.~Kono, ``A simple algebraic interface capturing scheme
  using hyperbolic tangent function,'' {\em International Journal for Numerical
  Methods in Fluids}, vol.~48, no.~9, pp.~1023--1040, 2005.

\bibitem{saincher2022}
S.~Saincher and V.~Sriram, ``An efficient operator-split {CICSAM} scheme for
  three-dimensional multiphase-flow problems on {C}artesian grids,'' {\em
  Computers \& Fluids}, vol.~240, p.~105440, 2022.

\bibitem{weymouth2010}
G.~D. Weymouth and D.~K.-P. Yue, ``Conservative volume-of-fluid method for
  free-surface simulations on {C}artesian-grids,'' {\em Journal of
  Computational Physics}, vol.~229, no.~8, pp.~2853--2865, 2010.

\bibitem{parker1992}
B.~Parker and D.~Youngs, {\em Two and three dimensional Eulerian simulation of
  fluid flow with material interfaces}.
\newblock Atomic Weapons Establishment, 1992.

\bibitem{francois2006}
M.~M. Francois, S.~J. Cummins, E.~D. Dendy, D.~B. Kothe, J.~M. Sicilian, and
  M.~W. Williams, ``A balanced-force algorithm for continuous and sharp
  interfacial surface tension models within a volume tracking framework,'' {\em
  Journal of Computational Physics}, vol.~213, no.~1, pp.~141--173, 2006.

\bibitem{cummins2005}
S.~J. Cummins, M.~M. Francois, and D.~B. Kothe, ``Estimating curvature from
  volume fractions,'' {\em Computers \& {S}tructures}, vol.~83, no.~6-7,
  pp.~425--434, 2005.

\bibitem{gottlieb1998}
S.~Gottlieb and C.-W. Shu, ``Total variation diminishing {R}unge-{K}utta
  schemes,'' {\em Mathematics of computation}, vol.~67, no.~221, pp.~73--85,
  1998.

\bibitem{shu1988}
C.-W. Shu and S.~Osher, ``Efficient implementation of essentially
  non-oscillatory shock-capturing schemes,'' {\em Journal of computational
  physics}, vol.~77, no.~2, pp.~439--471, 1988.

\bibitem{gottlieb2009}
S.~Gottlieb, D.~I. Ketcheson, and C.-W. Shu, ``High order strong stability
  preserving time discretizations,'' {\em Journal of Scientific Computing},
  vol.~38, no.~3, pp.~251--289, 2009.

\bibitem{caiden2001}
R.~Caiden, R.~P. Fedkiw, and C.~Anderson, ``A numerical method for two-phase
  flow consisting of separate compressible and incompressible regions,'' {\em
  Journal of Computational Physics}, vol.~166, no.~1, pp.~1--27, 2001.

\bibitem{bassano2003}
E.~Bassano, ``Numerical simulation of thermo-solutal-capillary migration of a
  dissolving drop in a cavity,'' {\em International journal for numerical
  methods in fluids}, vol.~41, no.~7, pp.~765--788, 2003.

\bibitem{parameswaran2023}
S.~Parameswaran and J.~Mandal, ``A stable interface-preserving reinitialization
  equation for conservative level set method,'' {\em European Journal of
  Mechanics-B/Fluids}, vol.~98, pp.~40--63, 2023.

\bibitem{liovic2006}
P.~Liovic, M.~Rudman, J.-L. Liow, D.~Lakehal, and D.~Kothe, ``A 3{D}
  unsplit-advection volume tracking algorithm with planarity-preserving
  interface reconstruction,'' {\em Computers \& {F}luids}, vol.~35, no.~10,
  pp.~1011--1032, 2006.

\bibitem{popinet2009}
S.~Popinet, ``An accurate adaptive solver for surface-tension-driven
  interfacial flows,'' {\em Journal of Computational Physics}, vol.~228,
  pp.~5838--5866, Sept. 2009.

\bibitem{kajzer2018_conference}
A.~Kajzer and J.~Pozorski, ``Diffuse interface models for two-phase flows in
  artificial compressibility approach,'' {\em Journal of Physics: Conference
  Series}, vol.~1101, p.~012013, Oct. 2018.

\bibitem{garoosi2022}
F.~Garoosi and T.-F. Mahdi, ``Numerical simulation of three-fluid
  {R}ayleigh-{T}aylor instability using an enhanced volume-of-fluid ({VOF})
  model: New benchmark solutions,'' {\em Computers \& Fluids}, vol.~245,
  p.~105591, 2022.

\bibitem{martin1952}
J.~C. Martin, W.~J. Moyce, J.~Martin, W.~Moyce, W.~G. Penney, A.~Price, and
  C.~Thornhill, ``Part {IV}. {A}n experimental study of the collapse of liquid
  columns on a rigid horizontal plane,'' {\em Philosophical Transactions of the
  Royal Society of London. Series A, Mathematical and Physical Sciences},
  vol.~244, no.~882, pp.~312--324, 1952.

\bibitem{ling2019}
K.~Ling, S.~Zhang, P.-Z. Wu, S.-Y. Yang, and W.-Q. Tao, ``A coupled
  volume-of-fluid and level-set method ({VOSET}) for capturing interface of
  two-phase flows in arbitrary polygon grid,'' {\em International Journal of
  Heat and Mass Transfer}, vol.~143, p.~118565, 2019.

\bibitem{vanLeer1974}
B.~Van~Leer, ``Towards the ultimate conservative difference scheme. {II}.
  monotonicity and conservation combined in a second-order scheme,'' {\em
  Journal of Computational Physics}, vol.~14, no.~4, pp.~361--370, 1974.

\bibitem{hysing2009}
S.~Hysing, S.~Turek, D.~Kuzmin, N.~Parolini, E.~Burman, S.~Ganesan, and
  L.~Tobiska, ``Quantitative benchmark computations of two-dimensional bubble
  dynamics,'' {\em International Journal for Numerical Methods in Fluids},
  vol.~60, no.~11, pp.~1259--1288, 2009.

\bibitem{ayachit2015paraview}
U.~Ayachit, {\em The paraview guide: a parallel visualization application}.
\newblock Kitware, Inc., 2015.

\bibitem{strubelj2009}
L.~{\v{S}}trubelj, I.~Tiselj, and B.~Mavko, ``Simulations of free surface flows
  with implementation of surface tension and interface sharpening in the
  two-fluid model,'' {\em International Journal of Heat and Fluid Flow},
  vol.~30, no.~4, pp.~741--750, 2009.

\bibitem{saito2017}
S.~Saito, Y.~Abe, and K.~Koyama, ``Lattice {B}oltzmann modeling and simulation
  of liquid jet breakup,'' {\em Physical Review E}, vol.~96, no.~1, p.~013317,
  2017.

\bibitem{lee2013}
H.~G. Lee and J.~Kim, ``Numerical simulation of the three-dimensional
  {R}ayleigh-{T}aylor instability,'' {\em Computers \& Mathematics with
  Applications}, vol.~66, no.~8, pp.~1466--1474, 2013.

\bibitem{pan2022}
D.~Pan, ``A high-order finite volume method solving viscous incompressible
  flows using general pressure equation,'' {\em Numerical Heat Transfer, Part
  B: Fundamentals}, vol.~82, pp.~146--163, Nov. 2022.

\end{thebibliography}

\end{document}